\documentclass[journal]{vgtc}                     

\vgtccategory{Research}

\author{%
  Guoruizhe Sun\textsuperscript{*},
  Yueqiao Chen\textsuperscript{*},
  Emily Guo,
  Yutong Yao, and
  \authororcid{Dongyu Liu}{0000-0002-8915-2785}
}

\authorfooter{
  \item
  	All authors are with the University of California at Davis. E-mail: \{grzsun, yeqchen, emiguo, ytyao, dyuliu\}@ucdavis.edu; 
  \item \textsuperscript{*}Guoruizhe Sun and Yueqiao Chen contributed equally to this work.
}

\abstract{Searching for time-series segments that match user-defined patterns is important in domains such as finance, climate science, and healthcare. However, existing visual query tools often struggle to support vague, composite, or fuzzy pattern descriptions, often requiring users to express their intent through precise sketches or rigid structured filters. We present \system, a coordinated natural-language and sketch-based querying system for univariate time-series pattern search. Rather than treating text and sketch as a fused input stream, \system~uses them as complementary representations of analytic intent: natural language supports semantic and compositional pattern descriptions, while sketching supports direct geometric refinement. The two modalities are linked through a shared visual context, editable feature representations, and synchronized result views, enabling users to move between text and sketch during iterative query formulation. At its core is an LLM-based semantic parsing pipeline that translates free-form natural-language queries into interpretable and editable shape-feature constraints. We evaluate \system~through two usage scenarios, a user study with failure-case analysis, and an assessment of the LLM-based semantic parsing pipeline. The results show that \system~supports effective time-series pattern search, with natural language serving as an accessible entry point and sketching providing a complementary mechanism for refinement and recovery when textual specifications are insufficient.
}


\teaser{
  \centering
  \includegraphics[width=\textwidth]{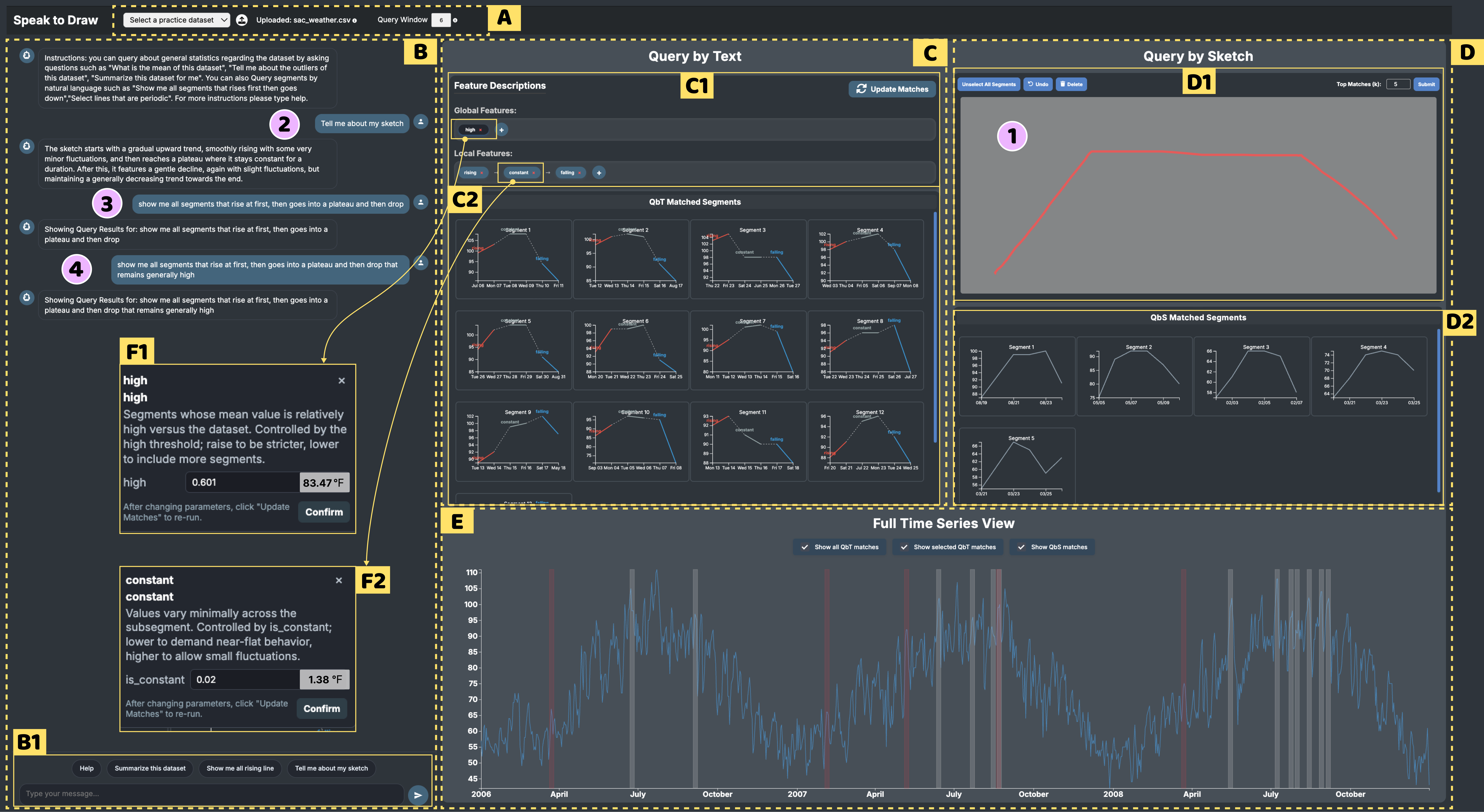}
  \caption{User interface of \system, shown with the Sacramento Weather dataset from 2006 to 2008. \textbf{Control Panel} (\textbf{A}), where users import a dataset and set the query window length. \textbf{Chat View} (\textbf{B}), which supports natural-language queries via an LLM; \textbf{B1} is the input box for user queries. \textbf{Query View by Text} (\textbf{C}), which shows results for natural-language and feature-based queries; \textbf{C1} is the Feature Panel for adjusting feature selections and \textbf{C2} displays snapshots of matched time-series windows. \textbf{Query View by Sketch} (\textbf{D}), which supports sketch-based queries; \textbf{D1} is the sketch canvas and \textbf{D2} shows matched windows for the sketch. \textbf{Full Time Series View} (\textbf{E}), which provides an overview of the entire time series and highlights matches from each query modality. \textbf{F1} presents the parameter-adjustment view for global feature, while \textbf{F2} is for local feature. The figure also illustrates a typical exploration loop for finding suitable weather to visit Sacramento: \textbf{1.} sketch to obtain initial matches, \textbf{2.} ask the LLM to describe the sketch, and \textbf{3\&4.} refine queries using the resulting natural-language description.}
  \label{fig:ui}
  \label{fig:teaser}
}

\graphicspath{{./}} 

\usepackage{tabu}                      
\usepackage{booktabs}                  
\usepackage{lipsum}                    
\usepackage{mwe}                       
\usepackage{ccicons}                   

\usepackage{mathptmx}                  

\usepackage{amsmath} 
\usepackage{xcolor}

\definecolor{dypink}{HTML}{ec008c}
\definecolor{dypurple}{HTML}{8654d1}

\usepackage{longtable}
\usepackage{listings}

\lstdefinestyle{codeblock}{
    backgroundcolor=\color{gray!10},
    basicstyle=\ttfamily\footnotesize,
    numbers=left,
    numbersep=12pt,
    xleftmargin=3em,
    keepspaces=true,
    tabsize=2,
    breaklines=true,
    breakatwhitespace=false,
    escapeinside={(*@}{@*)} 
}
\lstnewenvironment{CodeBlock}{
    \lstset{style=codeblock}
}{}

\definecolor{rzgreen}{HTML}{009900}

\newif\ifshowrev
\showrevfalse

\definecolor{gzpurple}{HTML}{B57EDC}

\newif\ifshowgz
\showgzfalse

\newcommand{\gz}[1]{%
  \ifshowgz
    {\color{gzpurple}#1}%
  \else
    {#1}%
  \fi
}

\usepackage{marginnote} 

\renewcommand*{\marginnote}[1]{}

\usepackage{xparse}
\newif\ifshowdelta
\showdeltafalse

\newcommand{\system}{ShapeTalk}

\title{ShapeTalk: Combining Natural Language and Sketch for Time-Series Pattern Querying}

\begin{document}
\maketitle


\section{introduction}
Time-series query refers to retrieving segments that match a user-defined temporal pattern and is a fundamental task in many domains. Analysts often search large time-series collections for specific patterns or anomalies in finance~\cite{EdwardsMageeBassetti2018}, transportation~\cite{li2015trend}, climate science~\cite{mudelsee2019trend}, healthcare~\cite{amini2021trend}, and beyond. For example, an investment analyst may ask for ``\textit{stocks that tanked in 2020},'' seeking trajectories that exhibit a sharp drop during that period. Broadly, analysts across domains rely on pattern search to identify salient temporal behaviors in large datasets.

A variety of interactive paradigms have been developed to support time-series pattern search, including query-by-sketch interfaces~\cite{correll2016semantics}, visual query builders based on examples and constraint parameter or relation settings~\cite{buono2005interactive, hochheiser2002interactive, liu2024relation}, and pattern-based search tools~\cite{yu2025noops,lekschas2020peax}.
These systems enable users to specify temporal patterns visually or through structured controls, but they often require users to translate their questions into precise sketches, parameterized filters, or low-level pattern specifications. This translation can be difficult when patterns are fuzzy, composite, or only partially formed in the user's mind.

Natural language (NL) offers a more flexible way to express such intent. A user can ask for ``\textit{segments that stay high for a while and then gradually decrease}'' without knowing a query syntax or predefined parameter set. However, interpreting such queries remains challenging because everyday language is often ambiguous or underspecified with respect to precise geometry. Terms such as ``\textit{gradual},'' ``\textit{bump},'' ``\textit{spike},'' or ``\textit{tanking}'' may imply different slopes, magnitudes, durations, or shapes depending on context. Mapping colloquial descriptions to rigorous time-series query constraints therefore requires both disambiguation and contextual understanding.

A common way to address this challenge is to define a controlled vocabulary and map NL queries to a predefined set of labels~\cite{shapesearch,Imani2021Qute,bendeck2024slopeseeker}. For example, SlopeSeeker~\cite{bendeck2024slopeseeker} uses a crowdsourced lexicon that links linguistic descriptors to quantifiable slope and shape properties, and supports search over visually salient events annotated with these labels. \gz{While effective, this approach can still place a burden on users to express their intent in terms the system already knows. In practice, users may describe temporal patterns using personal, domain-specific, or partially formed language, such as ``heatwave-like,'' or ``mostly stable with a small bump,'' which may not be within fixed set of labels.} Moreover, text-only interaction can be less intuitive than direct manipulation for iterative visual search. These limitations motivate a more flexible approach that supports open-ended pattern descriptions and iterative refinement across modalities.

\gz{Recent advances in LLMs create new opportunities for time-series pattern search by enabling richer forms of query specification. LLMs can parse free-form textual descriptions and extract structured attributes, such as direction, magnitude, duration, and repetition, which can then be operationalized as pattern constraints over time series. In this work, we present \emph{\textbf{\system}}, a coordinated natural-language and sketch-based querying system for univariate time-series pattern search.}

\gz{Rather than forcing these modalities into a single retrieval pipeline, \system~uses two complementary query pipelines: NL queries are translated into editable shape-feature constraints, whereas sketch queries are matched through direct shape similarity. The two modalities are coordinated through a shared interaction context and multi-view visualization interface, allowing users to move fluidly between them during query formulation and refinement.}

\gz{In this work, we use the term \emph{multimodal} to refer to a coordinated, cross-modal workflow rather than simultaneous multimodal fusion. \system~does not combine text and sketch into a single early-fusion model or require users to provide both inputs at the same time. Instead, it treats natural language and sketch as complementary ways of externalizing pattern intent. Natural language is useful for expressing semantic, relative, and compositional constraints, whereas sketching is useful for directly specifying or correcting geometric details. This design intentionally keeps the two retrieval pipelines separate while tightly linking their outputs through shared context, editable constraints, and coordinated visual feedback. Thus, multimodality in \system~is optional and opportunistic rather than mandatory.}

This design reflects a key observation: specifying a desired pattern remains an interpretive task regardless of modality. \gz{Users may begin with an imprecise visual concept, a partial verbal description, or both. Natural language does not eliminate this difficulty, and sketching alone does not always provide sufficient precision. \system~therefore supports pattern search as an iterative process of expressing, inspecting, and refining user intent across modalities. Users can begin with an NL query to obtain an initial set of candidate segments, then refine the results by editing extracted features or sketching over representative examples. Conversely, users who start with query-by-sketch can ask the system to ``describe my sketch'' in words, turning a visual query into an interpretable and reusable textual specification. Unlike approaches that rely on predefined linguistic labels or offline event libraries, \system~directly maps user utterances to editable feature constraints through an LLM-driven semantic parsing pipeline. As a result, \system~is adaptable to new datasets without requiring a fixed pattern vocabulary or pre-annotated events, supports fine-grained combinations of local and global shape characteristics, and produces explicit query representations that users can inspect and revise.}

The key contributions of this work are as follows:

\begin{itemize}[nosep]
\item 

\gz{\textbf{A coordinated natural-language and sketch-based interface for time-series pattern search} that supports cross-modal query refinement. \system~links two complementary query pipelines through shared visual context, editable feature constraints, and synchronized result views.}

\item \gz{\textbf{An inspectable LLM-driven semantic parsing pipeline} that translates free-form natural-language pattern descriptions into editable local and global shape-feature constraints, enabling users to review, correct, and refine the system's interpretation during exploration.}

\item \gz{\textbf{An evaluation} through usage scenarios, a user study, failure-case analysis, and parser assessment, showing how users employed natural language as an efficient entry point and used sketching or feature edits as complementary refinement mechanisms.}
\end{itemize}

\section{related work}

\subsection{Time Series Query}

Different interactive paradigms have been explored to specify queries for time-series retrieval. 
For example, \textbf{visual query builders} provide structured interactions such as ``timeboxes'' that constrain values over a time window, query-by-example systems that find patterns similar to a selected segment~\cite{buono2005interactive,hochheiser2002interactive,liu2024relation,querylines}, and workflows in which users annotate a small number of segments of interest and algorithms recommend additional candidates~\cite{yu2025noops,lekschas2020peax}. 
These approaches effectively leverage visual recognition, but they can make it difficult to specify complex, qualitative, or partially formed patterns when users cannot readily translate their intent into precise graphical constraints.

\textbf{Sketch-based interfaces} allow users to draw a desired pattern and retrieve similar shapes. This interaction is intuitive and direct, but sketches are inherently imprecise: the same drawing may admit multiple interpretations, and robust matching remains challenging~\cite{lee2019you}. 
Researchers have explored a variety of query-matching methods, including distance-based measures (e.g., Euclidean distance, DTW) and perception-oriented approaches, to capture patterns that are tolerant to scale or shifts and yield useful results for users~\cite{berndt1994using,correll2016semantics,deepsketch,mannino2018expressive,Fan2021SketchLSTM}.

\textbf{Text- and natural-language-based interfaces} provide a more flexible way to describe temporal patterns. Early systems such as Psaila et al.~\cite{psaila1995querying} supported declarative text-based queries, while more recent systems, including ShapeSearch~\cite{shapesearch}, Qute~\cite{Imani2021Qute}, and SlopeSeeker~\cite{bendeck2024slopeseeker}, define a time-series vocabulary or shape algebra (which we refer to collectively as \emph{shape features}) and map user descriptions onto these predefined constructs. 
These systems demonstrate the promise of language-based querying, but they typically rely on manually specified lexical categories and rule-based mappings. As a result, they can be brittle when users describe patterns in open-ended, compositional, or dataset-specific ways, and they often provide limited support for revising the system's interpretation once an initial query has been issued.

Our work builds on this line of research but differs in two key respects. First, we use an LLM-driven parser to translate free-form natural-language descriptions into an explicit, interpretable, and editable shape-feature representation, reducing reliance on manually enumerated vocabularies and rigid parsing templates. Second, we do not treat language as a standalone query modality, but integrate it with sketch-based interaction in a shared workflow. 

We do not claim an unbounded feature space;
rather, our contribution lies in enabling a more \emph{flexible specification and refinement process}. Natural-language input is mapped to explicit feature constraints, grounded in prior literature and calibrated through our pilot formative study, which users can inspect and edit. Sketch input then provides a complementary path for correcting or refining geometric details.

\subsection{Multimodal Visualization Interfaces}
Natural language interfaces (NLIs) have long been explored in visualization~\cite{kavaz2023chatbot,shen2023towards,Narechania2021nl4dv,wang2023towards} to support tasks such as chart generation, filtering, and querying. Earlier systems typically relied on traditional NLP techniques, including rule-based or template-driven parsers, which limited the coverage and flexibility of user input. Recent advances in LLMs have expanded the role of language in visualization, enabling more open-ended, interactive, and assistive workflows~\cite{vazquez2024are,ye2024generative,Agarwal2025ARO,Hoque2025NLGViz}. These systems support a wide range of tasks, including chart authoring, captioning, insight generation, and visual analytics assistance~\cite{dibia-2023-lida,Tian2025ChartGPT,wang2023llm4vis,li2024prompt4vis,chen2025interchat,tang2023vistext,shen2024data,sultanum2023datatales,shi2024nl2color,Xiao2024284,zhao2024leva,zhao2024lightva,Qiu2025SmartMLVsLM,lee2025videe}. Our work contributes to this space by studying an LLM-powered language interface specialized for time-series pattern search and tightly coupling it with sketch-based interaction.

More broadly, multimodal interaction has long been used to support more natural and effective human-computer interaction by combining complementary input channels~\cite{Turk2014MultimodalIA}. Early systems such as ``Put-That-There''~\cite{bolt1980put} demonstrated the value of integrating speech and pointing, and later work extended this idea through combinations of voice, gesture, gaze, and touch in mobile, immersive, and visualization settings~\cite{alhargan2017multimodal,kim2021data,yang2022hybridtrak,brone2023mobile}. 
Recent LLM-based systems further suggest the potential of language-centered multimodal workflows~\cite{masson2024directgpt,chen2025interchat}. Our work extends this direction to time-series search by treating language and sketch not as redundant alternatives, but as complementary modalities for query specification and iterative refinement.

\subsection{Time Series Visualization}
A wide range of visualization techniques has been developed for time-series data, differing in how they encode temporal structure and support analytic tasks~\cite{aigner2011visualization,brehmer2016timelines}. Line and area charts remain the most common representations because they provide a direct view of temporal trends, while alternative designs support more specialized needs, such as revealing cyclic patterns with spiral layouts~\cite{weber2001visualizing}, emphasizing dates with calendar-based views~\cite{van1999cluster}, and comparing multiple series through superposed lines, braided graphs, or small multiples~\cite{javed2010graphical}. Other approaches use glyphs to provide compact summaries of dynamic behavior~\cite{fuchs2013evaluation} or employ multiple coordinated views to support interactive exploration from complementary perspectives~\cite{liu2022mtv}. Building on these foundations, visual analytics systems for time-series data further integrate task-specific interactions to help users identify salient events, compare temporal patterns, and relate signals to contextual information~\cite{ruta2019sax,Yu2025InclusiViz,wang2024visual}.

Our work builds on this literature by focusing not on new visual encodings, but on multimodal query specification and refinement for time-series pattern search within a coordinated \mbox{visualization interface.}

\section{Design Goals and Rationale}
\label{sec:design_goals}
\gz{The following design requirements were derived from three sources: limitations identified in prior time-series query and natural-language visualization systems, observations from two formative pilot studies (see \cref{sec:nl_pipeline}), and iterative prototyping of \system. We do not present these requirements as novel design principles in themselves. Rather, they specify the interaction and system properties needed to support inspectable natural-language and sketch-based pattern querying in our target setting.}

\vspace{0.5em}
\noindent
\gz{\textbf{DG1. Open-Ended Natural Language Pattern Queries.}
\system~should allow users to describe desired time-series patterns in their own words, without requiring them to learn a fixed command syntax, select from a predefined vocabulary, or translate an early-stage idea into low-level parameter settings. This goal is motivated by prior text-based time-series query systems that rely on shape vocabularies or rule-based mappings~\cite{shapesearch,Imani2021Qute,bendeck2024slopeseeker}. These systems demonstrate the value of language for pattern search, but their coverage is bounded by manually specified labels and parsing rules; for example, SlopeSeeker~\cite{bendeck2024slopeseeker} maps terms such as ``tanking'' and ``spiking'' to crowdsourced slope descriptors, but users may still express similar intent through domain-specific, metaphorical, or partially formed phrases. Our formative pilot further showed that participants used varied descriptions for visually similar patterns, including terms such as \textit{oscillating}, \textit{sine wave}, and \textit{keeps going up and down}. DG1 therefore specifies the desired interaction property---open-ended linguistic specification---rather than a particular parsing method. In \system, we later instantiate this goal with an LLM-based parser because it can map varied phrasings to an explicit feature representation while keeping the interpretation inspectable and editable.}

\vspace{0.5em}
\noindent

\gz{\textbf{DG2. Composite Pattern Specification Across Local and Global Characteristics.}
\system~should support pattern queries that combine local temporal behaviors with segment-level context when users need both. We use \emph{local} characteristics to refer to behaviors over subregions of a query window, such as rising, falling, flattening, or changing curvature, and \emph{global} characteristics to refer to properties of the whole retrieved segment, such as being generally high, low, typical, or unusual relative to the dataset. This goal is motivated by the representational limits of existing query paradigms rather than by the claim that every task requires both kinds of constraints. Sketch-based interfaces are effective for specifying approximate geometry, but they do not naturally expose dataset-relative constraints such as ``high'' or ``typical.'' Conversely, lexicon- or rule-based systems can support named shape features~\cite{shapesearch,Imani2021Qute,bendeck2024slopeseeker}, but often make it cumbersome to combine ordered local stages with whole-window context. In our formative examples and pilot descriptions, users mixed these levels of intent, for example by describing a sequence such as ``rise, plateau, then drop'' while also wanting it to occur at a high temperature or within a typical range. DG2 therefore focuses on supporting this compositional form of pattern intent when it arises, while leaving the concrete feature vocabulary and matching strategy to the system implementation.}

\vspace{0.5em}
\noindent
\gz{\textbf{DG3. Cross-Modal and Iterative Refinement of Queries.}
\system~should support iterative movement between natural-language and sketch-based query representations. We do not aim to enforce simultaneous text--sketch input or to fuse both modalities into a single interpretation model. Instead, our goal is to let users choose the modality that best matches their current state of intent: natural language for semantic and compositional descriptions, sketching for direct geometric specification, and feature editing for precise adjustment. This cross-modal workflow supports refinement when one representation is ambiguous, incomplete, or difficult to control.}

\vspace{0.5em}
\noindent
\gz{\textbf{DG4. Transparent and Editable Query Interpretation.}
\system~should make its interpretation of user queries explicit, interpretable, and editable. Opaque handling of natural language can undermine trust and make it difficult for users to diagnose errors or refine their queries. Prior multimodal systems highlight the value of clearly linking user inputs to system outputs to improve interpretability and usability~\cite{chen2025interchat}. Accordingly, the system should expose an intermediate query representation, such as shape features and constraints over trend direction, magnitude, duration, and segment-level context, that users can inspect and adjust. Treating this intermediate representation as a first-class artifact allows analysts to correct misinterpretations and better understand how the system operationalizes their descriptions, independent of the specific parsing technique used to produce it.}

\vspace{0.5em}
\noindent
\textbf{DG5. Coordinated Visual Feedback for Query and Results.}
\system~should provide tightly coordinated textual and visual feedback to help users interpret results and maintain context during refinement. Prior work shows that linking query specifications to highlighted patterns reduces cognitive effort and supports insight, for example by connecting textual trend labels to chart regions~\cite{bendeck2024slopeseeker,yu2025noops}. Accordingly, \system~should coordinate feature controls, retrieved segments, sketch views, and the full time-series overview so that edits in one view are reflected immediately in the others. Such feedback gives users a clear mapping between their queries, the system's interpretation, and the resulting matches.

\section{LLM-driven Shape Feature Extraction Pipeline}
\label{sec:nl_pipeline}

\begin{figure*}[htbp]
  \centering
  \includegraphics[width=0.85\textwidth]{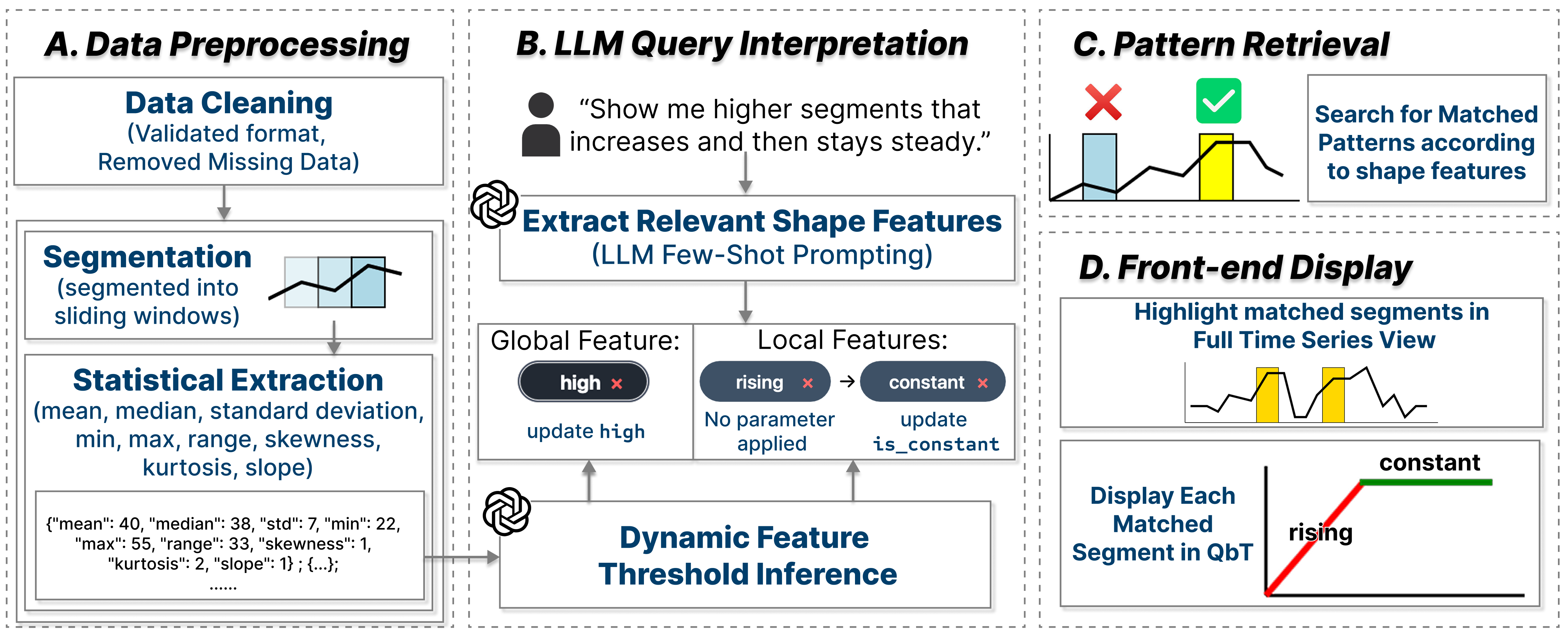}
      \caption{NL Query Pipeline. The pipeline preprocesses the data (A), uses LLMs to extract shape features and infer dynamic feature thresholds from the query and processed data (B), retrieves matching time-series patterns (C), and visualizes the results in the frontend (D).}
  \vspace{-0.2in}
  \label{fig:nl_workflow}
\end{figure*}

This section describes the internal mechanism of the NL processing pipeline in detail. The pipeline translates free-form queries into explicit, statistically defined shape features and then retrieves matching time-series segments (Figure~\ref{fig:nl_workflow}). This pipeline operationalizes \textbf{DG1} by allowing users to describe patterns in their own words, \textbf{DG2} by supporting both local and global features within a query, and \textbf{DG4}/\textbf{DG5} by producing explicit feature constraints that are tightly coordinated with visual feedback in the interface.

\vspace{0.5em}
\noindent\gz{\textbf{Scope and assumptions.}
\system~targets fixed-window pattern retrieval over univariate time series, instead of general time-series question answering or event discovery with unknown temporal extent. We make this choice deliberately for three reasons. First, fixed-length windows make retrieved segments comparable across Query by Text and Query by Sketch. Second, they provide users with an explicit temporal scale, which is important because the same linguistic descriptor can refer to different patterns at different time scales. Third, they keep matching interpretable and responsive enough for interactive refinement. \system~therefore does not aim to answer arbitrary natural-language questions over time-series data; instead, it focuses on helping users express, inspect, and refine qualitative shape-pattern constraints within a user-controlled temporal window.}

\subsection{Data Preprocessing}
Upon dataset upload, the system performs preprocessing that includes schema validation, data cleaning, and preparation for LLM-driven feature-threshold tuning. We assume a univariate time series with two columns: the first encodes the temporal index (e.g., date or timestamp), and the second contains the value of interest. The system verifies this schema and rejects incompatible inputs. To ensure consistent query behavior, rows with missing values in the target column are dropped. 
We also prepare the series for segmentation and statistical extraction used in our dynamic threshold inference step (Section~\ref{sec:threshold_tuning}). Specifically, we sort the time index and remove duplicates so that subsequent sliding-window operations yield well-defined, contiguous patches whose statistics  summarize the underlying temporal structure.

\subsection{LLM Query Interpretation}
When a user issues a NL query, the system uses an LLM to translate the description into an intermediate representation consisting of \emph{global} and \emph{local} shape features. This design addresses \textbf{DG1} by avoiding a fixed query syntax and \textbf{DG4} by surfacing the interpreted features for inspection and editing.

\gz{Within this scoped formulation, the role of the LLM is not to replace sketch-based search or to infer arbitrary temporal events from raw data. Rather, the LLM serves as a semantic compiler that maps colloquial pattern descriptions into explicit, editable constraints over a shape-feature space. This is useful even when some queries resemble sketchable shapes, because language can express aspects that are difficult to specify through sketch alone, including dataset-relative level constraints (e.g., high, low, typical, unusual), ordered semantic compositions, approximate durations, and reusable descriptions of a pattern. Sketching, in contrast, provides direct geometric control but does not by itself expose these semantic constraints.}

Formally, given a univariate time series $\{x_t\}_{t=1}^T$ and a user-specified window length $L$, we construct overlapping windows
\[
    w_i = (x_i, x_{i+1}, \dots, x_{i+L-1}), \quad i = 1,\dots, T-L+1.
\]
Each window $w_i$ is evaluated against a set of shape features $\mathcal{F} = \{f_1,\dots,f_K\}$, where each feature $f_k$ is a predicate
\[
    f_k(w_i; \theta_k) \in \{0,1\}
\]
parameterized by a threshold vector $\theta_k$ (e.g., slope cutoffs).
A window is considered a match if it satisfies all active features:
\[
    \text{match}(w_i) = 1 \iff \bigwedge_{f_k \in \mathcal{F}_\text{active}} f_k(w_i; \theta_k) = 1.
\]
Here, $\mathcal{F}_\text{active} = (f_1, \dots, f_m)$ is an \textbf{ordered list} of features derived from the user's description. We enforce that these features occur in the same temporal order within the window.

\vspace{0.5em}
\noindent\textbf{LLM few-shot prompting.}
To derive $\mathcal{F}_\text{active}$ from a free-form query, we use a few-shot prompt that enumerates our shape-feature vocabulary and provides example query--feature pairs. We distinguish between \emph{local} features, which describe segment-level characteristics (e.g., ``rising,'' ``falling,'' and ``concave''; see the full set in Appendix~A), and \emph{global} features, which describe overall segment-level classifications and are restricted to four labels: ``high,'' ``low,'' ``typical,'' and ``unusual.'' Users can phrase queries in colloquial language (e.g., ``show me higher segments that increase and then stay steady,'' Figure~\ref{fig:nl_workflow}-B) without memorizing keywords or syntax (\textbf{DG1}). Our initial shape-feature definitions are adapted from QUTE~\cite{Imani2021Qute}. The LLM (GPT-4o in our case) returns a Python dictionary specifying global features (e.g., \texttt{"typical"}) and \textbf{ordered} groups of local features (e.g., \texttt{("smooth")}, \texttt{("falling")}). This structure directly encodes composite, multi-stage patterns that combine local and global characteristics (\textbf{DG2}).

\vspace{0.5em}
\noindent\textbf{Pilot study to refine prompts.}
We conducted a pilot study to refine our few-shot prompts. We recruited 9 participants. Two participants had few years of experience in visual analytics, while the remaining participants had no prior experience. The survey presented participants with images of time-series patterns and asked them to describe each pattern in their own words. Participants used a wide range of terms; for example, a periodic weather pattern was described as \textit{oscillating}, \textit{periodic}, \textit{sine wave}, \textit{keeps going up and down}, and \textit{regular seasonality}. 
These observations support our decision to allow flexible, free-form queries rather than constraining users to a predefined vocabulary or syntax. Of the 9 participants, 3 continued to a second phase. Their NL descriptions from the first phase were processed through our LLM-based parsing pipeline to generate shape features. After being introduced to our shape-feature parser, these participants rated how strongly they agreed or disagreed with the LLM's translations of their descriptions into formal features.
Feedback from this study guided refinement of our few-shot examples in our LLM prompts. 

Originally, we proposed 20 examples; in the final version, 6 were edited, 3 were removed, and 8 were added. The edited examples tended to be overly complex, combining multiple shape features and specialized terminology that non-experts would be unlikely to use, which often confused the LLM's classification process. The newly added examples were designed to better reflect users' natural language. We also revised prompts to clarify distinctions among similar shape features such as ``spiky,'' ``complex,'' and ``noisy,'' improving the LLM's consistency. Overall, these refinements improved the parser's ability to translate free-form descriptions into our interpretable local and global feature set, enabling flexible specification and iterative refinement. We assess this capability in Section~\ref{sec:llm-parsing-quality}.

\subsection{Dynamic Threshold Inference}
\label{sec:threshold_tuning}
A central challenge in operationalizing shape features is choosing thresholds that generalize across datasets with different magnitudes and volatilities. For example, a segment that appears ``constant'' in a highly volatile series may be considered ``noisy'' in a more stable one. \gz{Users may choose fixed thresholds when they have prior knowledge of a dataset or exact parameters for certain features; otherwise, the system can use the LLM to infer thresholds from dataset-level statistics rather than fixing them a priori.}

\vspace{0.5em}
\noindent\textbf{Problem formulation.}
For each feature, we infer dataset-specific parameters so that predicates such as
\[
    \text{is\_constant}(w_i; \theta_{\text{const}}) : \operatorname{Var}(w_i) < \theta_{\text{const}}
\]
reflect the local meaning of terms such as ``constant'' for the current dataset. Instead of hand-tuning these values, we estimate them from a compact statistical summary of the series.

\vspace{0.5em}
\noindent\textbf{Segmentation and statistical extraction.}
Sending the raw time series to the LLM is impractical because of context-window limits and the difficulty of reasoning over long numeric sequences. Instead, we adopt a patch-based summarization strategy inspired by ChartInsighter~\cite{wang2025chartinsighter}, which shows that LLMs can reason effectively over aggregated statistical descriptors of chart regions.
As shown in Figure~\ref{fig:nl_workflow}-A, we segment the time series into patches using a sliding patch. For each patch, we compute descriptive statistics including the mean, median, standard deviation, minimum, maximum, range, skewness, kurtosis, slope, volume, and length. Volume is used only for local features. 

To keep token usage manageable, we express numeric values in scientific notation with two decimal places and serialize each patch summary as a JSON object keyed by statistic name.
Because LLMs have bounded context, we cap the number of summarized patches at 300 for local feature thresholding and 100 for global feature thresholding. For a given dataset, we choose the patch size and step size so that: (1) the step size equals the patch size, yielding non-overlapping patches and avoiding redundant, highly correlated inputs; and (2) the number of patches remains within the cap while still covering the full series. Using different patch budgets for local and global features reflects that global features depend more on coarse segment-level trends, whereas local features require finer-grained variation (\textbf{DG2}). The capped patch budget also bounds token usage, helping keep LLM inference latency relatively stable across datasets of different sizes.

\vspace{0.5em}
\noindent\textbf{Prompting the LLM for thresholds.}
We then provide the patch summaries and the formal definitions of each feature to the LLM using two prompt templates, one for global thresholds and one for local thresholds (see Appendix~A and B for details). The prompts describe the meaning of each parameter in natural language and ask the LLM to return a JSON dictionary of suggested thresholds. Because the input consists of structured statistics rather than raw sequences, the model can reason about relative variability, skewness, and magnitude at the dataset level. To further reduce ambiguity, we also express each local feature as a small Python-style function, in addition to its NL description. This representation is inspired by Program-of-Thought prompting~\cite{ChenM0C23}, which uses code-like structures to make linguistic descriptions more explicit. In our setting, these code-style definitions anchor each feature in concrete statistical operations and yield more consistent, interpretable threshold proposals. The returned JSON is parsed to update the parameter set used by the feature predicates. \textbf{Threshold inference is performed only once offline for each dataset; under our maximum patch budgets, the global and local passes take on average 5~s and 14~s, respectively}, with GPT-4o, making interactive ingestion of new datasets feasible.

\vspace{0.5em}

\noindent \textbf{Dynamic threshold comparison with fix percentiles.}
LLM-based dynamic thresholding can offer practical advantages over fixed percentile thresholding in some cases, although we do not claim it is universally better. Because threshold selection itself rarely has a single ground truth, a meaningful cutoff often depends on the dataset, the query semantics, and the analyst's domain understanding. Fixed percentile methods require users or analysts to choose a cutoff in advance (e.g., top 10\%, 20\%, or 30\%), and this choice is often subjective and difficult to transfer across datasets. In contrast, our LLM-based approach infers thresholds from patch-level summary statistics together with the semantic definition of each feature, allowing the threshold to better reflect the data context without requiring manual tuning.
This context sensitivity is especially useful when the same feature should be interpreted differently under different scales, levels of noise, or distributional characteristics. Percentile thresholds are tied to rank within a particular dataset and may therefore under-select or over-select relevant segments when applied to another dataset with different variance or structure. By contrast, LLM-inferred thresholds are statistically informed rather than rank-based, which can yield more meaningful cutoffs for the current dataset. 
For several features, dynamic thresholding tended to produce highlighted regions that more closely matched the intended patterns than fixed percentile-based thresholds. In a step-function case, for instance, percentile cutoffs required manual trial and error to achieve a comparable result. Additional examples are provided in Appendix~G.
At the same time, because LLM outputs can be imperfect, our system exposes the inferred thresholds for user inspection and manual adjustment whenever the suggested values do not match analyst expectations.

\subsection{Pattern Retrieval and Display}

Once the active feature sequence $\mathcal{F}_\text{active} = (f_1, \dots, f_m)$ and their thresholds $\{\theta_k\}$ are determined, the system retrieves matching patterns via a sliding-window matcher (Figure~\ref{fig:nl_workflow}-C).

\vspace{0.5em}
\noindent\textbf{Sliding-window-based matching.}
We apply a sliding window of length $L$ with step size 1 to the time series, generating candidate windows $\{w_i\}$. The default window length is $L = 6$, which provides a practical balance between local detail and computational efficiency. Users can adjust $L$ in the control panel to suit their task.
When users are unsure what window length is appropriate, the system can also leverage the LLM-based query interpretation pipeline to recommend a suitable $L$ based on the described pattern and dataset context.
\gz{Because temporal scale strongly affects the interpretation of shape features, we expose the query window length as an explicit user-controlled parameter rather than hiding it inside the model. Appendix~H further illustrates that the same feature can retrieve substantially different patterns under different window lengths, reinforcing the need to make temporal scale visible during interaction.}
We treat $\mathcal{F}_\text{active}$ as an ordered list of features derived from the user's description (e.g., ``rise then fall''). For each candidate window $w_i$ and $m = |\mathcal{F}_\text{active}|$ ordered local features, our default strategy partitions $w_i$ into $m$ contiguous subsegments of equal length $L/m$. Subsegment $j$ is then evaluated only against feature $f_j$ using its thresholds $\theta_j$, enforcing the specified temporal order. Global constraints (e.g., overall level or variance) are evaluated on the full window. A window is marked as a hit, $\text{match}(w_i) = 1$, if all required local and global predicates are satisfied.

Equal-length partitioning serves as a simple and interpretable default for multi-part queries. For cases requiring finer control over the duration of individual stages (e.g., ``3 days of increase followed by 5 days of decrease''), the system also allows users to specify per-feature durations through the chat interface (Section~5.3), which are extracted by the LLM and used to customize how each window $w_i$ is partitioned before matching.
To prevent clutter and over-counting when many adjacent windows satisfy the same constraints, we apply an anti-collision algorithm (following~\cite{liu2022mtv}) that suppresses heavily overlapping windows and retains representative segments that best satisfy the feature criteria. This yields a cleaner, non-redundant set of result segments.

\vspace{0.5em}
\noindent\textbf{Pilot study on feature perception.}
We conducted an additional pilot study to evaluate whether the segments highlighted by our shape feature definitions aligned with users' expectations. We recruited 6 participants, all of whom were college students or recent graduates with no prior experience in visual analytics research. The survey presented static time-series plots in which certain segments were highlighted by our system. 
Participants viewed the full time series with these highlighted regions, rated their agreement on a 5-point Likert scale (1 = strongly disagree, 5 = strongly agree), and later provided general comments. The highlighted results covered 29 examples selected to promote pattern diversity. Overall, participant feedback was positive, suggesting that the highlighted segments generally aligned with user expectations.

The study further suggested that users expected multi-part queries such as ``rise'' then ``fall'' to correspond to distinct, temporally ordered subsegments rather than unconstrained matches anywhere within the window. This observation informed our ordered subsegment matching scheme and supported equal-length partitioning as the default interpretation for multi-part queries. For example, a ``rise'' then ``fall'' query over $L = 6$ is divided into two length-3 subsegments, which are passed in order to the ``rise'' and ``fall'' predicates. 

Because the highlighted examples in the pilot were generated using dynamically inferred thresholds rather than manually tuned rules, the results also provided an initial check that our LLM-inferred thresholds, combined with even partitioning, produced matches broadly consistent with user expectations. Participant feedback also motivated our support for explicit duration specifications through chat when users want finer-grained control over individual stages. Together, these design choices make the alignment between users' descriptions and the resulting highlights more predictable and interpretable, while leaving room for future work on more flexible alignment strategies.

\gz{\textbf{Flexible alignment.}
Our default strategy---partitioning a window into equal-length subsegments for ordered local features---provides a simple and interpretable baseline, and our pilot study suggests that it often matches user expectations for descriptions such as ``rise then fall.'' However, it cannot fully capture queries with uneven, overlapping, or explicitly timed stages. Exhaustive search over all possible alignments is one alternative, but its cost grows quickly with window length and the number of features, making it poorly suited to interactive use. It can also return many candidate matches that satisfy some alignment mathematically but do not correspond well to the user's intended pattern (appendix~I). Our current design instead uses natural language to strike a practical middle ground: users can specify feature-specific durations through chat, allowing the system to customize subsegment lengths without brute-force search.}

\vspace{0.5em}
\noindent\textbf{Frontend display of matches.}
Building on the retrieved matches, the frontend provides coordinated visual feedback across multiple views to support \textbf{DG5}. As shown in Figure~\ref{fig:nl_workflow}-D, the Full Time Series View marks each hit as a colored box on the global time series, while a small-multiples view presents each matched segment as a line chart with annotations indicating which subsegments correspond to which features. Additional interactions and coordination between views and queries are described in Section~\ref{sec:system}.

\section{\system}
\label{sec:system}
\gz{Guided by our design goals (Section~\ref{sec:design_goals}), we develop \system, a univariate time-series querying system that coordinates free-form natural language and sketch input for flexible pattern specification and refinement. For each NL query, the LLM backend derives explicit global and local shape features that users can inspect and edit, making query interpretation transparent and controllable.} Although Query-by-Text (\textbf{QbT}) and Query-by-Sketch (\textbf{QbS}) are supported by distinct query pipelines, they are presented as coordinated views linked through synchronized multi-view interactions that preserve global context and provide immediate visual feedback (Figure~\ref{fig:ui}).

\subsection{Overview and Bidirectional Exploration Workflow}
\system~comprises five coordinated views that support two complementary exploration workflows: (1) starting from a sketch to obtain concrete examples and then generalizing through natural language, and (2) starting from natural language to construct a feature-based query and then refining it visually through sketching.

In a sketch-first workflow, users draw a desired pattern in the QbS View (Figure~\ref{fig:ui}-D1). The system returns matched windows in D2 and highlights their locations in the Full Time Series View (Figure~\ref{fig:ui}-E), helping users assess how well the sketch captures their intent (\textbf{DG2}, \textbf{DG5}). Users can then ask the LLM in the Chat View (Figure~\ref{fig:ui}-B1) to describe the sketch or selected matches. The resulting natural-language description is translated into an explicit feature set and corresponding QbT results (Figure~\ref{fig:ui}-C), enabling users to generalize beyond a single sketch and refine the query through editable feature \mbox{primitives (\textbf{DG2}, \textbf{DG4}).}

Conversely, in a text-first workflow, users begin by describing a desired temporal pattern in natural language in the Chat View. The system interprets the query, generates feature constraints, and displays matched examples in the QbT View (Figure~\ref{fig:ui}-C2). Users can then select representative matches and overlay them in the sketch canvas (Figure~\ref{fig:ui}-D1) to support further sketch-based refinement. This bidirectional linkage between QbT and QbS supports fluid movement between verbal and visual reasoning (\textbf{DG3}) while maintaining global context in the Full Time Series View (\textbf{DG5}).

Sketch-based and NL-based queries are processed through distinct pipelines rather than serving as interchangeable frontends to a single backend. This design preserves the complementary strengths of each modality: geometric precision in sketch-based querying and semantic flexibility in natural-language querying. While the LLM-based language pipeline may introduce some variability due to probabilistic interpretation and adaptive thresholding, sketch-based matching remains deterministic given the same input and similarity metric. Maintaining separate pipelines therefore reflects the difference between geometric specifications and semantic descriptions, while supporting diverse user reasoning strategies without forcing both modalities into a unified retrieval process.
In the following, we describe how this workflow is supported by three groups of views: (i) data-driven views that provide global context, (ii) NL-query-driven views that support feature-based querying, and (iii) sketch-query-driven views that support visual \mbox{pattern specification}.

\subsection{Data--Driven Views}

The data-driven views provide the global temporal context within which both NL and sketch queries operate.

\textbf{Control Panel}. The Panel (Figure~\ref{fig:ui}-A) allows users to import a univariate time-series dataset and specify the query window length $L$. The window length determines the granularity at which the series is segmented for both NL-based and sketch-based matching. Using a shared window length keeps candidate segments comparable and supports consistent retrieval across modalities.

\textbf{Full Time Series View}. The View (Figure~\ref{fig:ui}-E) presents the complete time series as a single line chart and serves as the primary view for contextualizing results. Matched segments from NL-based queries are highlighted in grey (yellow when selected), and matched segments from sketch-based queries are highlighted in red (dark red when selected). Standard interactions such as zooming and panning allow users to inspect surrounding context and understand what precedes and follows each match (\textbf{DG5}). Because results from QbT and QbS appear on the same global timeline, users can directly compare how different query formulations retrieve similar or different patterns, supporting iterative refinement and cross-modal validation (\textbf{DG2}, \textbf{DG3}).

\subsection{NL Query--Driven Views}

\system~supports free-form textual descriptions of temporal behavior (\textbf{DG1}). Users enter an NL query in the Chat View, which is passed to the LLM-based interpretation module described in Section~\ref{sec:nl_pipeline}. The LLM parses the query into interpretable shape features and associated thresholds, taking into account both query semantics and dataset-level statistics. These features form a structured constraint set that can be inspected and refined in the Feature Panel (Figure~\ref{fig:ui}-C1). The system then applies these constraints to the segmented dataset, retrieves windows that satisfy them, and displays the matched segments in the QbT result view (Figure~\ref{fig:ui}-C2) and the Full Time Series View. This process operationalizes \textbf{DG2} and \textbf{DG4} by translating ambiguous natural language into a transparent, editable representation that users can iteratively refine, while keeping LLM feedback, feature representations, and visual matches synchronized across views (\textbf{DG5}).

\textbf{Chat View}. The Chat View (Figure~\ref{fig:ui}-B) provides a text-based interface for specifying queries and obtaining explanations in natural language (\textbf{DG1}). Users may describe desired patterns or request descriptive or statistical summaries of the dataset. After a query is submitted, the system can suggest an overall window length $L$ when users do not have a clear target. Users may also specify the duration of individual feature stages in natural language (e.g., ``\textit{3 hours flat and then decreasing over the next 5 hours}''). The system extracts this duration information and uses it to determine an appropriate window length and feature partitioning (e.g., $L=8$ when the unit is hours). Users can additionally ask the LLM to interpret selected sketch-based results or rephrase earlier queries, supporting progressive refinement through conversational interaction. By centralizing NL interaction in a familiar chat format, this view lowers the barrier to specifying complex temporal behavior without requiring formal query syntax.

\textbf{Query View by Text (QbT View) and Feature Panel}. NL queries issued in the Chat View are transformed by the LLM pipeline into a structured set of global and local features (\textbf{DG2}). These features are displayed in the \textit{Feature Panel} (Figure~\ref{fig:ui}-C1), where each feature is associated with a threshold inferred by the LLM from the feature definition and dataset statistics, yielding a precise and editable specification of the pattern. The Feature Panel supports human-in-the-loop refinement (\textbf{DG4}): users can toggle features on or off, adjust thresholds, inspect expanded definitions via hover (Figure~\ref{fig:ui}-F1 and F2), reorder local features, and add or remove features using inline controls. These edits immediately update the matched windows shown in the QbT result view (Figure~\ref{fig:ui}-C2), which presents matched segments as small-multiple line-chart cards annotated with the features they satisfy. This layout helps users assess how the interpreted feature set maps to concrete patterns and whether the NL query was interpreted as intended (\textbf{DG2}, \textbf{DG5}), before optionally moving to sketch-based refinement.

\subsection{Sketch Query--Driven Views}
Complementing the NL query--driven views, the sketch query--driven views support visual pattern specification when users reason in terms of shapes rather than words. 
When users draw in the Sketch Canvas, \system~resamples both the sketch and candidate windows to the target window length and applies z-normalization so that matching focuses on shape rather than absolute magnitude, which is important when similar patterns occur at different value ranges across months or years. The normalized sketch is then compared against each candidate window using Pruned Dynamic Time Warping (PDTW) with a Sakoe--Chiba band~\cite{geler2019dynamic}. We use a warping window size of 5, which limits local temporal warping, and Euclidean distance as the point-wise cost metric. PDTW computes an alignment cost while pruning unlikely warping paths, improving efficiency while still accommodating local temporal misalignments. Its sensitivity to local slope allows it to capture rising, falling, and oscillatory structures that users often express visually. The resulting distances are used to rank candidate windows; the top-$k$ matches are shown in the QbS result view and highlighted in the Full Time Series View. By emphasizing shape and local dynamics, this sketch-based matching complements feature-based \mbox{NL querying (\textbf{DG2}), \textbf{DG3}).}

\textbf{Query View by Sketch (QbS View) and Sketch Canvas}. The Query View by Sketch (Figure~\ref{fig:ui}-D) provides the main interface for sketch-based querying (\textbf{DG1}, \textbf{DG3}). The Sketch Canvas (Figure~\ref{fig:ui}-D1) is an interactive drawing area in which user strokes appear in red. When users select segments from QbT, their normalized trajectories are overlaid as a dark gray band summarizing their variation, providing a concrete visual reference for the system's current interpretation and guiding sketch refinement. The canvas supports basic editing operations (e.g., undo and clear) and controls for specifying the desired number of matches. The top-$k$ results identified by sketch-based matching are shown as line-chart cards in the QbS result view (Figure~\ref{fig:ui}-D2) and are simultaneously highlighted in the Full Time Series View, allowing users to assess both global coverage and local detail in a \mbox{coordinated manner (\textbf{DG5}).}

\textbf{Matched Segments and Cross-View Coordination}. Both QbT and QbS present matched windows in a consistent card-based layout, with linked interactions across views. Selecting a card in either QbT (Figure~\ref{fig:ui}-C2) or QbS (Figure~\ref{fig:ui}-D2) highlights the corresponding segment in the Full Time Series View (Figure~\ref{fig:ui}-E), supporting rapid assessment of match quality and context. Selected segments can also be sent back to the Sketch Canvas to guide new sketches or used as examples in follow-up NL queries. This tight coordination operationalizes \textbf{DG3} and \textbf{DG5} by enabling users to move seamlessly between modalities while maintaining awareness of where and how patterns occur in the dataset.

\section{Evaluation}
\subsection{Usage Scenarios}

\textbf{Finding Suitable Weather for Grazing in the Summer}
A livestock farmer in the Sacramento region is planning when to schedule outdoor grazing for his herd. He wants to avoid periods of extreme heat, which can stress and dehydrate animals and reduce weight gain. The farmer is unsure which summer month typically experiences fewer heatwaves. Although he is not an expert at describing time-series patterns, he understands a heatwave as a period in which the daily maximum temperature starts lower, rises to a high level, remains high for several days, and then returns to lower values. To illustrate this pattern, he sketches a line that begins low, rises, stays high for a while, and then drops again (Figure~\ref{fig:ui}-1). The sketch results display several segments that match this specific shape, but the farmer wants a more general query. He therefore asks the Chat View to describe his sketch so that he can learn appropriate terms for searching similar patterns more broadly (Figure~\ref{fig:ui}-2).
Based on the generated description, he submits a textual query in the QbT view (Figure~\ref{fig:ui}-3): ``\textit{show me all segments that rise at first, then go into a plateau, and then drop.}'' The system translates this description into the shape features \emph{rising}, \emph{constant}, and \emph{falling}, where the local feature \emph{constant} captures the plateau portion (Figure~\ref{fig:ui}-F2). The QbT matched-segments panel (Figure~\ref{fig:ui}-C2) displays a grid of candidate windows, allowing the farmer to quickly assess how well the retrieved segments align with his notion of a heatwave.

The farmer notices that several matches occur during cooler seasons and therefore would not pose heat risks to the animals. To focus on genuinely hot periods, he refines his query by emphasizing high temperatures, modifying it to ``\textit{[original query] that remains generally high}'' (Figure~\ref{fig:ui}-4). This refinement adds a global feature, \emph{high}, requiring the segment mean to exceed a threshold before the segment is considered hot (Figure~\ref{fig:ui}-C1). The farmer then tightens this threshold using the corresponding control (Figure~\ref{fig:ui}-F1). After updating the matches, the QbT results become concentrated in summer months. In the Full Time Series View (Figure~\ref{fig:ui}-E), June shows fewer highlighted segments than later months, leading the farmer to conclude that June is the better \mbox{time for grazing.}

\begin{figure}
  \centering
  \includegraphics[width=0.99\linewidth]{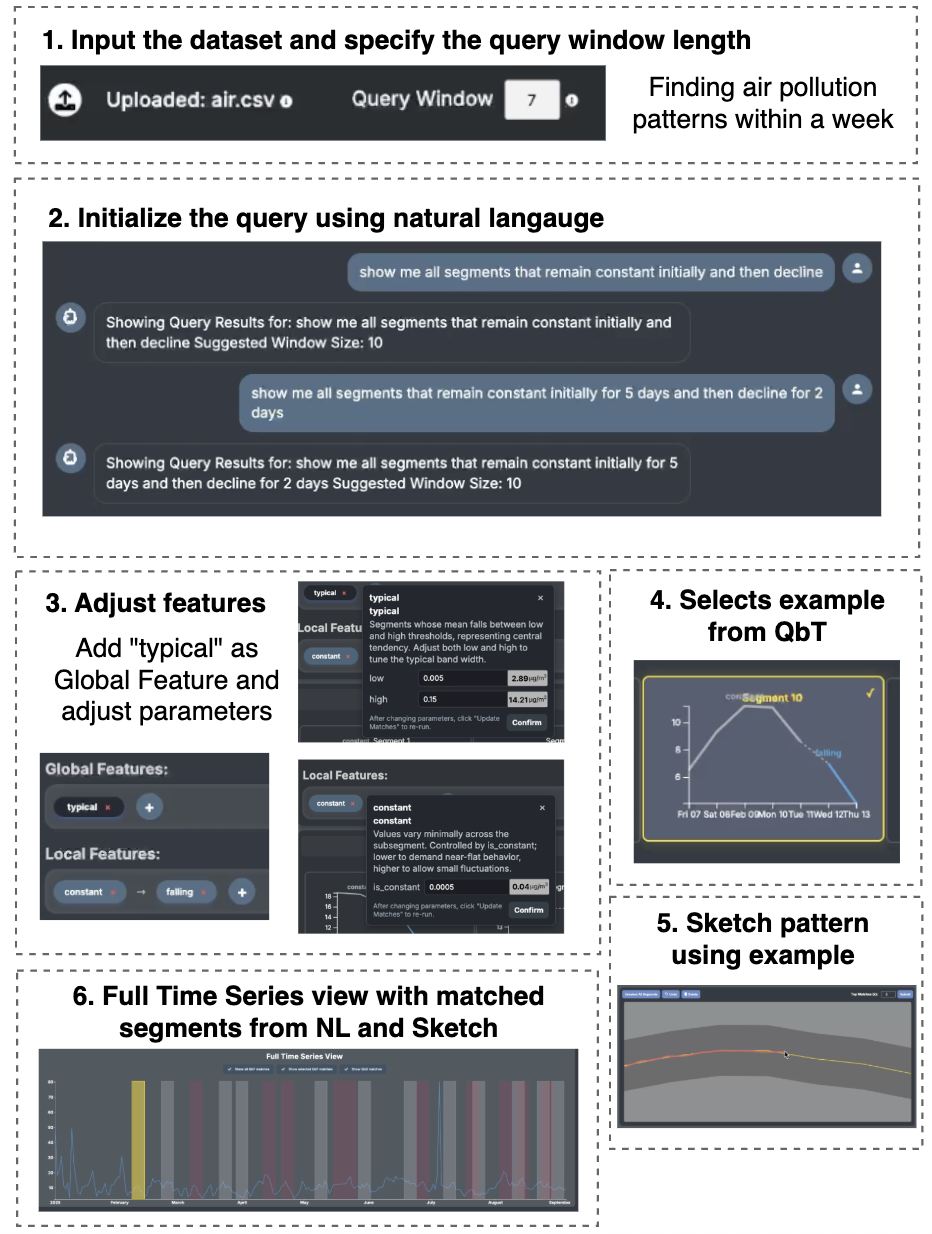}
  \caption{Exploring weekday--weekend PM2.5 patterns in Los Angeles. 
  }
  \vspace{-0.3in}
  \label{fig:case1}
\end{figure}

\vspace{0.5em}
\noindent\textbf{Weekly Air Quality.}
A climate scientist studying urban climate change examines short-term air pollution patterns in Los Angeles, California. Her goal is to identify weekly patterns of rising and falling PM2.5 concentrations. Previously, she manually inspected plots of the data, but she now seeks a more efficient way to find recurring temporal patterns. She begins by uploading her dataset and setting the query window length to 7 days ($L=7$) to focus on within-week behavior (Figure~\ref{fig:case1}-1). Using the chat interface, she issues an initial NL query (Figure~\ref{fig:case1}-2): ``\textit{Show me all segments that remain constant initially and then decline}.'' Her goal is to identify weeks in which PM2.5 levels stay high from Monday to Friday because of commuting, then decrease on Saturday and Sunday when fewer people travel to work. The system initially applies the default even-split interpretation of the query window, which does not match her intended weekday--weekend structure. She therefore refines the query to: ``\textit{Show me all segments that remain constant for 5 days and then decline for 2 days}.''

In the QbT results, she notices that some matched segments are anomalous and do not fit her intent. To filter out these atypical patterns, she adds \emph{typical} as a global feature, which retains segments whose mean falls within a central band. However, this setting yields too few matches, so she relaxes typical feature's thresholds by setting the lower and upper bounds to 0.005 and 0.15, respectively. She also adjust the \emph{constant} feature by setting \texttt{is\_constant} to 0.005, allowing small fluctuations while still capturing near-flat behavior (Figure~\ref{fig:case1}-3). After updating the matches, many segments follow the expected pattern: pollution starts of constant during Mondays and Tuesdays and declines during weekends, aligning with her commuting hypothesis.
At the same time, she also notices some segments in which pollution remain stable during the weekend and starts declining during weekdays. To investigate this alternative pattern, she selects the example of interest and turns to Query-by-Sketch (Figure~\ref{fig:case1}-4). She traces the example (Figure~\ref{fig:case1}-5) and submits it as a sketch query. The QbS results reveal additional segments matching this visual pattern, which she inspects in the Full Time Series View alongside the QbT matches (Figure~\ref{fig:case1}-6). These findings prompt her to further investigate possible explanations for weekend pollution episodes like special events or atypical \mbox{traffic patterns}.

\subsection{User Study}
We recruited 13 participants (P1--P13) for our user study. The study was determined to be exempt by the IRB at our institution. All participants provided informed consent prior to participation. Participation was voluntary, and participant anonymity was maintained. Participants had 1--6 years of experience using visualization tools and, on average, 2 years of experience with conversational LLMs. The study aimed to evaluate the effectiveness of Query-by-Text and sketch-based refinement, as well as the overall usability of the system.

\textbf{Tasks and procedure:} 
The study was conducted in person using the same device and environment to reduce variability. After a brief 10-minute hands-on training session, participants completed a pre-study survey to provide demographic information. In Task~1, each participant used \system~to answer eight analytical questions on the APPLE stock dataset (four easy and four hard), with question order counterbalanced across participants to assess task performance. Task~2 involved open-ended exploration of local city weather data, in which participants searched for self-defined interesting patterns to evaluate the flexibility and diversity of supported queries. This task emphasized more complex, fine-grained, and personally meaningful temporal behaviors. After both tasks, participants completed a post-study usability questionnaire. Full setup and results details are provided in Appendix~C and D.

\textbf{Results.}
\gz{At the participant level, most participants used both QbT and QbS at least once during the study: 92\% of participants used both in Task~1, and 85\% did so in Task~2. This suggests that both modalities were available and useful within the overall workflow. Participants also described the modalities as complementary.} As P1 noted, \textit{``It can use both modes to refine the search ... I say they enhance each other.''} P12 similarly highlighted the value of sketching: \textit{``The query-by-sketch thing is pretty convenient to craft a query if you don't know what you are doing.''}

Participants completed Task~1 questions in 1.36 minutes on average. In Task~2, participants identified an average of 3.3 interesting or personally meaningful patterns within a 10-minute window. These results suggest that \system~supports both efficient task-oriented search and more open-ended pattern exploration. In Task~1, participants were generally able to translate their intent into queries and locate target patterns quickly; in Task~2, they were able to use the system to iteratively explore and refine self-defined patterns in a less structured setting.

\begin{figure}
  \centering
  \includegraphics[width=0.99\linewidth]{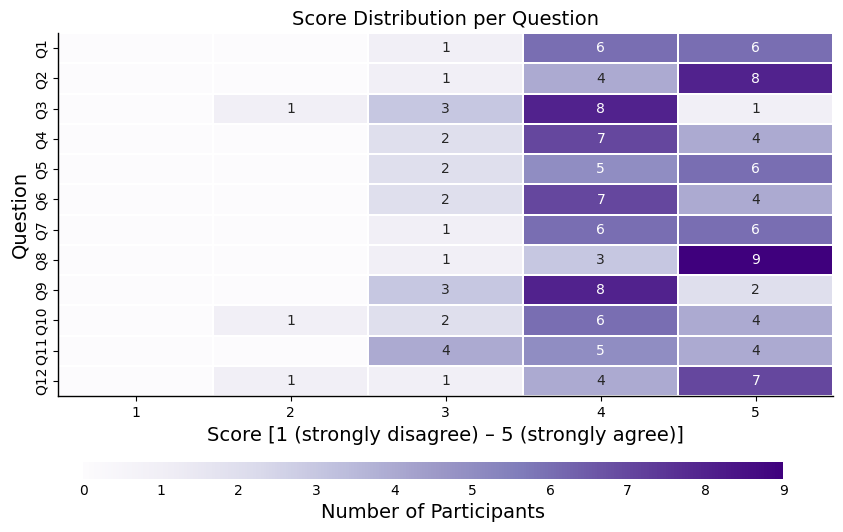}
  \caption{Post-study survey results. Each cell shows the number of participants selecting that score for each question.}
  \vspace{-0.2in}
  \label{fig:post_study}
\end{figure}

\gz{Regarding query behavior in Task~1, 84\% of participants began their search with QbT, using natural language, feature selection, or both. However, at the trial level, participants often relied on QbT alone: across the eight structured questions in Task~1, 68\% of trials were completed using only QbT. We interpret this result not as evidence that both modalities are equally necessary for every query, but as evidence that natural language and feature-based querying provide an efficient default entry point for many structured pattern-search tasks. Among text-based queries, 57\% were made using natural language alone, while 32\% relied solely on feature selection. Sketching was used more selectively, particularly when participants wanted to express a geometric shape directly, refine an example, or recover from an unsatisfactory textual query. Thus, \system{}'s multimodality is opportunistic rather than mandatory: users can complete simple or well-specified queries through text alone, while still having access to sketch-based refinement when textual descriptions become ambiguous or insufficient.} For example, P8 used only natural-language queries to complete all Task~1 questions successfully. At the same time, more experienced users valued direct feature control: P3, who had extensive experience with time-series data and visualization, preferred feature selection for clearer exploration, and P13 remarked that the feature panel \textit{``actually grounds you by showing the local features, so you can be sure that your prompt described the right thing.''}

The post-study survey included 12 questions (see Appendix~C for details), each rated on a 5-point Likert scale (1 = strongly disagree, 5 = strongly agree). The questions covered comfort with describing patterns in natural language (Q1), confidence in sketching patterns (Q2), perceived accuracy of text (Q3) and sketch translation (Q4), ease of use (Q5), interface intuitiveness (Q6), ease of learning (Q7), efficiency of sketching for certain queries (Q8), ease of adjusting feature parameters (Q9), effectiveness of combining language and sketching (Q10), perceived consistency (Q11), and support for iterative refinement (Q12). \gz{The score distribution (Figure~\ref{fig:post_study}) was consistently positive: participants found natural language intuitive, sketches reliable, and the cross-modal workflow effective.} Ratings for usability and learnability were also high, and participants agreed that the system provided consistent feedback and supported iterative refinement---two properties that are especially important for open-ended pattern discovery.

\subsection{Analysis of LLM Parsing Quality}
\label{sec:llm-parsing-quality}

To evaluate the quality of LLM-based parsing from natural language to shape features, we collected 138 real human queries describing time-series patterns and generated 42 synthetic queries. The synthetic queries, produced with GPT-5, were designed to broaden linguistic coverage beyond typical user inputs by including more complex adjectives, less common formulations, and feature descriptions that rarely appeared in the human-authored queries. One example is ``\textit{Find parts with almost no jaggedness}.'' We manually annotated both the human and synthetic queries according to our shape-feature framework and compared these annotations with the features produced by the LLM from the corresponding natural-language inputs. We ran the evaluation twice: once with few-shot prompting and once without. Accuracy was used as the evaluation metric, where a prediction was counted as correct only if the LLM output \textbf{exactly matched} the human annotation.

For real human queries, accuracy improved from 64\% without few-shot prompting to \textbf{86\%} with few-shot prompting. For synthetic queries, accuracy improved from 62\% to \textbf{83\%}. These results indicate that the LLM parser can reliably translate a wide range of natural-language descriptions into our shape-feature representation, and that few-shot prompting plays an important role in improving robustness and consistency. In particular, the gains on synthetic queries suggest that the parser generalizes beyond the phrasing patterns observed in everyday user input.
Without few-shot prompting, errors more often arose from format inconsistencies, unnecessary insertion of global features such as \textit{typical}, and confusion between related feature concepts such as \textit{high-amplitude} and \textit{spiky}. Few-shot examples help anchor both the expected output structure and the intended semantics of our feature vocabulary, leading to more stable and accurate translations.
\subsection{Analysis of Failure Cases}
For Task~1, we identified 27 individual failure cases across 103 trials. We define a failure case as an operation that did not immediately produce the outcome intended by the participant, requiring them to refine the query, retry the operation, or switch modalities to complete the task. Of these 27 cases, 14 involved unsatisfactory QbT results related to dynamic thresholding. Among these 14 cases, 8 returned no results because of a combination of overly large window lengths and overly strict thresholds. The remaining 6 returned results, but participants did not prefer the initial outputs. We also identified 3 cases involving errors in translating natural language into shape features, 4 cases in which the natural-language query fell outside the system's supported capabilities, 1 case with unsatisfactory sketch-based results, and 5 cases involving initial misinterpretation of the task. These cases suggest that difficulties arose not only from model interpretation, but also from the inherent ambiguity of open-ended pattern specification and the challenge of selecting thresholds that match user expectations in a dataset-dependent setting. 
The QbT-related cases mainly fell into three categories: in some instances, participants issued queries that did not fully match the task they intended to solve; in others, the LLM's interpretation did not completely capture the intended pattern; and in some cases, the parsed features were appropriate but the inferred thresholds were relatively restrictive, yielding fewer matching results.

\gz{Importantly, participants were usually able to recover from these situations through the system's cross-modal and editable workflow. In 15 failure cases, they switched modalities (e.g., from QbT to QbS or vice versa) to continue the task, whereas in the other 12 cases they iteratively refined the original modality. On average, failure cases required 1.63 more operations than non-failure cases. This recovery behavior reinforces one of our central motivations: when a single query formulation does not work as intended, users can inspect, revise, and re-express their intent through another modality rather than restarting from scratch. The task that required narrowing a large set of retrieved results exhibited the highest failure rate: 10 of the 13 participants encountered difficulty with Task~1 Q5, which was classified as hard. This finding suggests that result filtering and refinement in QbT may still require some user expertise and points to a promising direction for future work: supporting more direct natural-language control over result refinement (e.g., ``filter these results down to at most one'').}

\gz{The failure cases also clarify the role of sketching in the current workflow. Sketch was not used equally often across all tasks, nor was it required for many well-specified textual queries. Instead, it functioned as a complementary recovery and refinement mechanism. When QbT produced overly broad, overly strict, or semantically mismatched results, users could switch to QbS to re-express the desired geometry more directly. This suggests that the value of sketching lies less in replacing natural language and more in providing an alternative representation of intent when textual descriptions are difficult to specify precisely.}

\subsection{Analysis of Latency}
Our latency analysis suggests that \system~is practical for interactive use. Dynamic threshold inference is performed once offline for each dataset and, under capped patch budgets, takes about 5~s for global thresholds and 14~s for local thresholds with GPT-4o. Because the number of summarized patches is bounded, this offline cost remains stable even as the raw dataset grows. For online QbT queries, including both LLM parsing and matching, GPT-4o achieved the best latency among the tested models, averaging 651~ms on the Sacramento dataset, compared with 1745~ms for DeepSeek Rover V2 and 1620~ms for Devstral-Small-2505. This keeps typical NL query latency for medium-sized datasets well below one second.

Latency increases with dataset size and window configuration, but the trade-offs are predictable. Using GPT-4o, average QbT latency was 670~ms on the 1,977-point Sacramento weather dataset, 855~ms on the 2,713-point Bitcoin dataset, and 4027~ms on the 145,367-point energy dataset, where post-processing became more expensive because more candidate matches survived to anti-collision filtering. Window length showed a related effect on both latency and match count. On the Sacramento dataset, a broad query with a very small window ($L=2$) produced 946 matches and 936~ms latency, while $L=4$ and $L=8$ reduced this to 199 and 123 matches, with latency dropping to 662~ms and 666~ms, respectively. Under more realistic weekly-to-monthly settings, $L=7$ returned 131 matches at 766~ms, whereas $L=14$ returned only 8 matches at 568~ms. Once the window became too large for the queried pattern ($L \ge 21$ in this experiment), no matches were returned, and latency was dominated primarily by the scan over candidate windows rather than anti-collision filtering. Overall, these results suggest that very small windows tend to produce many overlapping matches and higher post-processing cost, whereas moderate window lengths ($L=6$--$14$) provide a more practical balance between responsiveness and expressive coverage for interactive query refinement. Full results are reported in Appendix~E.

\section{Discussion}
Our findings highlight several design implications in NL+X time-series querying and suggest directions for future work.

\gz{\textbf{Limits of the current multimodal integration.}
\system~currently supports coordinated cross-modal refinement rather than simultaneous multimodal fusion. Users can move between text, editable features, and sketches, but the system does not jointly infer a single query from concurrent text and sketch input, nor does it allow users to explicitly weight the relative influence of the two modalities. This design was intentional for the present system: natural language and sketch encode different forms of intent, and keeping their pipelines separate makes each interpretation inspectable and easier to revise. However, this choice also limits the kinds of multimodal ambiguity resolution that \system~can support. Future work could explore joint text--sketch interpretation, user-controlled modality weighting, and interaction techniques where a sketch constrains the geometric space while language specifies semantic filters or exceptions.}

\textbf{Ambiguity and underspecification in pattern queries.}
Some of the observed failure cases reflect limitations of the current system, but others stem from a more fundamental challenge: users' intended patterns are often ambiguous, incomplete, or context-dependent. For example, NL queries may admit multiple plausible interpretations, and users may implicitly expect the system to focus on specific temporal regions, such as winter months, even when this is not explicitly stated. These observations suggest that future systems should support more interactive disambiguation, for example by surfacing alternative interpretations, clarifying implicit assumptions, or allowing users to add contextual filters over time ranges or value ranges. More explicit support for constraint diagnosis could also help users understand why a query returns too many, too few, or no results, and how to \mbox{relax or revise it}.

\textbf{Matching as an iterative, inspectable process.}
Our results also suggest that pattern retrieval should not be treated as a one-shot matching problem. When users requested strict patterns over long windows, the system sometimes returned approximate matches or none at all, making it difficult to distinguish between mismatched intent, overly restrictive constraints, and genuine absence of the pattern in the data. This points to the need for more informative feedback about which constraints were satisfied or violated, and for interaction techniques that support progressive relaxation, reformulation, and comparison of alternative query specifications. In this sense, one important lesson is that successful time-series querying depends not only on matching quality, but also on how well the system helps users inspect and revise the interpretation of their queries.

\textbf{Toward more robust sketch-based interaction.}
Sketch-based querying introduced a different but important challenge. Because sketches are normalized, resampled, and matched using a similarity metric, small drawing imperfections can lead to unintended shape differences. Participants therefore expressed a desire for more support when authoring sketches, such as smoothing, snapping, or template-based guides. This suggests that future work should improve sketch-based querying not only through stronger matching algorithms, but also through richer interaction design. In particular, perception-oriented similarity metrics from prior work could be combined with drawing aids and editable intermediate representations to better align sketch input with users' intended shapes. Such support may make sketching more robust, interpretable, and complementary to NL querying.

\gz{\textbf{Lessons of using LLMs.}
Our findings suggest that the value of LLMs in \system~lies less in perfectly resolving user intent than in making open-ended NL pattern specification practical within an interactive workflow. Our goal is not to show that natural language uniformly outperforms sketch-based querying. Instead, \system~explores how language and sketch can support different stages of pattern specification: language provides a flexible entry point for semantic and compositional intent, while sketching supports direct geometric refinement. By translating free-form queries into visible shape features and corresponding matches, the system allows users to inspect, validate, and revise the model's interpretation rather than treating it as a black box. Because LLM-based parsing remains imperfect (hallucinations still exist) and can misread ambiguous or uncommon phrasing, it is most effective when paired with complementary, more controllable interaction mechanisms such as sketching. More broadly, this points to a key design principle for future systems: LLMs should function as interpretable and revisable mediators of user intent, embedded within cross-modal interfaces that support verification, correction, \mbox{and iterative refinement.}}

\gz{\textbf{Making temporal scale visible to users.}
Our results suggest that temporal scale is not merely a backend parameter, but an important part of how users formulate and interpret pattern queries. The same description can imply different intended behaviors depending on the time span over which it is applied. For example, a user asking for a \textit{complex} pattern may be looking for short-term fluctuations, broader multi-stage variation, or a long-term irregular trend. In our experiments, querying \textit{complex} with different window lengths produced substantially different results (Appendix~H): smaller windows emphasized localized variation and sometimes returned no matches, whereas larger windows captured broader segment-level structures.}

\gz{This sensitivity highlights why temporal scale must be surfaced as an explicit design consideration in pattern search, rather than treated as a hidden implementation detail.
In \system, we address this tension by allowing users to adjust $L$ directly and by using the LLM pipeline to suggest an initial window length when users do not have a clear target. This design lowers the barrier for novice users while preserving explicit user control. More broadly, our experience suggests that future time-series query systems should make temporal scale more visible and explorable, for example by recommending several plausible window lengths or allowing users to compare results across candidate scales within the same interface.}

\section{Conclusion}
\gz{We presented \system, a coordinated natural-language and sketch-based querying system for retrieving fixed-window patterns in univariate time series. By exposing an interpretable shape-feature space and linking Query-by-Text with Query-by-Sketch through coordinated visual feedback, \system~enables expressive pattern search without requiring strict syntax or advanced domain expertise, while keeping query representations transparent and editable. Our studies show that participants used natural language as an efficient entry point and used sketching or feature editing as complementary mechanisms for refinement and recovery. Overall, \system~offers a practical balance between usability, accuracy, and efficiency in interactive time-series pattern search and points toward future cross-modal visual querying systems powered by LLMs.}

\acknowledgments{%
This work was supported in part by the U.S. National Science Foundation under Grant No. IIS-2427770.
}

\bibliographystyle{abbrv-doi-hyperref}

\bibliography{main}

\appendix
\onecolumn

\begin{center}
{\LARGE \textbf{Appendix Table of Contents}}
\end{center}

\vspace{3em}

{\LARGE \textbf{Appendix A: Prompt Design}} \dotfill 14
\vspace{1em}

{\LARGE \textbf{Appendix B: Feature Definition}} \dotfill 20
\vspace{1em}

{\LARGE \textbf{Appendix C: User Study Setup}} \dotfill 22
\vspace{1em}

{\LARGE \textbf{Appendix D: User Study Results}} \dotfill 23
\vspace{1em}

{\LARGE \textbf{Appendix E: Latency Analysis}} \dotfill 24
\vspace{1em}

{\LARGE \textbf{Appendix F: Code for Review}} \dotfill 27
\vspace{1em}

{\LARGE \textbf{Appendix G: Dynamic Threshold Comparison With Fix Percentile Threshold}} \dotfill 28
\vspace{1em}

{\LARGE \textbf{Appendix H: Different Window Length Comparison}} \dotfill 31
\vspace{1em}

{\LARGE \textbf{Appendix I: Latency evaluation across different sub-feature splitting techniques}} \dotfill 32

\clearpage

\begin{center}
{\LARGE \textbf{Appendix A: Prompt Design}}
\end{center}

\noindent{\Large{ \textbf{Prompt for natural language query to formal features}}}
\begin{CodeBlock}
You are an AI assistant designed to support analysts and researchers in interpreting complex time series data from natural language queries.
Your only task is to extract relevant time series features explicitly or implicitly described in the input query.
Do not generate new queries, rephrase the input, or infer beyond what is provided. Return only the extracted features.

You must identify both local and global features as follows:

Local features (describing segment-level characteristics):
"rising", "falling", "concave", "convex", "linear", "non-linear", "constant", "smooth", "noisy", "complex", "simple", "spiky", "dropout", "periodic", "aperiodic", "symmetric", "asymmetric", "step", "no-step", "high-amplitude", "low-amplitude", "high-volume", "low-volume".

Global features (describing overall segment classification):
"high", "low", "typical", "unusual".


Example inputs and expected outputs for few-shot learning: (*@\textcolor{red}{[EXAMPLES (Few Shot Prompting Examples below)]}@*)

Your next response must be a valid Python dictionary in the format shown above.

Important Instructions:
1. Global features MUST be placed only in the "global" key in the dictionary at most 1 feature per query: "high", "low", "typical", "unusual"
2. Local features MUST be placed only in the "local" key in the dictionary: "rising", "falling", "concave", "convex", "linear", "non-linear", "constant", "smooth", "noisy", "complex", "simple", "spiky", "dropout", "periodic", "aperiodic", "symmetric", "asymmetric", "step", "no-step", "high-amplitude", "low-amplitude", "high-volume", "low-volume".
3. Global features must only appear under the 'global' key, and local features must only appear under the 'local' key.
"high", "low", "typical", "unusual" CANNOT appear in 'local' key like the following
4. Do not include any explanations, commentary, or reasoning.
5. Do not wrap the dictionary in any formatting (e.g., no triple backticks, no python label, no quotation marks around the entire dictionary).

After reading this instruction, do not generate new queries or paraphrase the input.
Your response must consist of only a valid Python dictionary containing the extracted features. Nothing more.
    
Query: (*@\textcolor{red}{[INSERT USER QUERY]}@*)
Expected Features:
\end{CodeBlock}

\noindent{\Large \textbf{Few Shot Prompting Examples}}
\begin{CodeBlock}
[
    {
        "query": "I need to pinpoint where trading volume spikes and then quickly drops within the same session. Show me those segments.",
        "features": {"global": (), "local": (("high-volume", "spiky"), ("falling",))}
    },
    {
        "query": "I'm examining temperature trends and need to identify phases where temperatures are unusually high but then show a steady decline over several days. Can you locate these for me?",
        "features": {"global": ("high",), "local": (("rising",),("falling",))}
    },
    {
        "query": "Track periods of high consumer interest followed by a sudden drop in engagement. These shifts are critical for our strategy adjustments.",
        "features": {"global": (), "local": (("high-volume",), ("falling",))}
    },
    {
        "query": "Identify when energy consumption are stable or constant",
        "features": {"global": (), "local": (("constant",),)}
    },
    {
        "query": "Identify segments with low volume",
        "features": {"global": (), "local": (("low-volume",),)}
    },
    {
        "query": "Show me the parts where there is a symmetrical increase and symmetric decrease with high amplitude.",
        "features": {"global": (), "local": (("symmetric", "rising"), ("symmetric", "falling", "high-amplitude"))}
    },
    {
        "query": "Highlight all the segments that are high relative to the data",
        "features": {"global": ("high",), "local": ()}
    },
    {
        "query": "Choose the lines that are comparatively low in the dataset",
        "features": {"global": ("low",), "local": ()}
    },
    {
        "query": "Find lines with low fluctuation or noise that is also relatively high",
        "features": {"global": ("high",), "local": (("smooth",),)}
    },    
    {
        "query": "smooth line throughout the segment with downward trend then upward trend",
        "features": {"global": (), "local": (("smooth","falling",),("smooth","rising",))}
    },
    {
        "query": "Identify segments that are quiet in amplitude.",
        "features": {"global": (), "local": (("low-amplitude",),)}
    },
    {
        "query": "Lines that create a upside down U shape",
        "features": {"global": (), "local": (("concave",),)}
    },
    {
        "query": "U shape or parabolic shape or parabola",
        "features": {"global": (), "local": (("convex",),)}
    },
    {
        "query": "Select all segments that are typical",
        "features": {"global": ("typical",), "local": ()}
    },
    {
        "query": "A sudden drop or V shape or outlier or straight line with a V decline",
        "features": {"global": (), "local": (("dropout",),)}
    },
    {
        "query": "Detect segments with high variability and a lack of symmetry",
        "features": {"global": (), "local": (("complex","asymmetric",),)}
    },
    {
        "query": "Search for regions with no repeating patterns, which are then followed by convex trends",
        "features": {"global": (), "local": (("aperiodic",),("convex",))}
    },
    {
        "query": "Select regions with continuous changes, elevated variance, and curvature that deviates from a straight line",
        "features": {"global": (), "local": (("no-step","noisy","non-linear",),)}
    },
    {
        "query": "Identify segments that are flat and lack any noticeable peaks or fluctuations",
        "features": {"global": (), "local": (("simple",),)}
    },
    {
        "query": "A high rise followed by decline",
        "features": {"global": ("high",), "local": (("rising",),("falling",))}
    },
    {
        "query": "Can you find me a pattern that is high low high",
        "features": {"global": ("high",), "local": ()}
    },
    {
        "query": "A line that rises then falls repeating this trend",
        "features": {"global": (), "local": (("periodic",),)}
    },
    {
        "query": "Many spikes with a falling tendency",
        "features": {"global": (), "local": (('spiky', 'falling'),)}
    },
    {
        "query": "A straight line",
        "features": {"global": (), "local": (('linear',),)}
    },
    {
        "query": "stable, then rise, then stable, then rise",
        "features": {"global": (), "local": (('rising','step'),)}
    },
]
\end{CodeBlock}

\noindent{\Large \textbf{Dynamic Threshold Inference}}

\noindent{\normalsize \textbf{Local Shape Feature Prompt}}
\begin{CodeBlock}
You are an AI assistant designed to support analysts and researchers in interpreting time series data from natural language queries.

Your ONLY task is to compute dynamic threshold values for key statistical features based on the provided data and suggest a window size for each feature. The window size is an integer. Do NOT generate or reformulate queries. Do NOT modify the input.

The input is a list of dictionaries, where each dictionary represents one segment of a time series with the following numerical features:
"mean", "median", "std", "min", "max", "range", "skewness", "kurtosis", "slope", volume

Here are the helper functions:
def slope(segment):
    return np.diff(segment)

def variance(segment):
    return np.var(segment)

def amplitude(segment):
    return np.max(segment) - np.min(segment)

def volume(segment):
    return np.sum(np.abs(segment))

Your job is to dynamically determine VALUE based on the definition of these functions:
The function is_linear identifies segments where the rate of change varies very little compared to other segments, suggesting a relatively consistent and straight trend.
def is_linear(segment):
    return variance(slope(segment)) < is_linear_VALUE

The function is_constant identifies segments where the values show minimal variation compared to other segments, indicating a relatively flat or unchanging pattern.
def is_constant(segment):   
    return variance(segment) < is_constant_VALUE

The function is_smooth identifies segments with relatively low variation in values compared to others, indicating a gentle and stable pattern without sharp fluctuations.
def is_smooth(segment):
    return variance(segment) < is_smooth_VALUE

The function is_noisy identifies segments with relatively high variation in values compared to others, indicating a pattern with frequent or abrupt fluctuations.
def is_noisy(segment):
    return variance(segment) > is_noisy_VALUE

The function is_complex identifies segments that show both high variability and multiple peaks compared to other segments, indicating a pattern with rich or intricate fluctuations.
def is_complex(segment):
    return variance(segment) > is_complex_VALUE and len(find_peaks(segment)[0]) > 2

The function is_simple identifies segments with low variability and no prominent peaks compared to other segments, indicating a relatively plain and uniform pattern.
def is_simple(segment):
    return variance(segment) < is_simple_VALUE and len(find_peaks(segment)[0]) == 0
    
The function is_spiky identifies segments that contain at least one peak and show sudden sharp changes in value. It identifies segments that contain sudden, isolated spikes.
def is_spiky(segment):
    peaks, _ = find_peaks(segment)
    return len(peaks) > 0 and np.any(np.abs(np.diff(segment)) > is_spiky_VALUE)

The function is_dropout identifies segments that contains at least one point x such that x is significantly smaller than the typical magnitude of the data
def is_dropout(segment):
    return any(x < is_dropout_VALUE for x in segment)

The function is_periodic identifies segments that exhibit noticeable repeating frequency components, indicating a relatively regular or cyclical pattern compared to other segments.
def is_periodic(segment):
    freq = np.abs(fft(segment))
    return np.any(freq > is_periodic_VALUE)

The function is_step identifies segments that contain at least one sudden and significant jump in value, indicating a relatively abrupt change or shift in the data.
def is_step(segment):
    return any(np.abs(np.diff(segment)) > is_step_VALUE)

The function is_high_amplitude identifies segments with relatively large differences between their highest and lowest values, indicating strong fluctuations or pronounced peaks and troughs.
def is_high_amplitude(segment):
    return amplitude(segment) > is_high_amplitude_VALUE

The function is_high_volume detects segments that exhibit relatively strong intensity or activity, based on their overall magnitude or energy being higher than typical segments.
def is_high_volume(segment):
    return volume(segment) > is_high_volume_VALUE

The function is_low_volume identifies segments with a relatively small magnitude or energy, indicating a weak or subdued pattern compared to other segments.
def is_low_volume(segment):
    return volume(segment) < is_high_volume_VALUE

Checks whether the values in the segment consistently increase over time (overall upward trend).
def is_rising(segment):
    return all(slope(segment) > 0)

Checks whether the values in the segment consistently decrease over time (overall downward trend).
def is_falling(segment):
    return all(slope(segment) < 0)

Checks whether the slope of the segment steadily decreases, forming a curve that bends downward (concave shape).
def is_concave(segment):
    return np.all(np.diff(slope(segment)) < 0)

Checks whether the slope of the segment steadily increases, forming a curve that bends upward (convex shape).
def is_convex(segment):
    return np.all(np.diff(slope(segment)) > 0)

Checks whether the segment is mirror-symmetric around its center by comparing the first half with the reversed second half.
def is_symmetric(segment):
    n = len(segment)
    return np.allclose(segment[:n // 2], segment[-(n // 2):][::-1])

Checks whether the segment is not symmetric around its center (i.e., it does not mirror itself).
def is_asymmetric(segment):
    return not is_symmetric(segment)

All thresholds must be derived directly from the provided feature values using consistent statistical logic.
All window must be derived directly based on provided feature functions and data. Try to suggest window size so apparent patterns show up.

Return your output as a plain dictionary. Do NOT use markdown, code formatting, or any extra text.

Use the following format exactly:

{"is_linear": is_linear_VALUE, "is_linear_explanation": "Explain your choice in one sentence using only words.", "linear_window_size": integer suggestion for is_linear feature,
"is_constant": is_constant_VALUE, "is_constant_explanation": "Explain your choice in one sentence using only words.", "constant_window_size": integer suggestion for is_constant feature,
"is_smooth": is_smooth_VALUE, "is_smooth_explanation": "Explain your choice in one sentence using only words.", "smooth_window_size": integer suggestion for is_constant feature,
"is_noisy": is_noisy_VALUE, "is_noisy_explanation": "Explain your choice in one sentence using only words.", "noisy_window_size": integer suggestion for is_noisy feature,
"is_complex": is_complex_VALUE, "is_complex_explanation": "Explain your choice in one sentence using only words.", "complex_window_size": integer suggestion for is_complex feature,
"is_simple": is_simple_VALUE, "is_simple_explanation": "Explain your choice in one sentence using only words.", "simple_window_size": integer suggestion for is_simple feature,
"is_spiky": is_spiky_VALUE, "is_spiky_explanation": "Explain your choice in one sentence using only words.", "spiky_window_size": integer suggestion for is_spiky feature,
"is_dropout": is_dropout_VALUE, "is_dropout_explanation": "Explain your choice in one sentence using only words.", "dropout_window_size": integer suggestion for is_dropout feature,
"is_periodic": is_periodic_VALUE, "is_periodic_explanation": "Explain your choice in one sentence using only words.", "periodic_window_size": integer suggestion for is_periodic feature,
"is_step": is_step_VALUE, "is_step_explanation": "Explain your choice in one sentence using only words.", "step_window_size": integer suggestion for is_step feature,
"is_high_amplitude": is_high_amplitude_VALUE, "is_high_amplitude_explanation": "Explain your choice in one sentence using only words.",  "high_amplitude_window_size": integer suggestion for is_high_amplitude feature,
"is_high_volume": is_high_volume_VALUE, "is_high_volume_explanation": "Explain your choice in one sentence using only words.",  "high_volume"_window_size": integer suggestion for is_high_volume" feature,
"is_low_volume": is_low_volume_VALUE, "is_low_volume_explanation": "Explain your choice in one sentence using only words.", "low_volume_window_size": integer suggestion for is_low_volume feature,
"rising_window_size": integer suggestion for is_rising feature,
"asymmetric_window_size": integer suggestion for is_asymmetric feature,
"falling_window_size": integer suggestion for is_falling feature,
"concave_window_size": integer suggestion for is_concave feature,
"convex_window_size": integer suggestion for is_convex feature,
"symmetric_window_size": integer suggestion for is_symmetric feature}


Here is the data: (*@\textcolor{red}{[INSERT STATISTICAL SUMMARY]}@*)
\end{CodeBlock}

\noindent{\normalsize \textbf{Global Shape Feature Prompt}}
\begin{CodeBlock}
You are an AI assistant designed to support analysts and researchers in interpreting time series data from natural language queries.

Your ONLY task is to compute dynamic threshold values for key statistical features based on the provided data. Do NOT generate or reformulate queries. Do NOT modify the input.

The input is a list of dictionaries, where each dictionary represents one segment of a time series with the following numerical features:
"mean", "median", "std", "min", "max", "range", "skewness", "kurtosis", and "slope".

Your job is to determine:
"high": a threshold such that approximately 70% of all "mean" values are below this value.
"low": a threshold such that approximately 30% of all "mean" values are below this value.
"variance": a threshold based on "std" that distinguishes segments with tightly clustered values from the rest.

All thresholds must be derived directly from the provided feature values using consistent statistical logic (e.g., percentiles or variance spread).

Return your output as a plain dictionary. Do NOT use markdown, code formatting, or any extra text.

Use the following format exactly:

{"high": VALUE, "high_explanation": "Explain your choice of high in one sentence using only words.",
"low": VALUE, "low_explanation": "Explain your choice of low in one sentence using only words.",
"variance": VALUE, "variance_explanation": "Explain your choice of variance in one sentence using only words."}

Here is the data: (*@\textcolor{red}{[INSERT STATISTICAL SUMMARY]}@*)
\end{CodeBlock}

\noindent{\normalsize \textbf{Statistical Summary Example}}

\begin{CodeBlock}
[{
    'segment 0': {
        'mean': 0.00353, 
        'median': 0.00344, 
        'std': 0.000272, 
        'min': 0.00322, 
        'max': 0.00414, 
        'range': 0.000928, 
        'skewness': 0.951, 
        'kurtosis': -0.00107, 
        'slope': -3.06e-05, 
        'volume': 0.0353,
        'length': 6}
},
{
    'segment 1': {
        ...
        ...
        ...
\end{CodeBlock}

\clearpage

\begin{center}
\noindent{\LARGE \textbf{Appendix B: Feature Definition}}
\end{center}

All features inferred through dynamic thresholding include an additional parameter, VALUE, in the function call. This VALUE is determined by the LLM inference pipeline.

\begin{longtable}{|
p{1.7cm}|
>{\centering\arraybackslash}m{3cm}|
>{\centering\arraybackslash}m{3cm}|
>{\raggedright\arraybackslash}m{9cm}|
}
\caption{Local Feature Definitions} \\
\hline
\multicolumn{1}{|c|}{\textbf{Feature}} & 
\textbf{Shape Segment} & 
\textbf{Explanation} & 
\multicolumn{1}{c|}{\textbf{Code}} \\
\hline


Rising &
\includegraphics[width=0.15\textwidth]{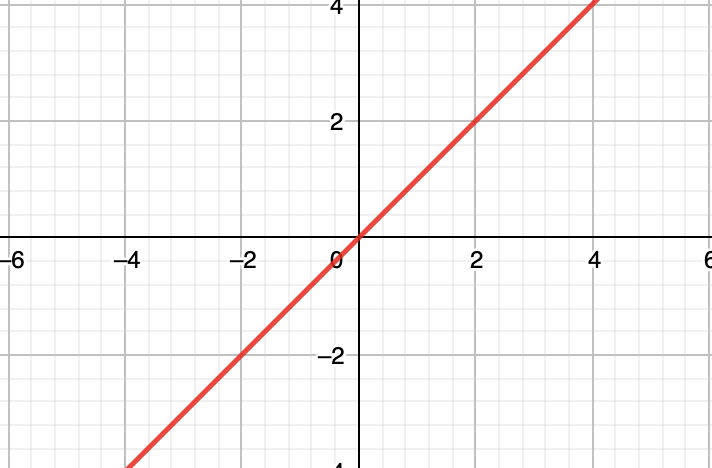} &
Segments that rise monotonically &
\begin{minipage}[t]{5cm}
\vspace*{-\baselineskip}    
{\normalsize
\lstset{
  aboveskip=0pt,
  belowskip=0pt
}
\begin{lstlisting}
def is_rising(segment):
    return all(slope(segment) > 0)
\end{lstlisting}}
\end{minipage}
\\
\hline

Falling &
\includegraphics[width=0.15\textwidth]{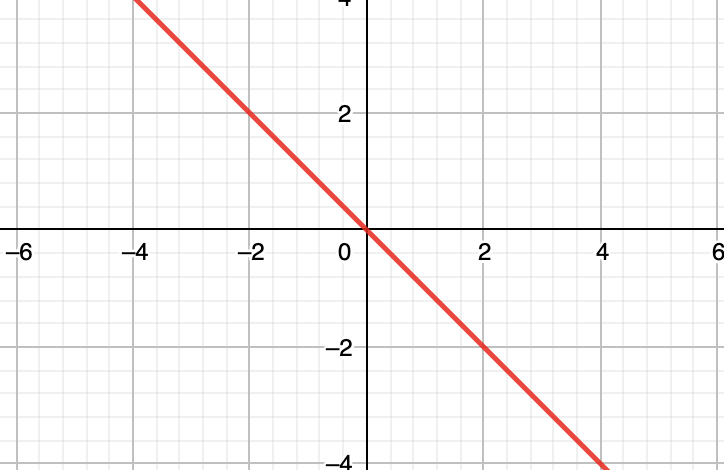} &
Segments that rise monotonically &
\begin{minipage}[t]{5cm}
\vspace*{-\baselineskip}    
{\normalsize
\lstset{
  aboveskip=0pt,
  belowskip=0pt
}
\begin{lstlisting}
def is_falling(segment):
    return all(slope(segment) < 0)
\end{lstlisting}}
\end{minipage}
\\
\hline

Concave &
\includegraphics[width=0.15\textwidth]{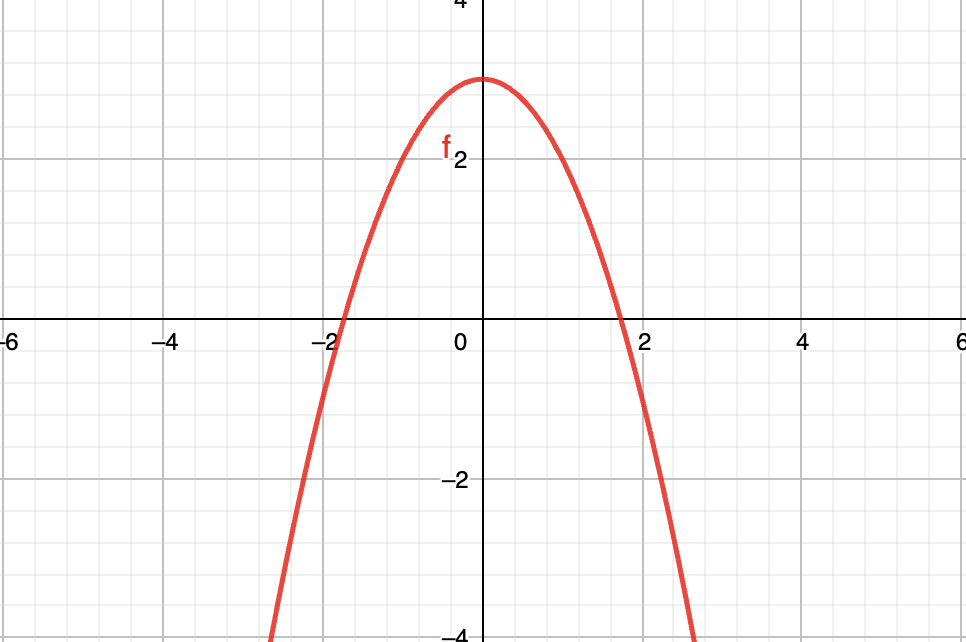} &
Segments that are curving downward &
\begin{minipage}[t]{5cm}
\vspace*{-\baselineskip}    
{\normalsize
\lstset{
  aboveskip=0pt,
  belowskip=0pt
}
\begin{lstlisting}
def is_concave(segment):
    return np.all(np.diff(slope(segment)) < 0)
\end{lstlisting}}
\end{minipage}
\\
\hline

Convex &
\includegraphics[width=0.15\textwidth]{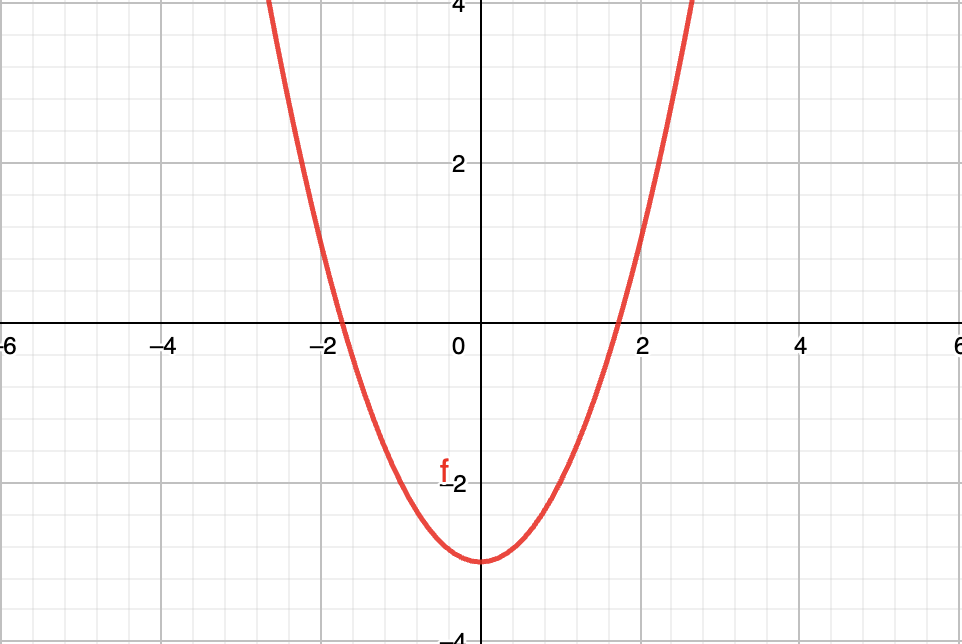} &
Segments that are curving upward &
\begin{minipage}[t]{5cm}
\vspace*{-\baselineskip}    
{\normalsize
\lstset{
  aboveskip=0pt,
  belowskip=0pt
}
\begin{lstlisting}
def is_convex(segment):
    return np.all(np.diff(slope(segment)) > 0)
\end{lstlisting}}
\end{minipage}
\\
\hline

Linear &
\includegraphics[width=0.15\textwidth]{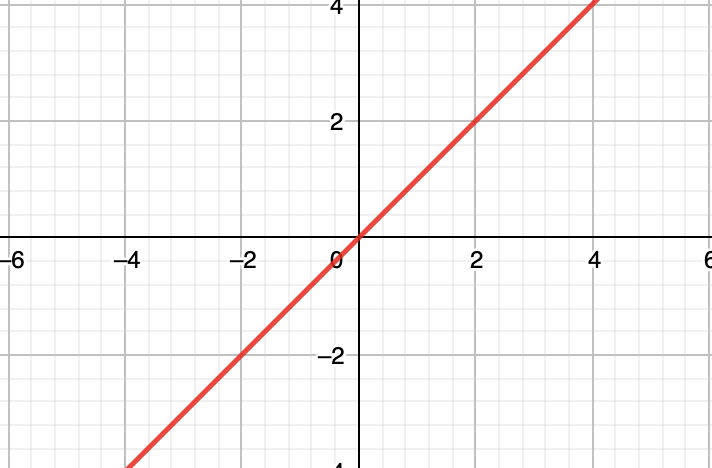} &
Segments where the rate of change varies very little compared to others &
\begin{minipage}[t]{5cm}
\vspace*{-\baselineskip}    
{\normalsize
\lstset{
  aboveskip=0pt,
  belowskip=0pt
}
\begin{lstlisting}
def is_linear(segment, VALUE):
    return variance(slope(segment)) < VALUE
\end{lstlisting}}
\end{minipage}
\\
\hline

Non Linear &
\includegraphics[width=0.15\textwidth]{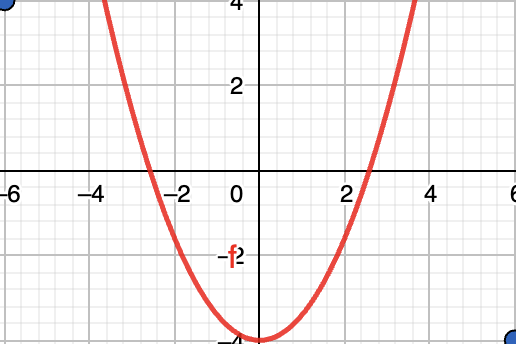} &
Segments where the rate of change varies a lot compared to others &
\begin{minipage}[t]{5cm}
\vspace*{-\baselineskip}    
{\normalsize
\lstset{
  aboveskip=0pt,
  belowskip=0pt
}
\begin{lstlisting}
def is_non_linear(segment, VALUE):
    return variance(slope(segment)) > VALUE
\end{lstlisting}}
\end{minipage}
\\
\hline

Constant &
\includegraphics[width=0.15\textwidth]{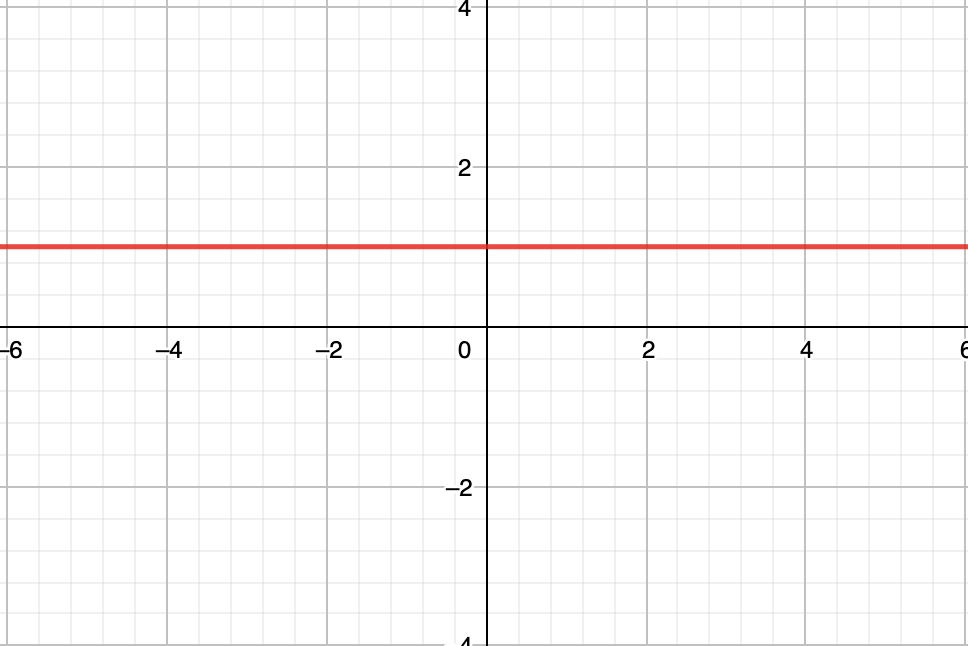} &
Segments where the values show minimal variation &
\begin{minipage}[t]{5cm}
\vspace*{-\baselineskip}    
{\normalsize
\lstset{
  aboveskip=0pt,
  belowskip=0pt
}
\begin{lstlisting}
def is_constant(segment, VALUE):
    return variance(segment) < VALUE
\end{lstlisting}}
\end{minipage}
\\
\hline

Smooth &
\includegraphics[width=0.15\textwidth]{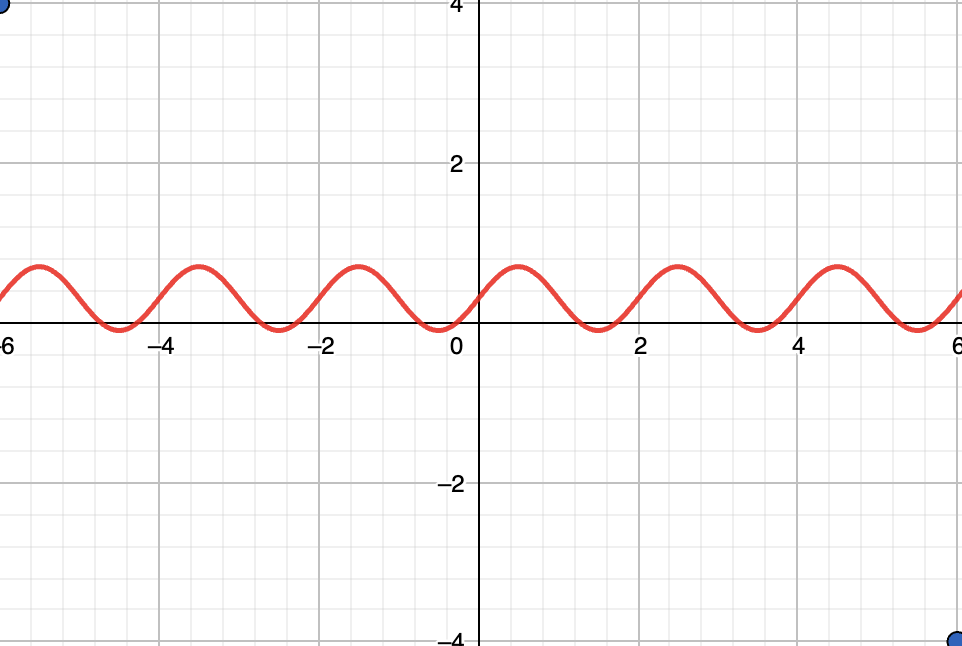} &
Segments with relatively low variation in values compared to others &
\begin{minipage}[t]{5cm}
\vspace*{-\baselineskip}    
{\normalsize
\lstset{
  aboveskip=0pt,
  belowskip=0pt
}
\begin{lstlisting}
def is_smooth(segment, VALUE):
    return variance(segment) < VALUE
\end{lstlisting}}
\end{minipage}
\\
\hline

Noisy &
\includegraphics[width=0.15\textwidth]{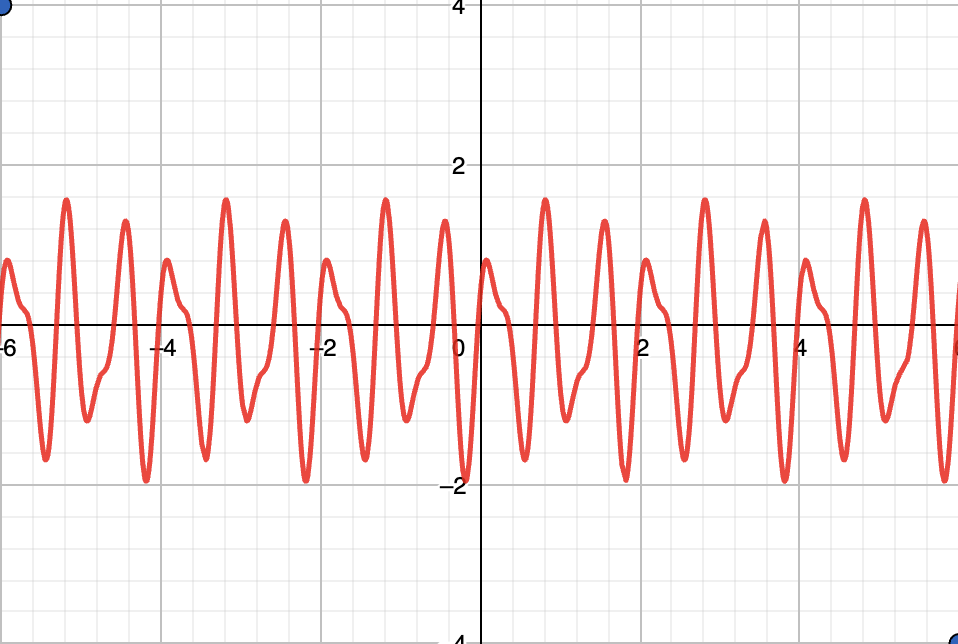} &
Segments with relatively high variation in values compared to others &
\begin{minipage}[t]{5cm}
\vspace*{-\baselineskip}    
{\normalsize
\lstset{
  aboveskip=0pt,
  belowskip=0pt
}
\begin{lstlisting}
def is_noisy(segment, VALUE):
    return variance(segment) > VALUE
\end{lstlisting}}
\end{minipage}
\\
\hline

Complex &
\includegraphics[width=0.15\textwidth]{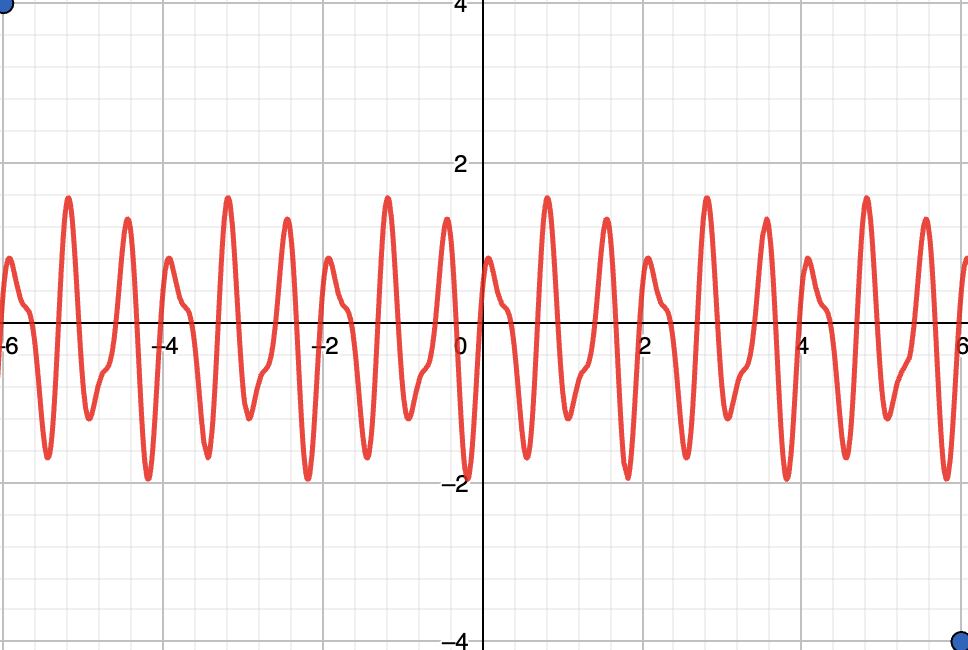} & 
segments that show both high variability and multiple peaks & 
\begin{minipage}[t]{5cm}
\vspace*{-\baselineskip}    
{\normalsize
\lstset{
  aboveskip=0pt,
  belowskip=0pt
}
\begin{lstlisting}
def is_complex(segment, VALUE):
    return variance(segment) > VALUE and \
           len(find_peaks(segment)[0]) > 2
\end{lstlisting}}
\end{minipage}
\\
\hline

Simple & 
\includegraphics[width=0.15\textwidth]{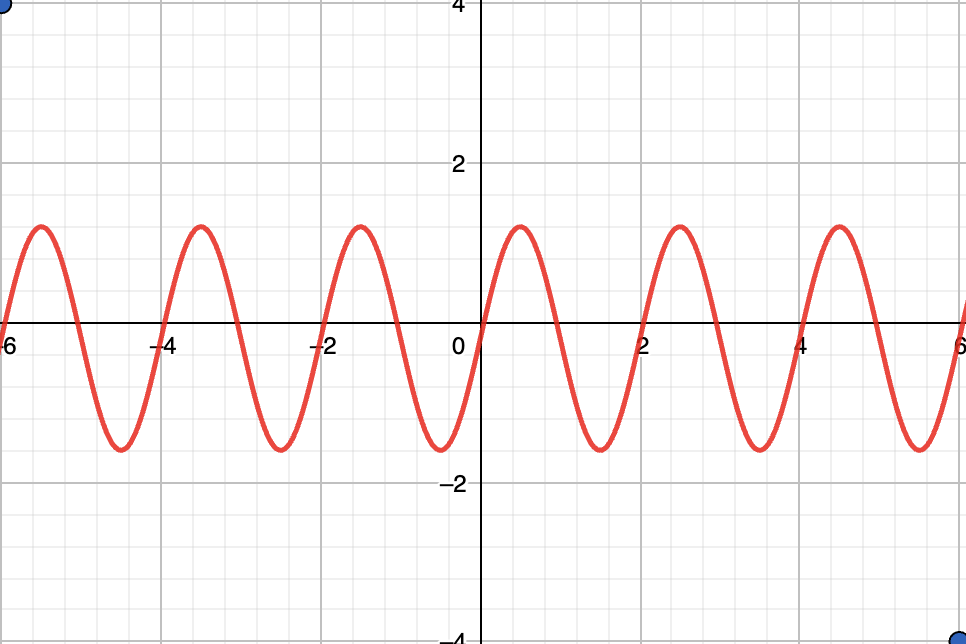} & 
segments with low variability and no prominent peaks & 
\begin{minipage}[t]{5cm}
\vspace*{-\baselineskip}    
{\normalsize
\lstset{
  aboveskip=0pt,
  belowskip=0pt
}
\begin{lstlisting}
def is_simple(segment, VALUE):
    return variance(segment) < VALUE and \
           len(find_peaks(segment)[0]) == 0
\end{lstlisting}}
\end{minipage}
\\
\hline
Spiky & 
\includegraphics[width=0.15\textwidth]{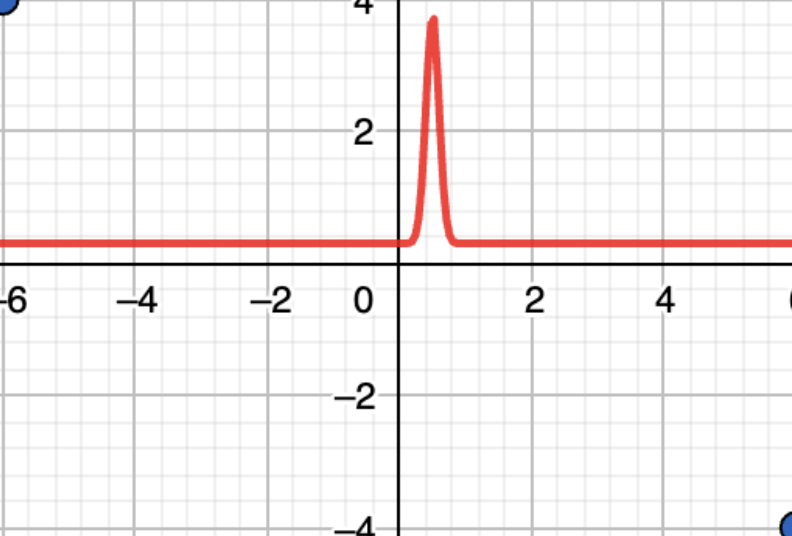}& 
segments that contain at least one peak and show sudden sharp changes in value & 
\begin{minipage}[t]{5cm}
\vspace*{-\baselineskip}    
{\normalsize
\lstset{
  aboveskip=0pt,
  belowskip=0pt
}
\begin{lstlisting}
def is_spiky(segment, VALUE):
    peaks, _ = find_peaks(segment)
    return len(peaks) > 0 and \
    np.any(np.abs(np.diff(segment)) > VALUE)
\end{lstlisting}}
\end{minipage}
\\
\hline

Dropout & \includegraphics[width=0.15\textwidth]{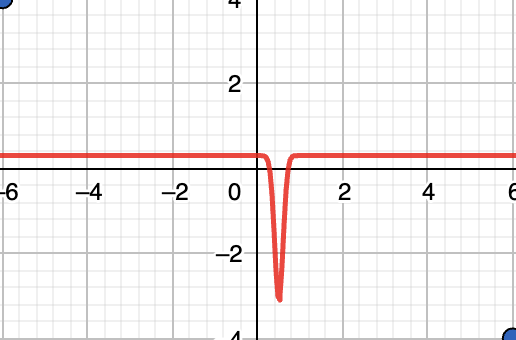} & 
segments that contains at least one point x such that x is significantly smaller than the rest of the data in that segment &
\begin{minipage}[t]{5cm}
\vspace*{-\baselineskip}    
{\normalsize
\lstset{
  aboveskip=0pt,
  belowskip=0pt
}
\begin{lstlisting}
def is_dropout(segment, VALUE):
    return any(x < VALUE for x in segment)
\end{lstlisting}}
\end{minipage}
\\
\hline

Periodic & 
\includegraphics[width=0.15\textwidth]{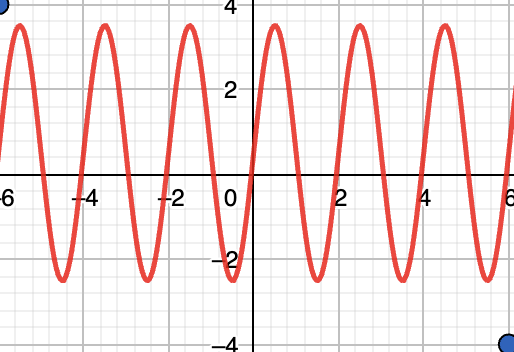} & 
segments that exhibit noticeable repeating frequency components & 
\begin{minipage}[t]{5cm}
\vspace*{-\baselineskip}    
{\normalsize
\lstset{
  aboveskip=0pt,
  belowskip=0pt
}
\begin{lstlisting}
def is_periodic(segment, VALUE):
    freq = np.abs(fft(segment))
    return np.any(freq > VALUE)
\end{lstlisting}}
\end{minipage}
\\
\hline

Aperiodic & 
\includegraphics[width=0.15\textwidth]{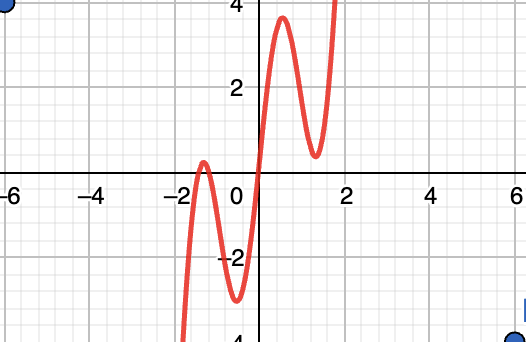} & 
segments that does not exhibit noticeable repeating frequency components & 
\begin{minipage}[t]{5cm}
\vspace*{-\baselineskip}    
{\normalsize
\lstset{
  aboveskip=0pt,
  belowskip=0pt
}
\begin{lstlisting}
def is_aperiodic(segment, VALUE):
    freq = np.abs(fft(segment))
    return np.all(freq < VALUE)
\end{lstlisting}}
\end{minipage}
\\
\hline

Symmetric & 
\includegraphics[width=0.15\textwidth]{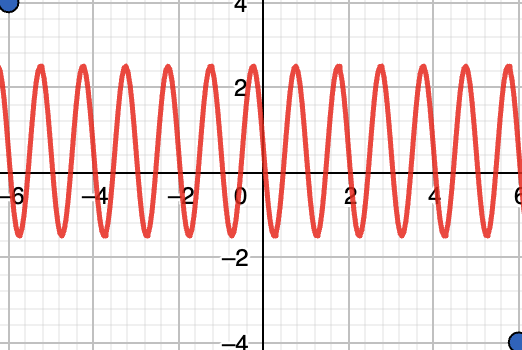} & 
segment mirrors itself around the center & 
\begin{minipage}[t]{5cm}
\vspace*{-\baselineskip}    
{\normalsize
\lstset{
  aboveskip=0pt,
  belowskip=0pt
}
\begin{lstlisting}
def is_symmetric(segment):
    n = len(segment)
    return np.allclose(segment[:n // 2], 
    segment[-(n // 2):][::-1])
\end{lstlisting}}
\end{minipage}
\\
\hline

Asymmetric & 
\includegraphics[width=0.15\textwidth]{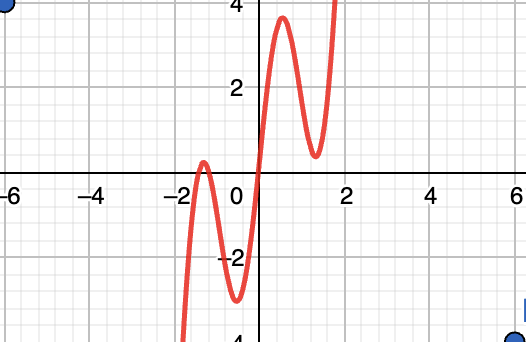} & 
segment does not mirrors itself around the center & 
\begin{minipage}[t]{5cm}
\vspace*{-\baselineskip}    
{\normalsize
\lstset{
  aboveskip=0pt,
  belowskip=0pt
}
\begin{lstlisting}
def is_asymmetric(segment):
    return not is_symmetric(segment)
\end{lstlisting}}
\end{minipage}
\\
\hline

Step & 
\includegraphics[width=0.15\textwidth]{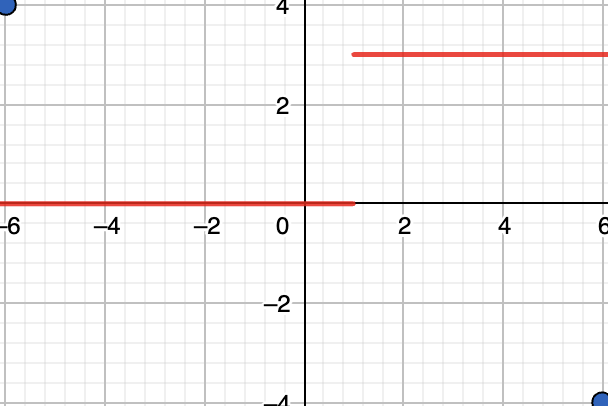} & 
segments that contain at least one sudden and significant jump in value & 
\begin{minipage}[t]{5cm}
\vspace*{-\baselineskip}    
{\normalsize
\lstset{
  aboveskip=0pt,
  belowskip=0pt
}
\begin{lstlisting}
def is_step(segment, VALUE):
    return any(np.abs(np.diff(segment)) > VALUE)
\end{lstlisting}}
\end{minipage}
\\
\hline

No Step & 
\includegraphics[width=0.15\textwidth]{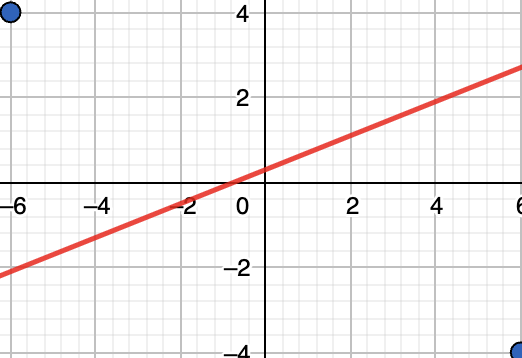} & 
segments that does not contain at sudden  and significant jump in value & 
\begin{minipage}[t]{5cm}
\vspace*{-\baselineskip}    
{\normalsize
\lstset{
  aboveskip=0pt,
  belowskip=0pt
}
\begin{lstlisting}
def is_no_step(segment, VALUE):
    return all(np.abs(np.diff(segment)) < VALUE)
\end{lstlisting}}
\end{minipage}
\\
\hline

High Amplitude & 
\includegraphics[width=0.15\textwidth]{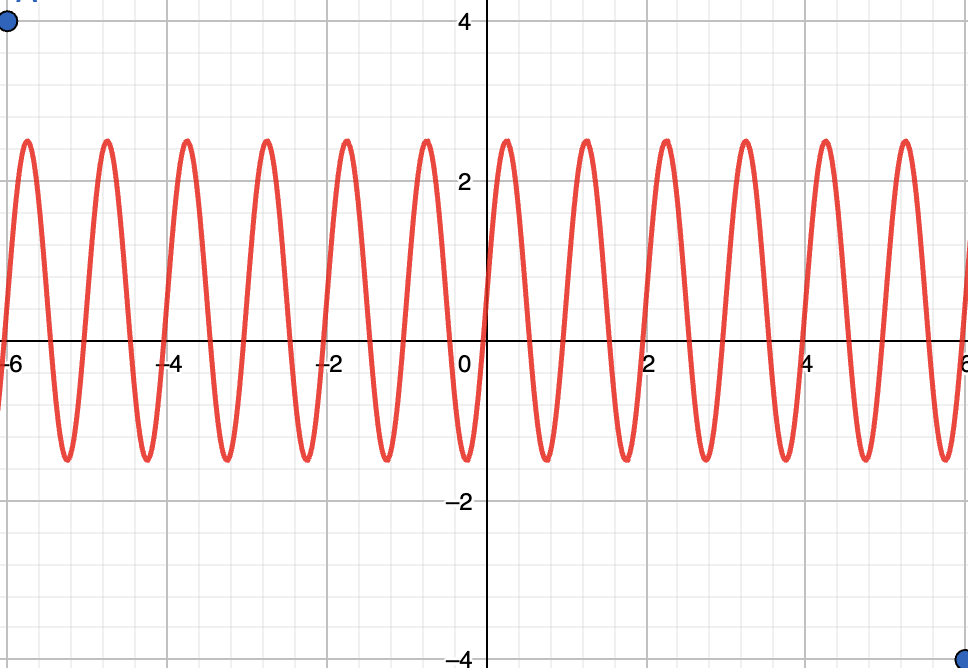} & 
segments with relatively large differences between their highest and lowest values & 
\begin{minipage}[t]{5cm}
\vspace*{-\baselineskip}    
{\normalsize
\lstset{
  aboveskip=0pt,
  belowskip=0pt
}
\begin{lstlisting}
def is_high_amplitude(segment, VALUE):
    return amplitude(segment) > VALUE
\end{lstlisting}}
\end{minipage}
\\
\hline

Low Amplitude & 
\includegraphics[width=0.15\textwidth]{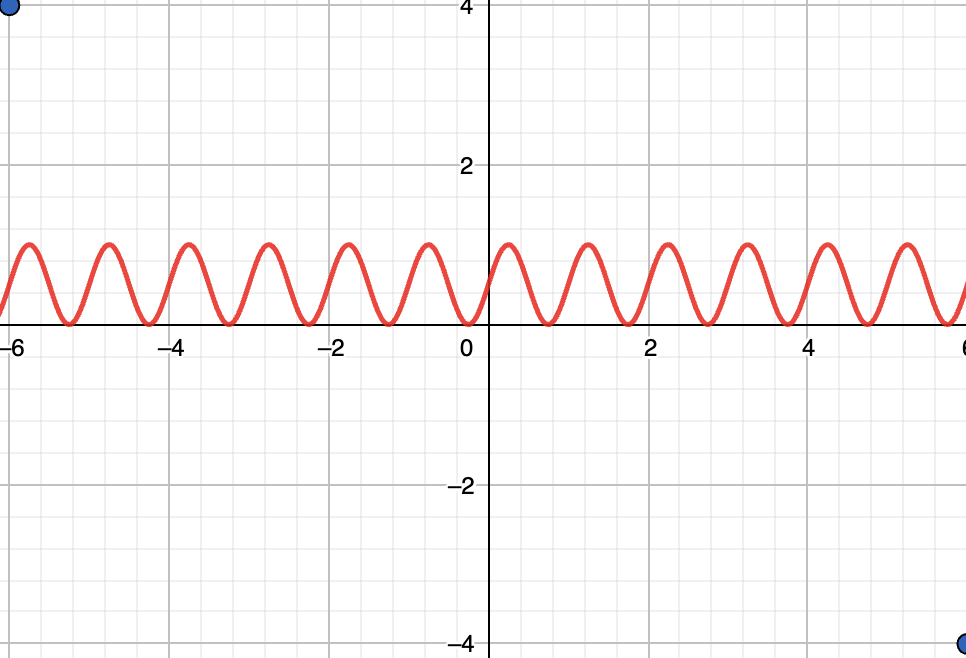} & 
segments with relatively large small between their highest and lowest values & 
\begin{minipage}[t]{5cm}
\vspace*{-\baselineskip}    
{\normalsize
\lstset{
  aboveskip=0pt,
  belowskip=0pt
}
\begin{lstlisting}
def is_low_amplitude(segment, VALUE):
    return amplitude(segment) < VALUE
\end{lstlisting}}
\end{minipage}
\\
\hline

High Volume & 
\includegraphics[width=0.15\textwidth]{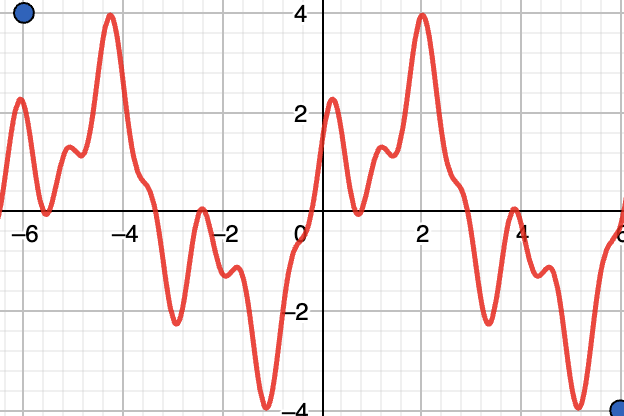} & 
segments that exhibit relatively strong intensity, based on their overall magnitude &
\begin{minipage}[t]{5cm}
\vspace*{-\baselineskip}    
{\normalsize
\lstset{
  aboveskip=0pt,
  belowskip=0pt
}
\begin{lstlisting}
def is_high_volume(segment, VALUE):
    return volume(segment) > VALUE
\end{lstlisting}}
\end{minipage}
\\
\hline

Low Volume & 
\includegraphics[width=0.15\textwidth]{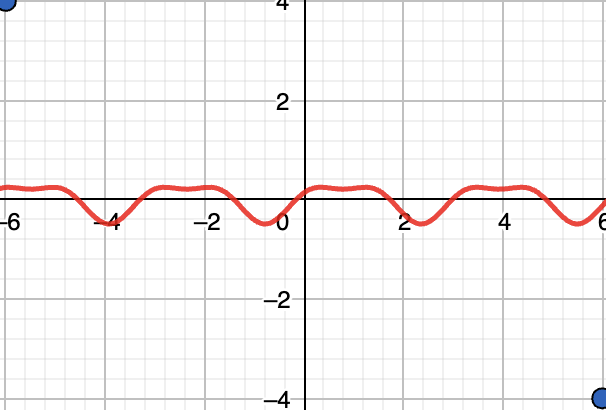} & 
segments that exhibit relatively small magnitude &
\begin{minipage}[t]{5cm}
\vspace*{-\baselineskip}    
{\normalsize
\lstset{
  aboveskip=0pt,
  belowskip=0pt
}
\begin{lstlisting}
def is_low_volume(segment, VALUE):
    return volume(segment) < VALUE
\end{lstlisting}}
\end{minipage}
\\
\hline

\end{longtable}

\clearpage
\begin{center}
\noindent{\LARGE \textbf{Appendix C: User Study Setup}}
\end{center}

\noindent Before the users official begin their tasks, we ask them to fill out a pre user study survey which includes numerical answers as well as rating from 1-5. 1 being strongly disagree while 5 is strongly agree. \\

\noindent{\Large \textbf{Pre User Study Questions}}

\begin{enumerate}[label=\textbf{Q\arabic*.}]
    \item Years of experience using data visualization tools [Numeric Answer]
    \item Years of experience using ChatGPT or other AI dialogue systems [Numeric Answer]
    \item I frequently work with time-series data [Rating 1-5]
    \item I feel comfortable describing patterns or shapes in natural language
 [Rating 1-5]
    \item I feel confident sketching patterns or shapes(e.g., rising, falling, U-shapes, peaks).
 [Rating 1-5]\\

\end{enumerate}

\noindent{\Large \textbf{Task 1}}

\noindent Users are provided with the AAPL stock price of the last 3 years and were asked to use our system to answer the following questions.

\begin{enumerate}[label=\textbf{Q\arabic*.}]
    \item Count the number of 6 day periods where AAPL’s price increases consecutively each day
    \item Locate a three-day window during which AAPL’s prices are relatively high compared to the rest of the time series
    \item Locate a pattern with a window size of 10 in AAPL stock have a pattern that looks like a wave
    \item Suppose you are going to buy AAPL stocks and sell them after a week. Show me segments where it would be a bad time to buy them
    \item Select a continuous 30-day window where the stock price remains mostly steady over time. Follow up, Now there are many results. Can you narrow down the results to 1 result?
    \item Find me segments of length 5 that are shaped like a V and in a good time to buy in
    \item Find out whether AAPL has more 7-day streaks of price increases or 7-day streaks of price declines.
    \item Find all 5-day segments where AAPL’s price continuously declines. Then choose one interesting pattern. Then, identify another 5-day pattern looks similar to the chosen one. \\

\end{enumerate}

\noindent{\Large \textbf{Task 2}}

\noindent Users are provided with the daily temperature of [our local city] since 2021 and were asked to perform free exploration to find anything that seems interesting to them.

\begin{enumerate}[label=\textbf{Q\arabic*.}]
    \item Your job is to do free exploration and find anything that seems interesting or is a fact \\
\end{enumerate}

\noindent{\Large \textbf{Post User Study Questions}}

\noindent After users completed the 2 tasks, they will conclude their user study by filling out a post user study survey. All questions on this survey are rating based from 1 to 5. 1 being strongly disagree and 5 being strongly agree. \\

\begin{center}
\centering
\begin{tabular}{p{0.95\linewidth}}
\textbf{Q1.} I feel comfortable describing patterns or shapes in natural language using \system \\[6pt]
\textbf{Q2.} I feel confident sketching patterns or curves using \system(e.g., rising, falling, U-shapes, peaks). \\[6pt]
\textbf{Q3.} I felt the system translated my text query into visual results accurately \\[6pt]
\textbf{Q4.} The sketch input matched the results I expected \\[6pt]
\textbf{Q5.} I thought the system was easy to use \\[6pt]
\textbf{Q6.} The interface was intuitive and well-organized. \\[6pt]
\textbf{Q7.} I would imagine that most people would learn to use this system very quickly. \\[6pt]
\textbf{Q8.} I felt sketching was more efficient for certain types of queries \\[6pt]
\textbf{Q9.} I found it easy to adjust the feature parameters \\[6pt]
\textbf{Q10.} I think that using natural language to generate trends and sketching to refine results is effective \\[6pt]
\textbf{Q11.} I thought there was very few inconsistency in this system \\[6pt]
\textbf{Q12.} I can iteratively refine my query whenever the initial query does not return the expected results \\
\end{tabular}
\end{center}

\clearpage
\vspace{1cm}
\begin{center}
\noindent{\LARGE \textbf{Appendix D: User Study Results}}
\end{center}


\noindent{\Large \textbf{Pre User Study Results}}

\begin{center}
\begin{tabular}{|p{1.5cm}|p{1cm}|p{1cm}|p{1cm}|p{1cm}|p{1cm}|}
\hline
\centering Participants & \centering Q1 & \centering Q2 & \centering Q3 & \centering Q4 & \centering Q5 \tabularnewline
\hline
1 & 2 & 1 & 1 & 4 & 5 \\
\hline
2 & 5 & 3 & 2 & 3 & 4 \\
\hline
3 & 5 & 2 & 5 & 3 & 4 \\
\hline
4 & 6 & 2 & 4 & 5 & 5 \\
\hline
5 & 6 & 2.5 & 3 & 4 & 5 \\
\hline
6 & 5 & 2 & 2 & 4 & 4 \\
\hline
7 & 1 & 2 & 3 & 4 & 4 \\
\hline
8 & 4 & 2 & 5 & 4 & 4 \\
\hline
9 & 2 & 3 & 5 & 4 & 4 \\
\hline
10 & 2 & 3 & 3 & 4 & 4 \\
\hline
11 & 3 & 3 & 1 & 4 & 4 \\
\hline
12 & 4 & 3 & 4 & 4 & 4 \\
\hline
13 & 4 & 3 & 4 & 4 & 5 \\
\hline
\end{tabular}
\label{pre-study}
\end{center}

\vspace{0.5in}
\noindent{\Large \textbf{Post User Study Results}}

\begin{center}
\begin{tabular}{|p{0.8cm}|p{0.8cm}|p{0.8cm}|p{0.8cm}|p{0.8cm}|p{0.8cm}|p{0.8cm}|p{0.8cm}|p{0.8cm}|p{0.8cm}|p{0.8cm}|p{0.8cm}|p{0.8cm}|}
\hline
PID & Q1 & Q2 & Q3 & Q4 & Q5 & Q6 & Q7 & Q8 & Q9 & Q10 & Q11 & Q12 \\
\hline
1 & 3 & 5 & 4 & 5 & 4 & 4 & 4 & 5 & 4 & 4 & 5 & 5 \\
\hline
2 & 4 & 4 & 4 & 4 & 3 & 3 & 3 & 4 & 3 & 4 & 4 & 5 \\
\hline
3 & 4 & 4 & 4 & 4 & 5 & 5 & 4 & 4 & 4 & 3 & 3 & 4 \\
\hline
4 & 5 & 5 & 3 & 4 & 5 & 4 & 4 & 4 & 4 & 5 & 3 & 4 \\
\hline
5 & 5 & 3 & 3 & 3 & 4 & 4 & 4 & 3 & 4 & 3 & 4 & 3 \\
\hline
6 & 4 & 4 & 4 & 4 & 4 & 4 & 5 & 5 & 4 & 4 & 3 & 4 \\
\hline
7 & 4 & 5 & 2 & 3 & 3 & 4 & 5 & 5 & 3 & 2 & 5 & 5 \\
\hline
8 & 5 & 5 & 4 & 5 & 5 & 5 & 5 & 5 & 5 & 5 & 4 & 5\\
\hline
9 & 5 & 4 & 4 & 5 & 5 & 4 & 5 & 5 & 4 & 5 & 4 & 5 \\
\hline
10 & 4 & 5 & 3 & 4 & 5 & 4 & 4 & 5 & 4 & 4 & 3 & 4\\
\hline
11 & 5 & 5 & 4 & 5 & 4 & 3 & 4 & 5 & 3 & 4 & 4 & 2\\
\hline
12 & 4 & 5 & 4 & 4 & 5 & 5 & 5 & 5 & 4 & 5 & 5 & 5 \\
\hline
13 & 5 & 5 & 5 & 4 & 4 & 5 & 5 & 5 & 5 & 4 & 5 & 5 \\
\hline
\end{tabular}
\end{center}

\vspace{0.5in}
\noindent{\Large \textbf{Task 1 Modality}}

\begin{tabular}{|
    >{\raggedright\arraybackslash}p{1.5cm}|
    >{\raggedright\arraybackslash}p{1.5cm}|
    >{\raggedright\arraybackslash}p{1.5cm}|
    >{\raggedright\arraybackslash}p{1.5cm}|
    >{\raggedright\arraybackslash}p{1.5cm}|
    >{\raggedright\arraybackslash}p{1.5cm}|
    >{\raggedright\arraybackslash}p{1.5cm}|
    >{\raggedright\arraybackslash}p{1.5cm}|
    >{\raggedright\arraybackslash}p{1.5cm}|
}
\hline

Participants & Q1 & Q2 & Q3 & Q4 & Q5 & Q6 & Q7 & Q8 \\
\hline
1 & NL& NL& Feature & NL & NL, Sketch, Feature & NL & NL & NL, Sketch\\
\hline
2 & NL & NL & NL &NL, Feature & NL, Sketch, Feature & NL & NL, Feature& NL  \\
\hline
3 & Feature & Feature &Sketch&Feature&Feature&Sketch&Feature&Feature \\
\hline
4 &Feature&NL, Feature&Sketch&Sketch&Feature, Sketch&Feature, Sketch&Feature&Feature, Sketch \\
\hline
5 &NL&Feature&NL&NL&Feature, Sketch&Sketch&Feature&Feature, Sketch \\
\hline
6 &NL&Feature&NL&Feature&NL, Sketch&NL, Feature&NL&NL, Sketch  \\
\hline
7 &NL&NL, Feature&Sketch&NL&NL, Sketch&Sketch&NL&NL  \\
\hline
8 &NL&NL&NL&NL&NL&NL&NL&NL  \\
\hline
9 &NL&Feature&NL&NL&Feature, Sketch&NL&NL&NL, Sketch \\
\hline
10 &NL&NL&Feature&NL, Sketch&Feature, Sketch&Feature, Sketch&Feature&Feature \\
\hline
11 &Feature&Feature&Feature, Sketch&NL&Feature, Sketch&NL, Feature, Sketch&Feature&Feature, Sketch\\
\hline
12 & NL &NL&NL&NL&NL, Sketch, Feature&NL&NL&NL, Sketch  \\
\hline
13 &NL, Sketch&NL&Sketch&Sketch&NL, Sketch&Sketch&NL&NL  \\
\hline
\end{tabular}

\clearpage
\vspace{1cm}
\begin{center}
\noindent{\LARGE \textbf{Appendix E: Latency Analysis}}
\end{center}

\noindent{\textbf{Latency Testing and Analysis}}
These experiments focus on the efficiency and responsiveness of the LLM-driven shape feature extraction (QbT) pipeline (Figure~\ref{fig:nl_workflow}) under different model choices, dataset sizes, and window lengths. Together, they answered: \textit{What is the system’s latency performance across different models and dataset sizes, and is it sufficient for real-time interaction?}

\vspace{0.5em}
\noindent{\normalsize \textbf{Experimental Considerations}} All performance tests were conducted on a standard personal laptop equipped with an AMD Ryzen 7 5800HS CPU, an NVIDIA GeForce RTX 3060 GPU, 32~GB RAM, and a stable internet connection for all model API calls. We report backend latency measured from the time a natural-language query is submitted to the time the matched segments are returned from the server, including both LLM calls and the NL-to-segment matching pipeline.

We compare three language models: GPT-4o (our default), DeepSeek Rover V2, and Devstral-Small-2505.
Each model is queried through the same backend API and receives identical few-shot prompts for feature extraction, ensuring that differences in latency reflect model and infrastructure behavior rather than prompt design.

For datasets, we use three univariate time series: monthly maximum temperature in Sacramento from 1743--2013, daily Bitcoin closing prices from 2014--2022, and hourly energy consumption data with 145{,}367 observations. These datasets span different magnitudes, seasonal behaviors, and lengths, providing a range of realistic usage scenarios. Detailed descriptions are as follows:

\begin{itemize}
\item \textbf{Sacramento weather.} This dataset captures a historical record of monthly temperature (in Celsius) from January 1849 to September 2013. The original dataset contains entries from multiple cities worldwide; we filter and preprocess it to retain only the 1{,}977 entries corresponding to Sacramento, CA.

\item \textbf{Bitcoin.} This dataset records daily Bitcoin closing prices in USD, with 2{,}713 observations. The original data are multivariate, including open, high, low, close, adjusted close, and volume. For our experiments, we preprocess the dataset to a univariate series containing only the closing price.

\item \textbf{Energy consumption.} This time series consists of hourly energy load measurements from PJM Interconnection LLC, a regional transmission organization that manages electrical transmission systems across Delaware, Illinois, Indiana, Kentucky, Maryland, Michigan, New Jersey, North Carolina, Ohio, Pennsylvania, Tennessee, Virginia, West Virginia, and the District of Columbia. Energy consumption is measured in megawatts (MW), with 145{,}367 observations in total. 
    
\end{itemize}

\vspace{0.5em}
\noindent{\normalsize \textbf{Model Comparison Latency Test}} The first experiment compares the response times of different language models in the QbT pipeline. For each model, we measure how quickly it can extract shape feature specifications for the following natural-language queries:

\begin{enumerate}
    \item Find segments where volume is high and then falls.
    \item Identify phases of steady temperature rise followed by a sudden drop.
    \item Track periods of high consumer interest followed by a sudden drop.
    \item Identify when energy consumption spikes to high levels and stabilizes.
    \item Show parts where there is a symmetrical rise and fall.
\end{enumerate}

We use a medium-sized time-series dataset, Sacramento weather dataset. Each query is executed four times per model (20 runs per model in total). We record the total end-to-end latency per run and then compute the mean, minimum, maximum, and standard deviation (SD).

\begin{table}[h]
\centering
\caption{Model comparison results for NL query latency (20 trials per model).}
\begin{tabular}{lcccc}
\toprule
\textbf{Model} & \textbf{Avg Latency (ms)} & \textbf{Min Latency (ms)} & \textbf{Max Latency (ms)} & \textbf{SD (ms)} \\
\midrule
GPT-4o & 651.59 & 481.78 & 992.97 & 137.93 \\
DeepSeek Rover V2 & 1745.00 & 535.27 & 2474.68 & 359.95 \\
Devstral-Small-2505 & 1620.73 & 1281.08 & 2052.86 & 186.25 \\
\bottomrule
\end{tabular}
\label{tab:model-latency}
\end{table}

As shown in Table~\ref{tab:model-latency}, GPT-4o consistently achieves the lowest latency across all queries, with a mean of 651.59~ms and latencies ranging from 481.78 to 992.97~ms. The largest latencies occur for broader queries that generate many candidate segments prior to anti-collision filtering, which must be resolved before results can be returned. For this experiment, we fix the window length at $L=5$ to isolate model-level differences. The relatively small SD for GPT-4o indicates that its latencies are tightly clustered around the mean, yielding predictable and stable behavior.

DeepSeek Rover V2 and Devstral-Small-2505 are both slower than GPT-4o but complete all queries without failures. DeepSeek Rover V2 exhibits the highest average latency and the largest SD, with occasional spikes up to 2474.68~ms, indicating less stable performance for real-time interaction. Devstral-Small-2505 has a lower SD than DeepSeek but a higher mean latency than GPT-4o, placing it between the two in terms of the trade-off between speed and stability.

Overall, these results confirm that GPT-4o is a suitable default for interactive use in our system, and we use it for the remaining experiments in this appendix. At the same time, the architecture supports swapping models when users or deployments prioritize other criteria such as cost or on-premise hosting.

\vspace{0.5em}
\noindent{\normalsize \textbf{Dataset Scalability Test}} The second experiment evaluates how the QbT pipeline scales with increasing dataset size. We use GPT-4o and the fixed query:

\begin{quote}
\emph{“Find segments where data is generally low with an increase and then a decrease.”}
\end{quote}

For each dataset, we hold the window length and query constant, and we measure:

\begin{itemize}
    \item Average latency (ms) over 20 runs,
    \item Average number of matched segments returned per run,
    \item Standard deviation of latency (ms).
\end{itemize}

\begin{table}[ht]
\centering
\caption{Dataset scaling performance results for the QbT pipeline (20 trials per dataset).}
\begin{tabular}{lcccc}
\toprule
\textbf{Dataset} & \textbf{Dataset Size} & \textbf{Avg Latency (ms)} & \textbf{Avg Matches} & \textbf{SD (ms)} \\
\midrule
Sacramento Weather & 1{,}977 & 670.95  & 1    & 143.81 \\
Bitcoin            & 2{,}713 & 855.78  & 298  & 188.10 \\
Energy Consumption & 145{,}367 & 4027.66 & 1938 & 238.99 \\
\bottomrule
\end{tabular}
\label{tab:scaling-performance}
\end{table}

Table~\ref{tab:scaling-performance} shows a clear trend: larger datasets lead to higher average latency. The Sacramento weather and Bitcoin series have comparable lengths, and their mean latencies are also of similar magnitude. In contrast, the energy consumption dataset is orders of magnitude longer and produces substantially more initial matches (on average 1938 segments), resulting in a mean latency of about 4~s.

The growth in latency is not strictly linear in the number of points. A substantial factor is the cost of post-processing, particularly the anti-collision routine that suppresses heavily overlapping segments. When many windows satisfy the feature predicates, this overlap resolution step becomes more expensive, as seen in the Bitcoin and energy consumption datasets. Despite this increase in absolute latency, the SD remains relatively small across all datasets, indicating that latency per dataset is stable and predictable rather than erratic. For our target real-time analysis scenarios, this behavior is desirable: users can anticipate a consistent level of responsiveness given a dataset of similar size and complexity.

These results suggest that our pipeline can handle datasets with tens of thousands of points while still providing interactive feedback, especially for medium-sized datasets.

\vspace{0.5em}
\noindent{\normalsize \textbf{Window Length Sensitivity Analysis}} Finally, we investigate how the window length $L$ influences latency and the number of matches in the QbT pipeline. Although our system also exposes a top-$k$ parameter in the sketch-based (QbS) pipeline, in practice we observe that window length is a primary driver of performance for NL-based matching because it directly controls the number and size of candidate segments.

Using the Sacramento weather dataset, we consider two sets of window lengths:

\begin{itemize}
    \item \textbf{Realistic window lengths:} $L \in \{7, 14, 21, 30, 60\}$, corresponding to weekly to roughly bi-monthly periods.
    \item \textbf{Exponentially scaling window lengths:} $L \in \{2, 4, 8, 16, 32, 64\}$.
\end{itemize}

For both sets, we reuse a single query:

\begin{quote}
\emph{“Find segments where there is an increase and then a decrease.”}
\end{quote}

As before, we measure the average latency and the number of matched segments returned by the backend.

\begin{table}[ht]
\centering
\caption{Latency and match count for realistic window lengths on the Sacramento weather dataset.}
\begin{tabular}{lcc}
\toprule
\textbf{Window Length} & \textbf{Latency (ms)} & \textbf{Matches} \\
\midrule
7  & 766.33  & 131 \\
14 & 567.71  & 8   \\
21 & 873.21  & 0   \\
30 & 545.31  & 0   \\
60 & 1176.81 & 0   \\
\bottomrule
\end{tabular}
\label{tab:window-length-realistic}
\end{table}

\begin{table}[ht]
\centering
\caption{Latency and match count for exponentially scaling window lengths on the Sacramento weather dataset.}
\begin{tabular}{lcc}
\toprule
\textbf{Window Length} & \textbf{Latency (ms)} & \textbf{Matches} \\
\midrule
2   & 936.40  & 946 \\
4   & 661.62  & 199 \\
8   & 666.17  & 123 \\
16  & 624.82  & 0   \\
32  & 564.09  & 0   \\
64  & 1226.96 & 0   \\
\bottomrule
\end{tabular}
\label{tab:window-length-exponential}
\end{table}

As shown in Table~\ref{tab:window-length-realistic}, realistic window sizes (7--60 days) exhibit two notable behaviors. First, there are no rising-and-falling patterns that span 21 days or longer for this dataset under our feature thresholds, resulting in zero matches for $L \ge 21$ and correspondingly lower costs in collision handling. Second, latency is driven by a combination of the number of candidate windows and the number of matches: for $L=7$, the query produces 131 matches and a latency of 766.33~ms, whereas for $L=14$ the match count drops to 8 and latency decreases accordingly.

A similar trend appears in Table~\ref{tab:window-length-exponential}. The smallest window length ($L=2$) yields 946 matches, leading to higher latency (936.40~ms) than $L=4$ or $L=8$, despite the shorter window size. This is because the broad query combined with tiny windows produces many overlapping hits that must be filtered by the anti-collision routine. As $L$ increases beyond 8, the pattern becomes harder to satisfy within a single window, the match count drops to zero, and latency is dominated by a single pass over the candidate windows.

These findings highlight a trade-off: very small windows can produce many broad matches and higher post-processing costs, whereas very large windows may miss shorter structures entirely but offer faster matching. In practice, we observe that moderate window sizes (e.g., 6–14 points) balance expressiveness and performance for interactive exploration. The ability to adjust $L$ directly in the control panel allows users to tune this trade-off based on their analysis goals.

\vspace{0.5em}
\noindent{\normalsize \textbf{Discussion and Experimental Implications}} Across the three experiments—model comparison, dataset scaling, and window-length sensitivity—our system demonstrates robust and predictable latency for the NL-based QbT pipeline under a range of technical and user-configurable parameters. The results show that, with GPT-4o as the underlying model, typical end-to-end query latencies for medium-sized datasets remain well below one second, while even very large datasets produce stable, though higher, response times.

These quantitative findings complement our qualitative evaluation and support that \emph{\system} can sustain near-real-time interaction for interactive time-series exploration across different models, dataset sizes, and window settings. They also illustrate how system parameters (e.g., model choice, dataset size, and window length) influence the trade-off between responsiveness and coverage, reinforcing our modular design that encourages users to explore and adapt these configurations to their analytic needs.

We do not benchmark Query-by-Sketch (QbS) latency separately in this appendix. Our backend implementation uses Pruned Dynamic Time Warping (PDTW) with a Sakoe--Chiba band for sketch matching, a method that is widely regarded as efficient and robust for time-series similarity search in visualization and data mining research. Given this well-established behavior, we choose to focus on the LLM-driven NL pipeline.

\vspace{0.5em}
\noindent{\normalsize \textbf{Dynamic Threshold Latency Test}}

\begin{table}[h]
\centering
\renewcommand{\arraystretch}{1.3} 
\begin{tabular}{c c c c}
\hline
\textbf{Runs (ms)} & \textbf{Sacramento Weather} & \textbf{Bitcoin} & \textbf{Power Consumption} \\
\hline
Run 1 & 19928.261 & 3378.066 & 10426.545 \\
Run 2 & 3770.389 & 4660.419 & 3477.049 \\
Run 3 & 3534.618 & 3481.128 & 3582.057 \\
Run 4 & 3658.744 & 3776.069 & 4034.661 \\
Run 5 & 3595.858 & 4228.401 & 3635.872 \\
Run 6 & 4498.815 & 3874.44 & 4451.632 \\
Run 7 & 8962.33 & 3764.074 & 3898.044 \\
Run 8 & 3786.021 & 3965.911 & 4123.548 \\
Run 9 & 3367.199 & 4012.65 & 5833.246 \\
Run 10 & 3919.085 & 3831.468 & 4029.175 \\
\hline
Avg (ms) & 5902.132 & 3897.2626 & 4749.1829 \\
Std (ms) & 5200.468268 & 363.6385863 & 2104.51757 \\

\hline
\end{tabular}
\caption{Global Feature Dynamic Threshold Inference Latency Test}
\label{tab:example}
\end{table}

\begin{table}[h]
\centering
\renewcommand{\arraystretch}{1.3} 
\begin{tabular}{c c c c}
\hline
\textbf{Runs (ms)} & \textbf{Sacramento Weather} & \textbf{Bitcoin} & \textbf{Power Consumption} \\
\hline
Run 1 & 47585.769 & 8463.898 & 15427.556 \\
Run 2 & 10024.17& 10984.649 & 8468.051 \\
Run 3 & 8439.493 & 8397.402 & 8395.343 \\
Run 4 & 9099.21 & 8278.842 & 13259.835 \\
Run 5 & 21419.721 & 9249.771 & 8019.842 \\
Run 6 & 28655.181 & 14858.809 & 10278.496 \\
Run 7 & 14592.406 & 8975.597 & 9383.246\\
Run 8 & 14947.945 & 39282.809 & 12714.622 \\
Run 9 & 11146.323 & 7700.612 & 11727.422 \\
Run 10 & 9436.645 & 10887.378 & 10369.598 \\

\hline

Avg (ms) & 17534.6863 & 12707.9767 & 10804.4011 \\
Std (ms) & 12355.16908 & 9570.244789 & 2435.952977 \\

\hline
\end{tabular}
\caption{Local Feature Dynamic Threshold Inference Latency Test}
\label{tab:example}
\end{table}

\textbf{Discussion on LLM based Dynamic Threshold Inference Latency}
We conducted experiments on three datasets of varying sizes 
using GPT-4o. Patch size affects computational cost: for example, using 100 patches for global feature inference is faster than using 300 patches for local feature inference. While patch size contributes to performance differences, our results indicate that the primary factor influencing computation time for both global and local feature threshold inference is not the dataset size itself, but the network latency involved in sending data to and receiving responses from the LLM. We observed several instances in which response times were unusually long, even for smaller datasets, suggesting that latency variability dominates overall runtime. These findings imply that network latency, rather than dataset size, is the main bottleneck for this part in our system.

\clearpage
\vspace{1cm}
\begin{center}
\noindent{\LARGE \textbf{Appendix F: Code}}
\end{center}

link: \url{https://github.com/via-cs/shape-talk}

\clearpage
\vspace{1cm}

\begin{center}
\noindent{\LARGE \textbf{Appendix G: Dynamic Threshold Comparison With Fix Percentile Threshold}}
\end{center}

\begin{center}
\textbf{Comparison of LLM derived threshold vs Fix Percentile using Step function on Federal Reserve Interest Rates}
\end{center}

\includegraphics[width=1\textwidth]{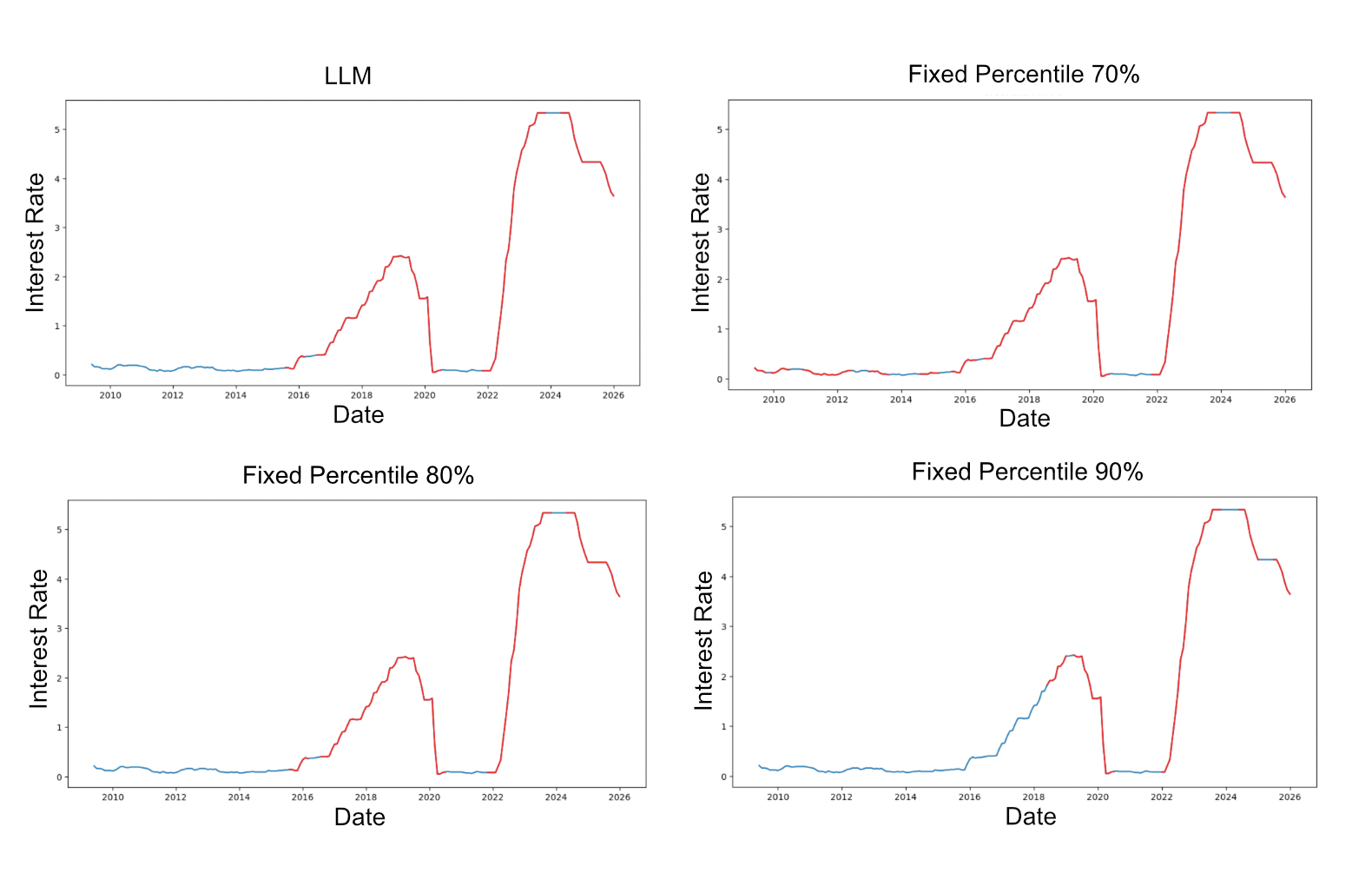}

\begin{center}
\textbf{Comparison of LLM derived threshold vs Fix Percentile using Step function on Electricity Consumption}
\end{center}

\includegraphics[width=1\textwidth]{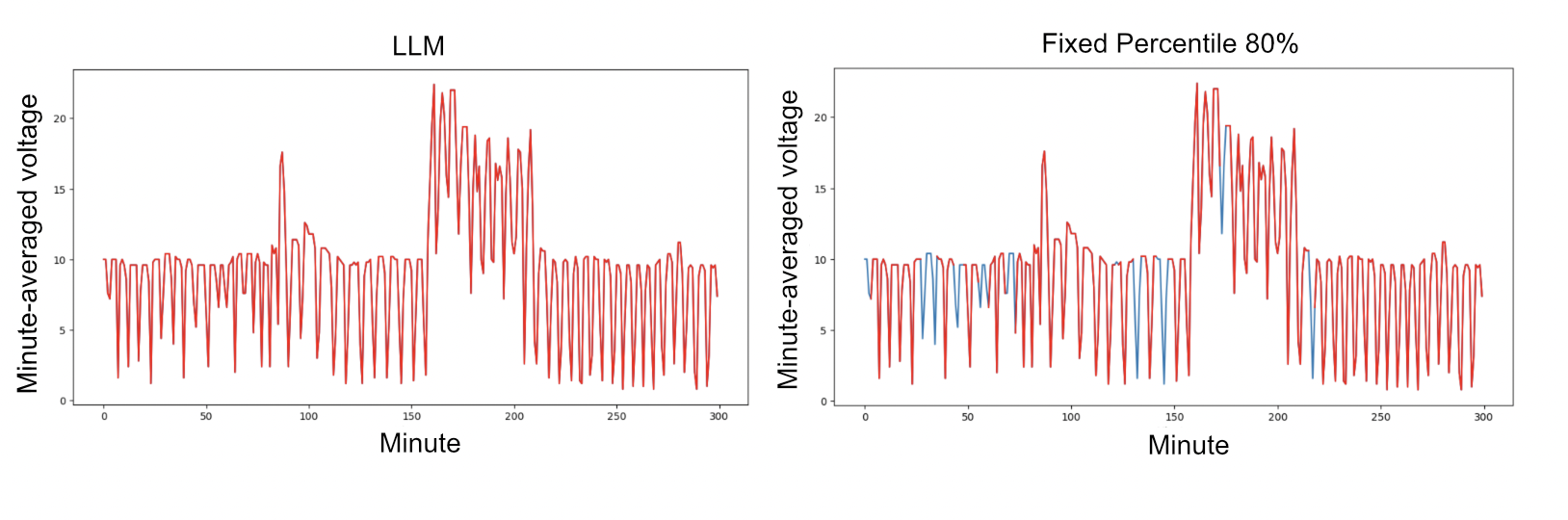}

The highlighted regions in red indicate segments that satisfy the definition of a step function.

LLM-derived thresholds provide both contextual and statistical awareness, whereas fixed percentile thresholds do not. In the Federal Reserve example, using fixed thresholds of 70\% and 90\% results in both over-selected and missed regions. In contrast, the LLM initially proposes a reasonable threshold that effectively highlights step-function segments. This does not imply that fixed percentiles such as 70\% or 90\% are inherently incorrect; rather, it shows that selecting an appropriate percentile often requires domain expertise.

Furthermore, a single fixed threshold does not generalize well across datasets. Although an 80\% threshold appears reasonable for the Federal Reserve dataset, applying the same threshold to an electricity consumption dataset causes several relevant regions to be missed. This limitation highlights the advantage of LLM-derived thresholds, which adapt to the statistical characteristics of the data distribution and provide more context-aware thresholding.

We also provide some further comparisons between LLM derived threshold and fix percentile threshold in the table below. The dataset used for the table below is the price of bitcoin. Percentile were fixed 80\% or its negation 20\%.

\begin{longtable}{|
p{2.5cm}|
>{\centering\arraybackslash}m{6.8cm}|
>{\centering\arraybackslash}m{6.8cm}|
}

\hline
\textbf{Feature} & \textbf{LLM Derived Threshold} & \textbf{Fix Percentile Threshold} \\
\hline
\endfirsthead

Linear & \includegraphics[width=0.4\textwidth]{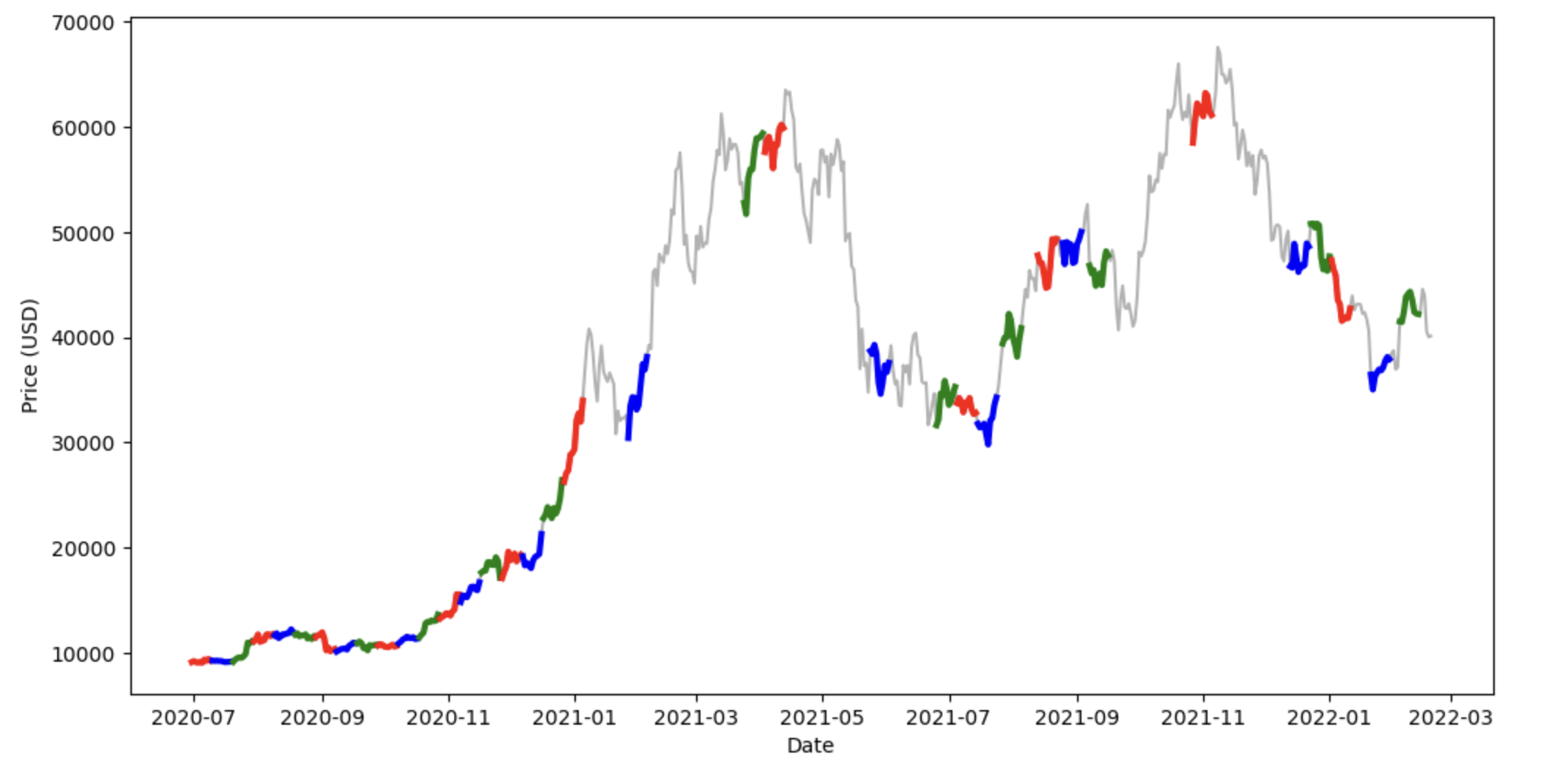} & \includegraphics[width=0.4\textwidth]{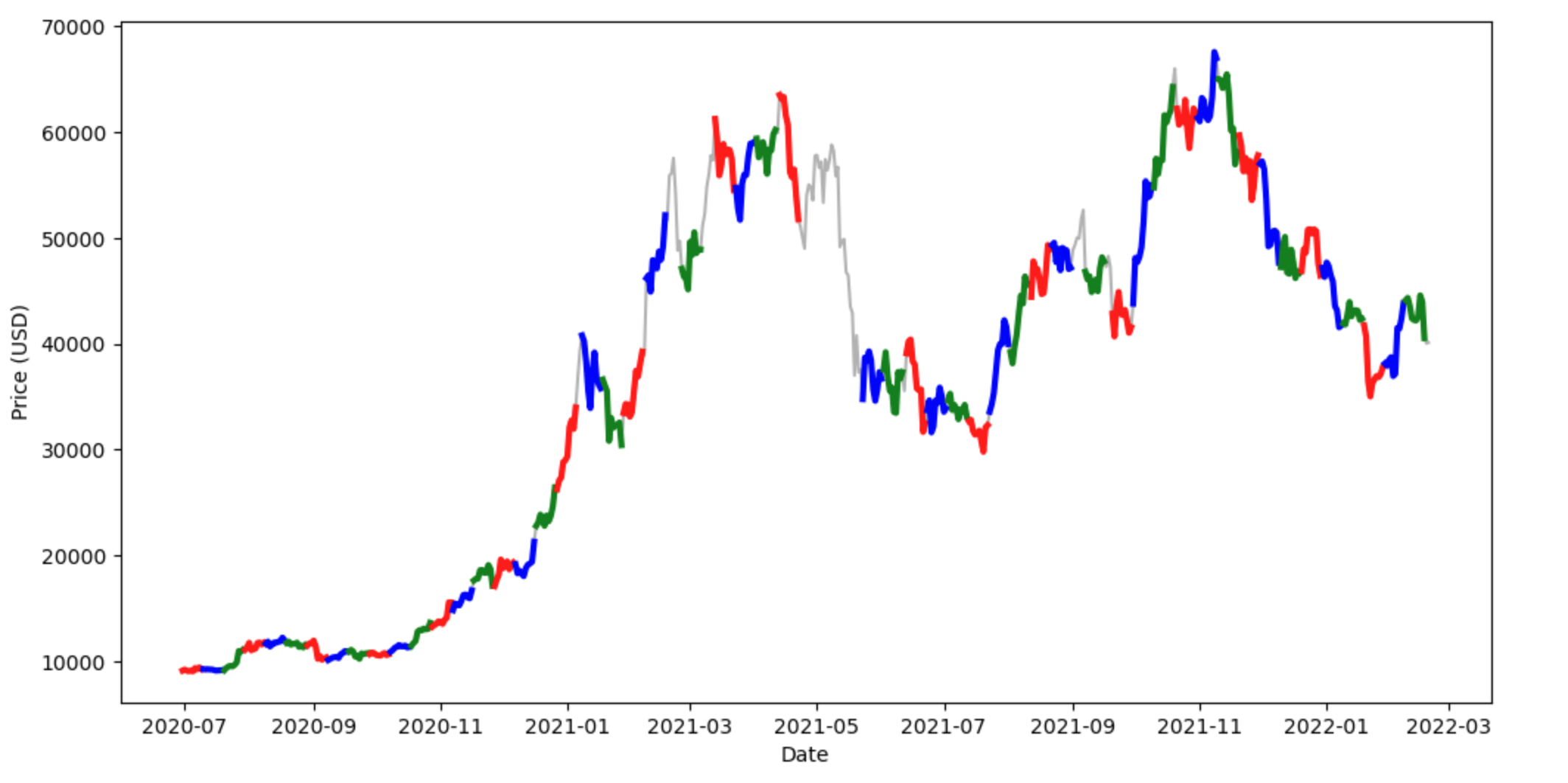} \\
\hline

Constant & \includegraphics[width=0.4\textwidth]{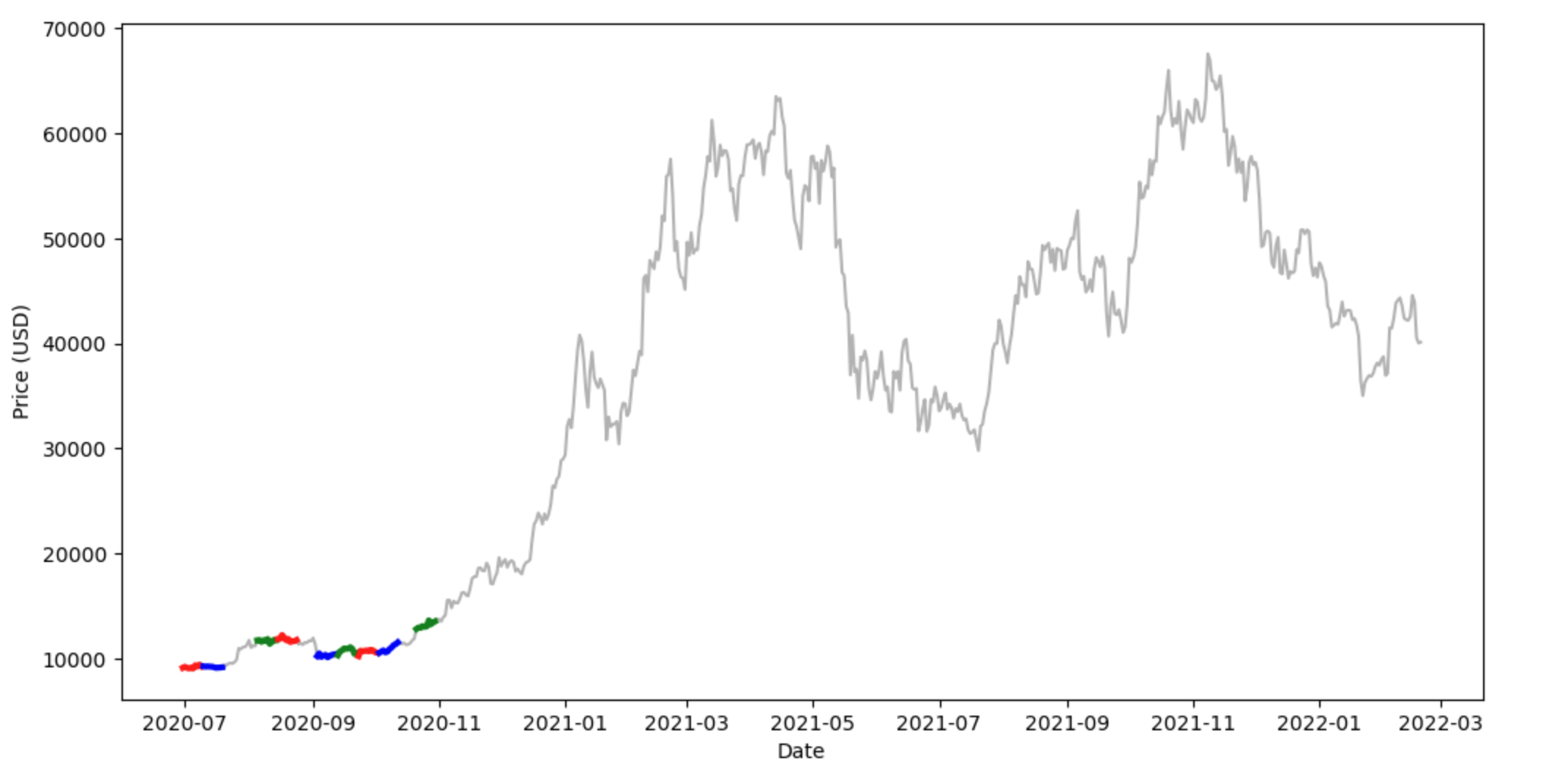} & \includegraphics[width=0.4\textwidth]{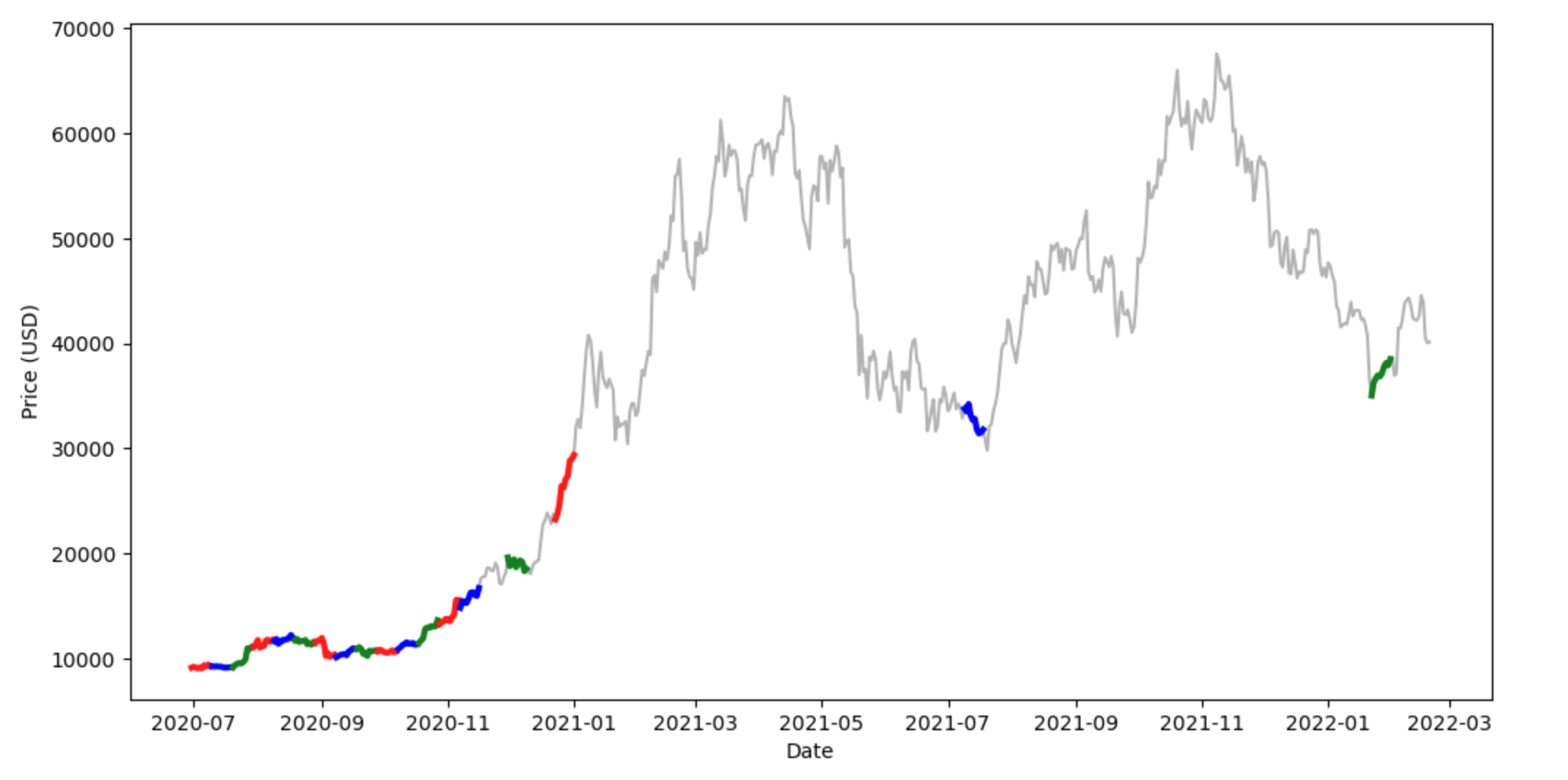} \\
\hline

Noisy & \includegraphics[width=0.4\textwidth]{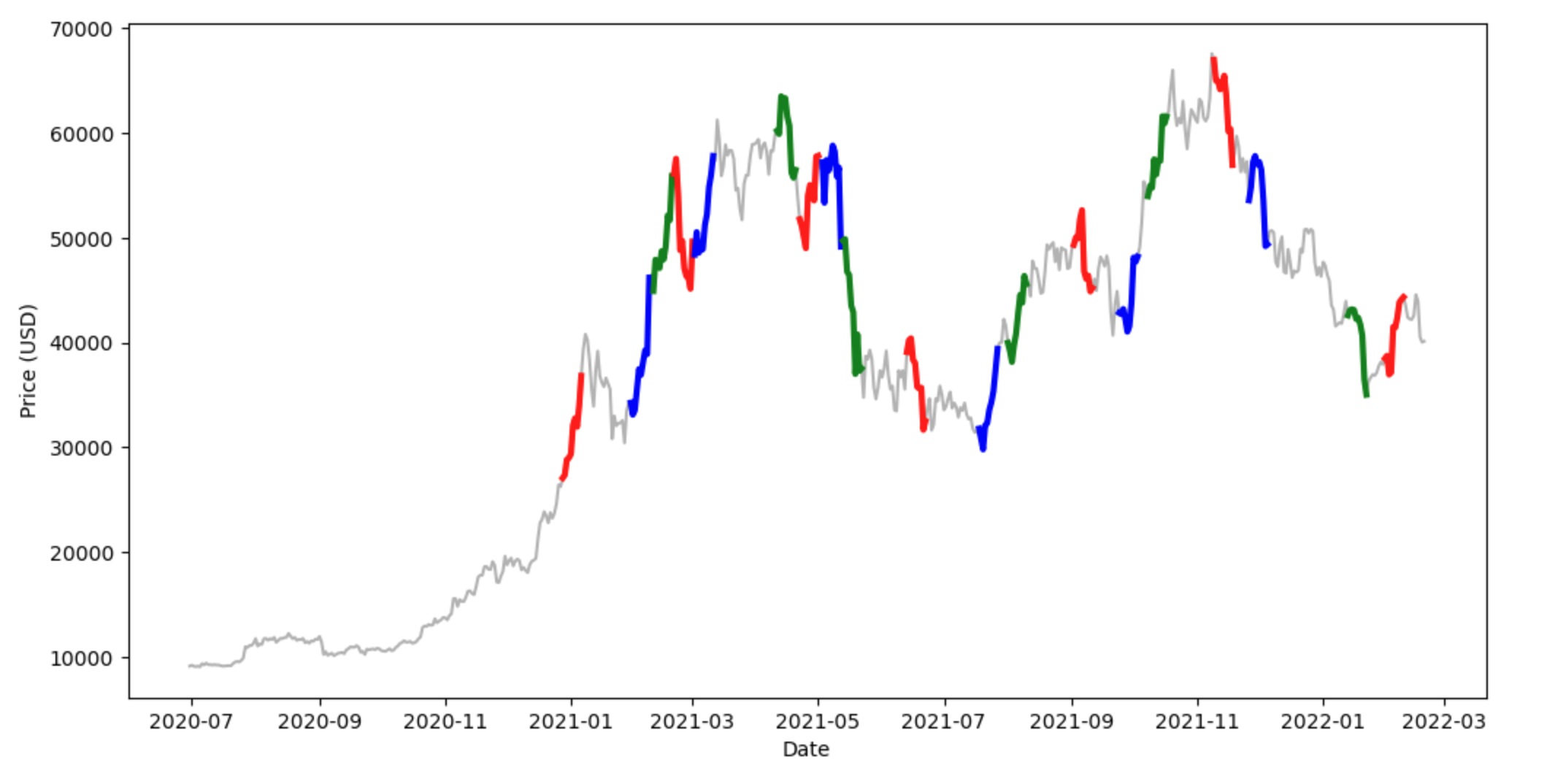} & \includegraphics[width=0.4\textwidth]{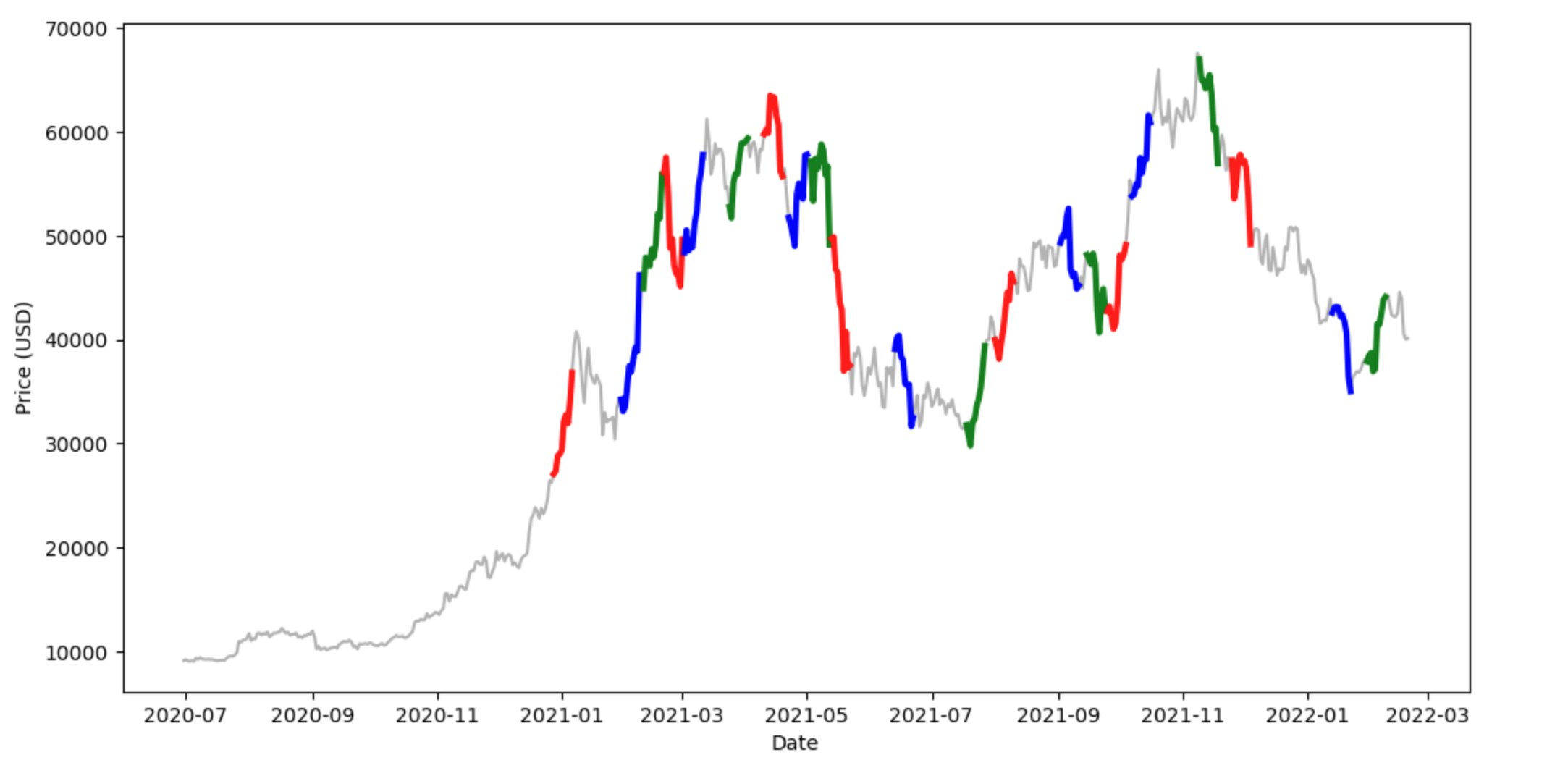} \\
\hline

Complex & \includegraphics[width=0.4\textwidth]{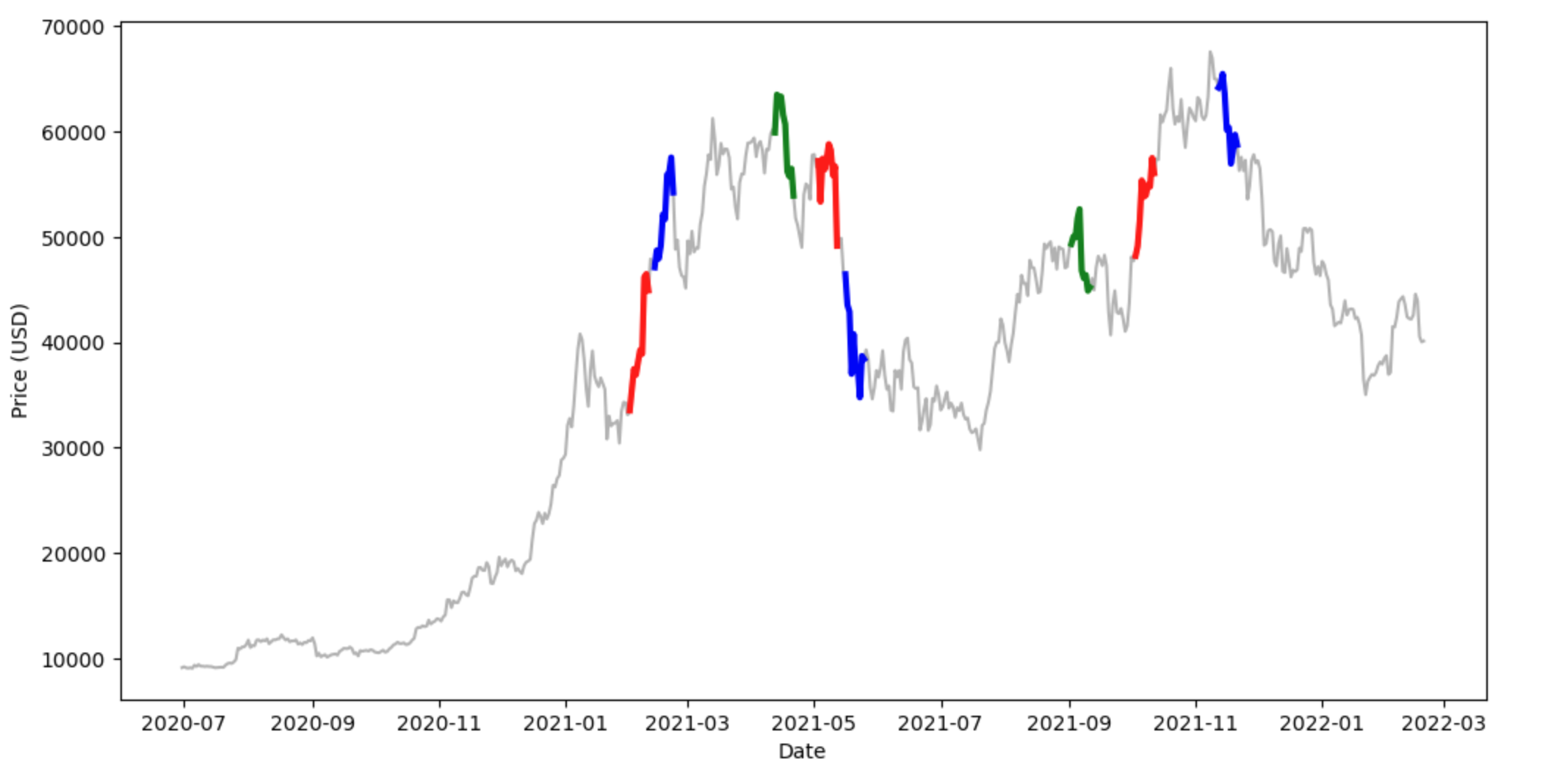} & \includegraphics[width=0.4\textwidth]{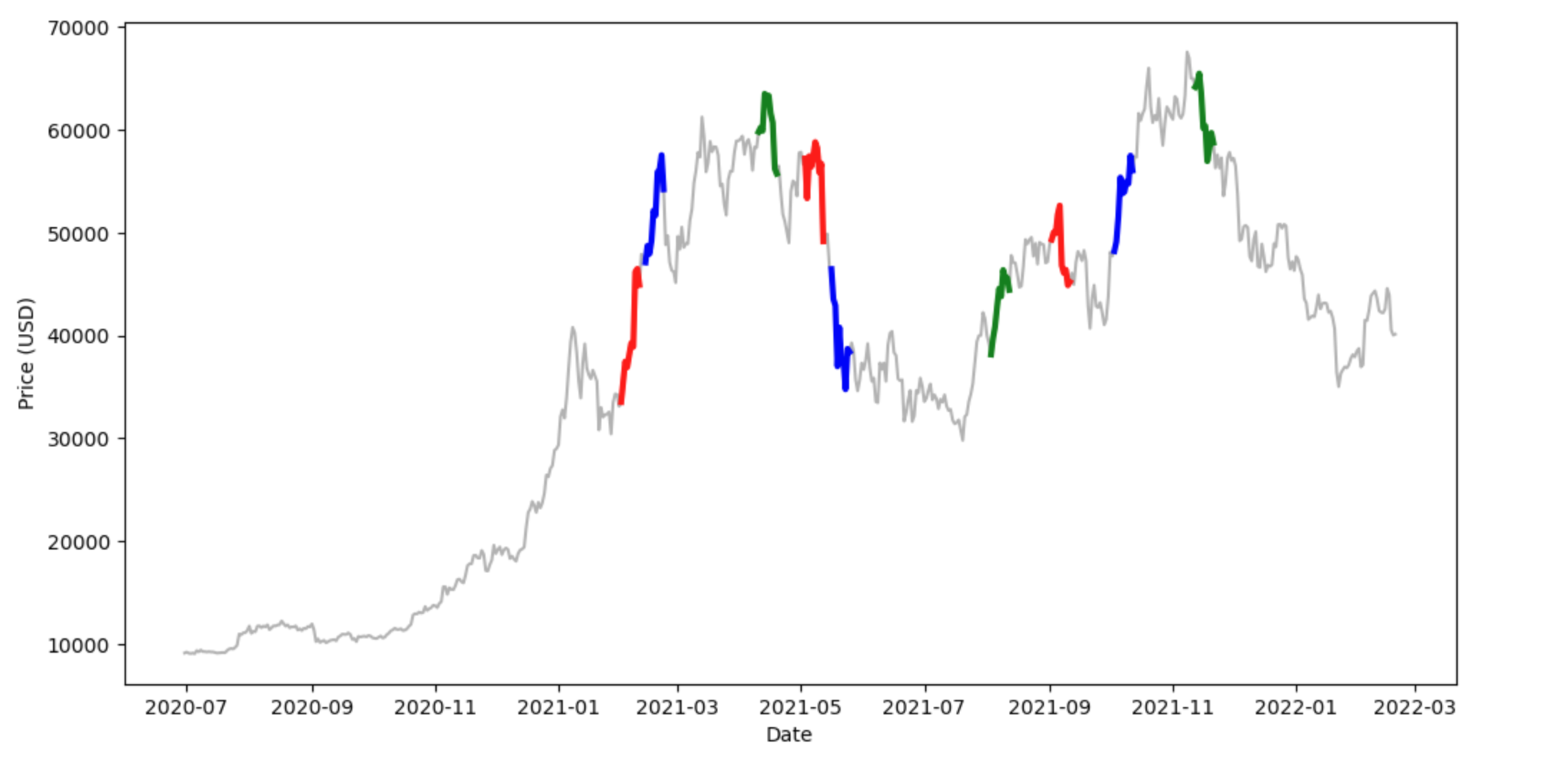} \\
\hline

Step & \includegraphics[width=0.4\textwidth]{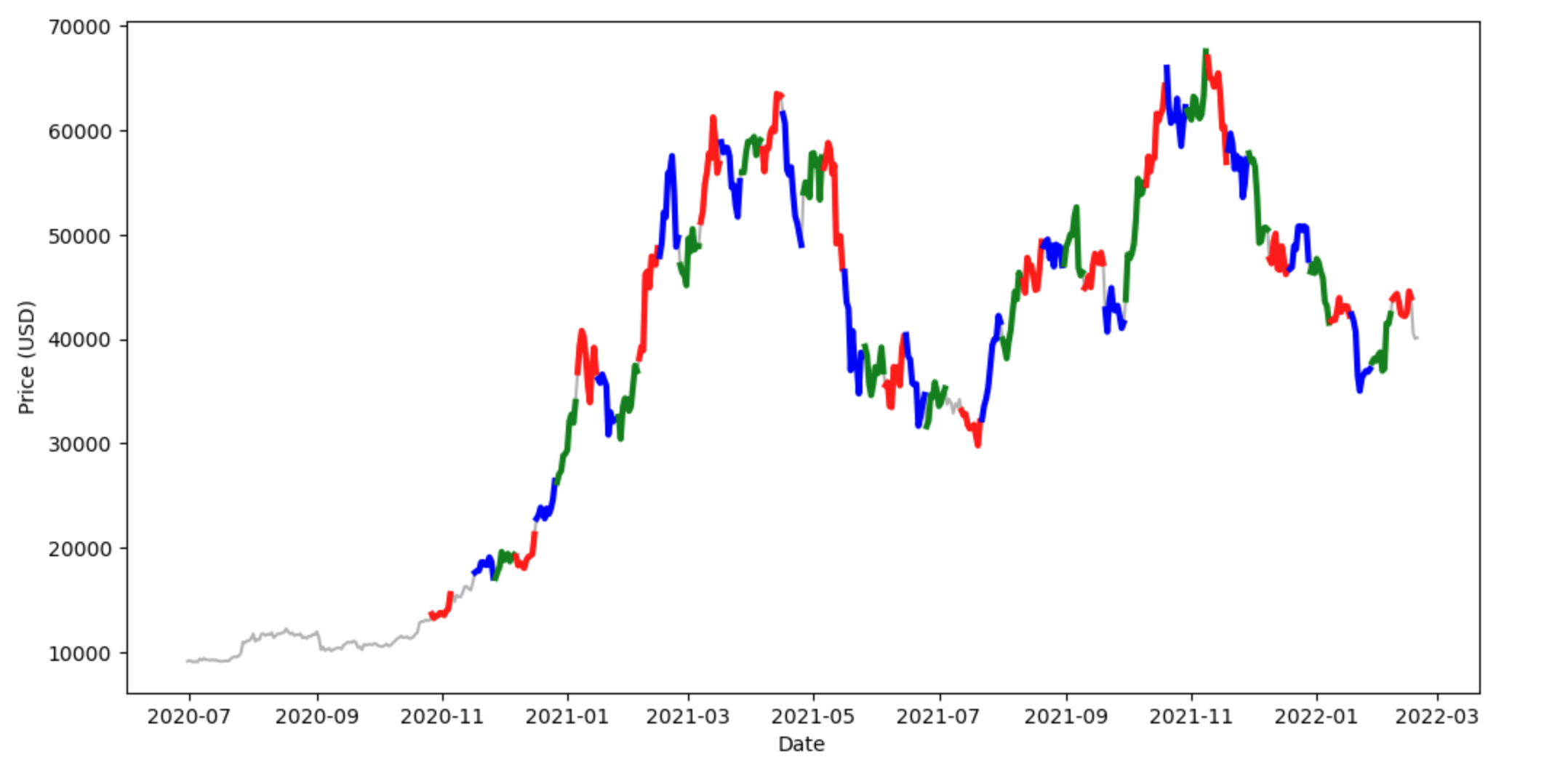} & \includegraphics[width=0.4\textwidth]{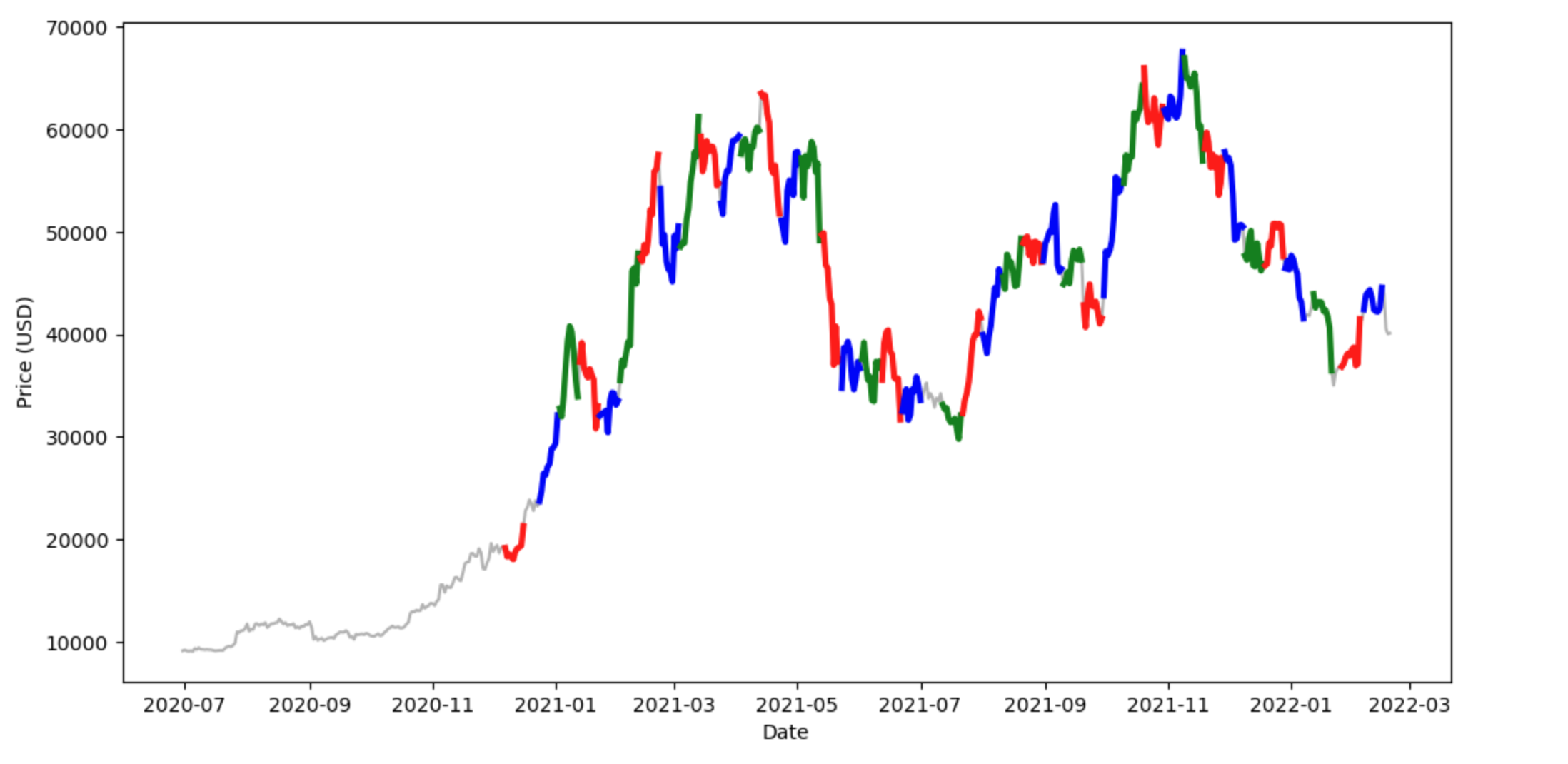}\\
\hline

Simple & \includegraphics[width=0.4\textwidth]{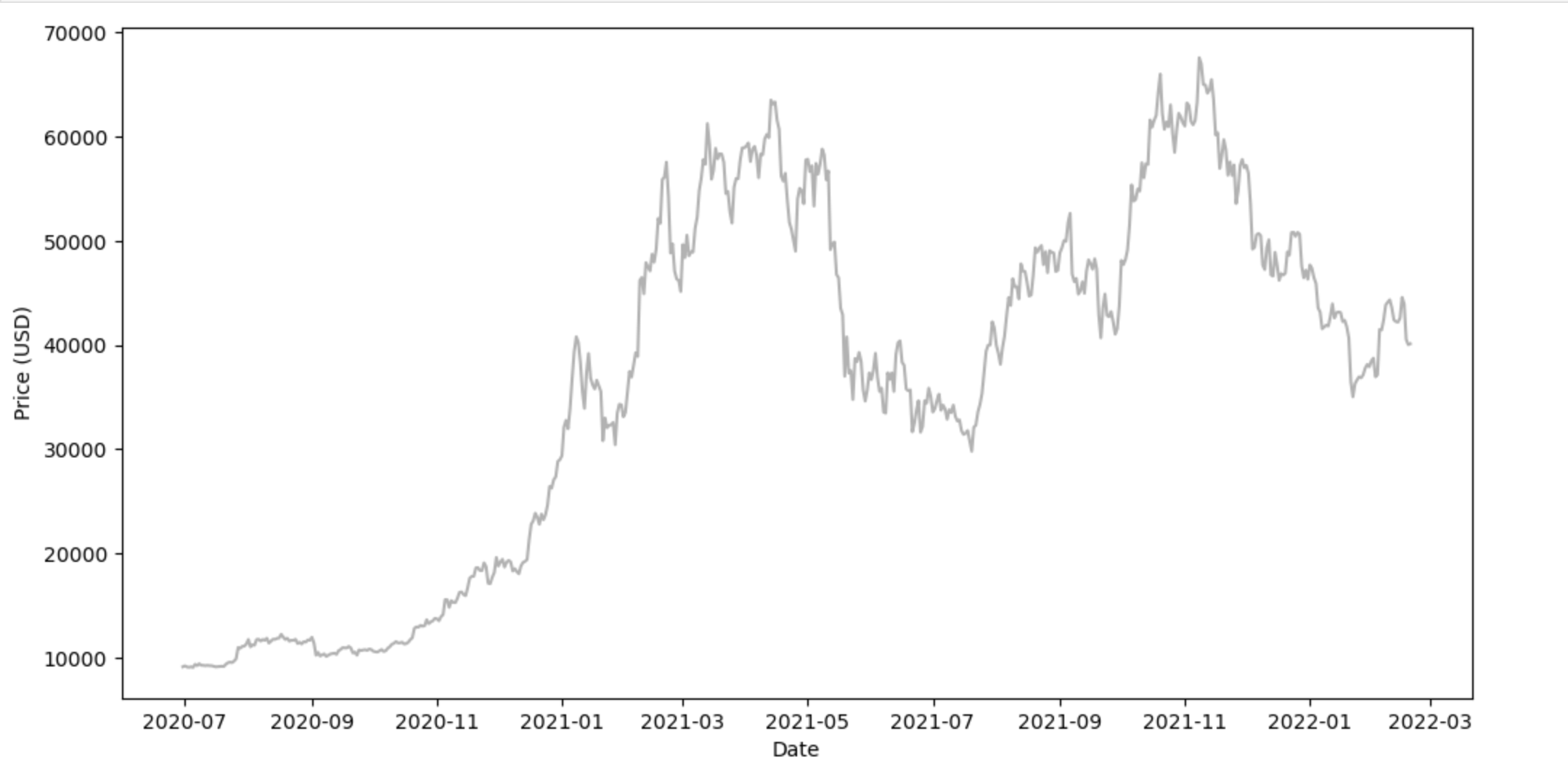} & \includegraphics[width=0.4\textwidth]{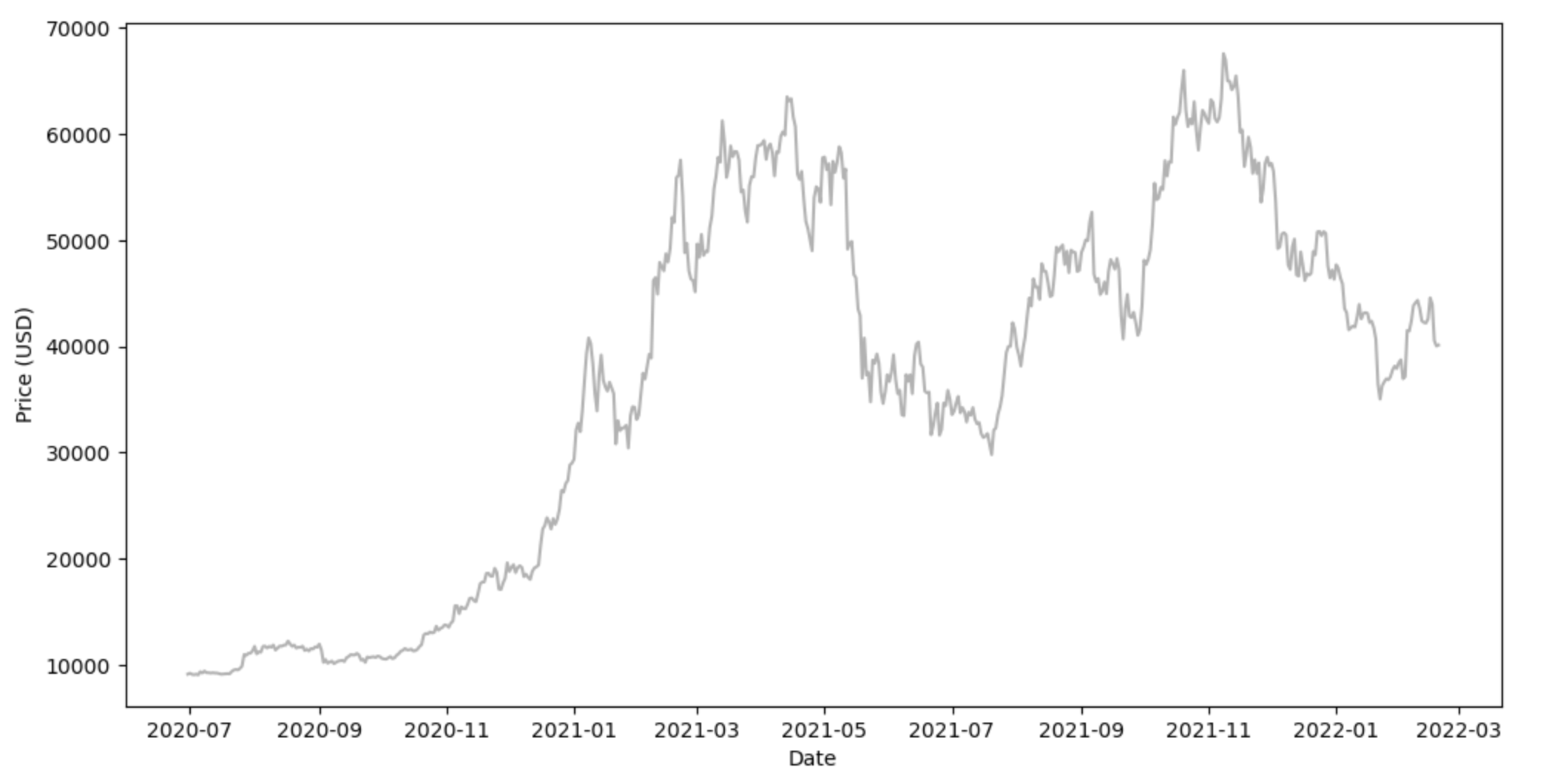} \\
\hline

High Amplitude & \includegraphics[width=0.4\textwidth]{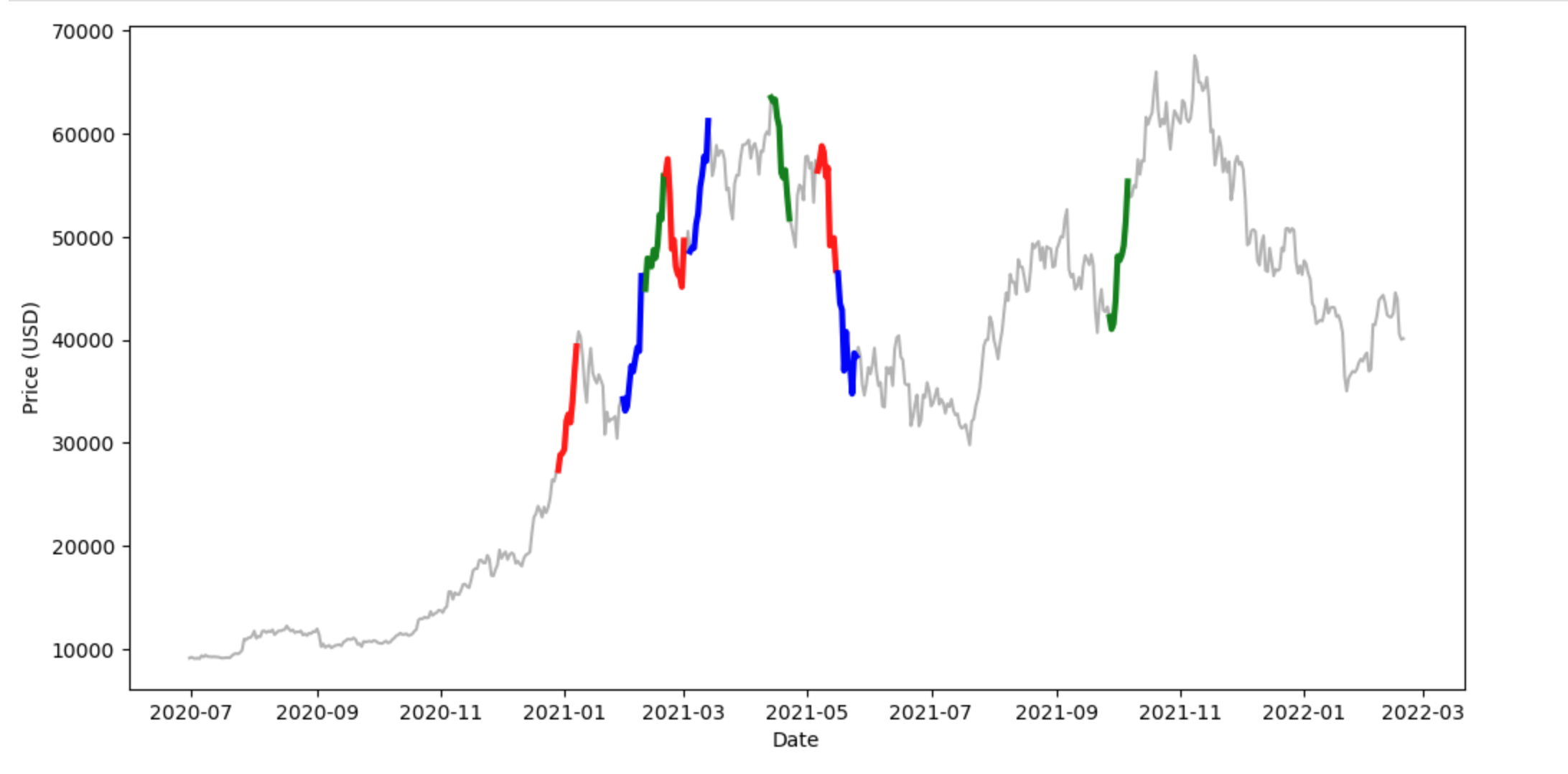} & \includegraphics[width=0.4\textwidth]{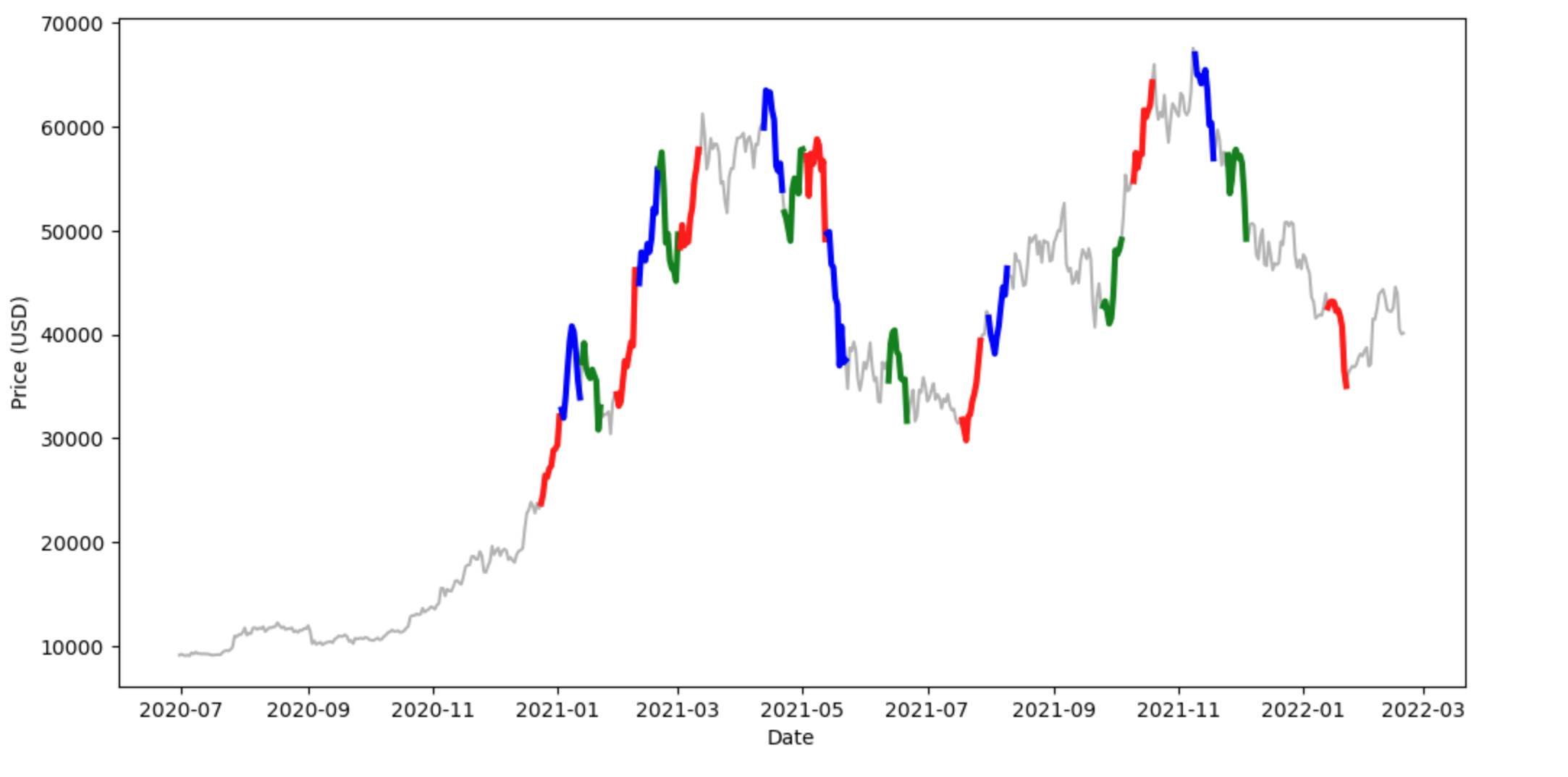} \\
\hline
\end{longtable}

\clearpage
\vspace{1cm}

\begin{center}
\noindent{\LARGE \textbf{Appendix H: Different Window Length Comparison}}
\end{center}

\begin{center}
\textbf{Comparison of "Complex" for Different Window Size using Bitcoin Price Data}
\end{center}

\begin{longtable}{|>{\centering}m{3cm}|p{15cm}|}
\hline
\textbf{Window Length} & \multicolumn{1}{c|}{\textbf{Highlighted Regions}} \\
\hline
\LARGE{5} & \includegraphics[width=0.75\textwidth]{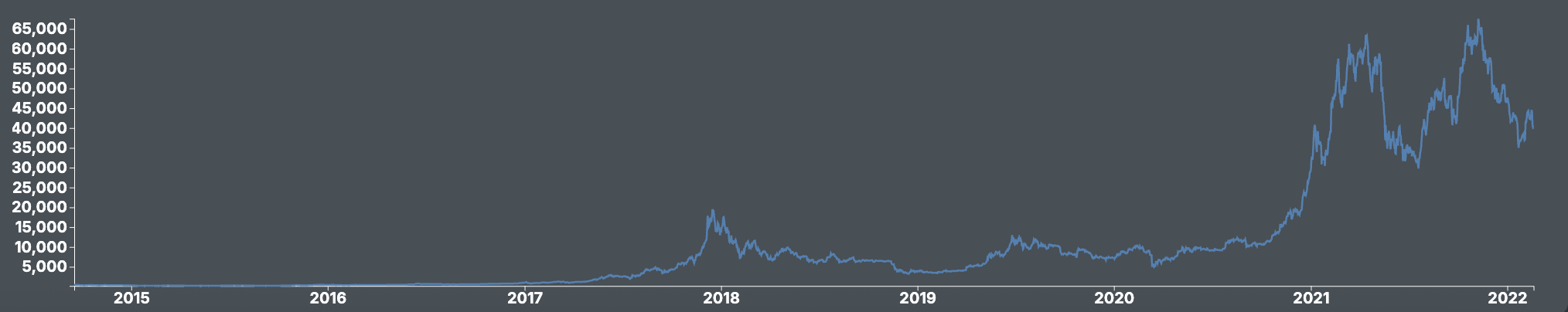} \\
\hline
\LARGE{10} &  \includegraphics[width=0.75\textwidth]{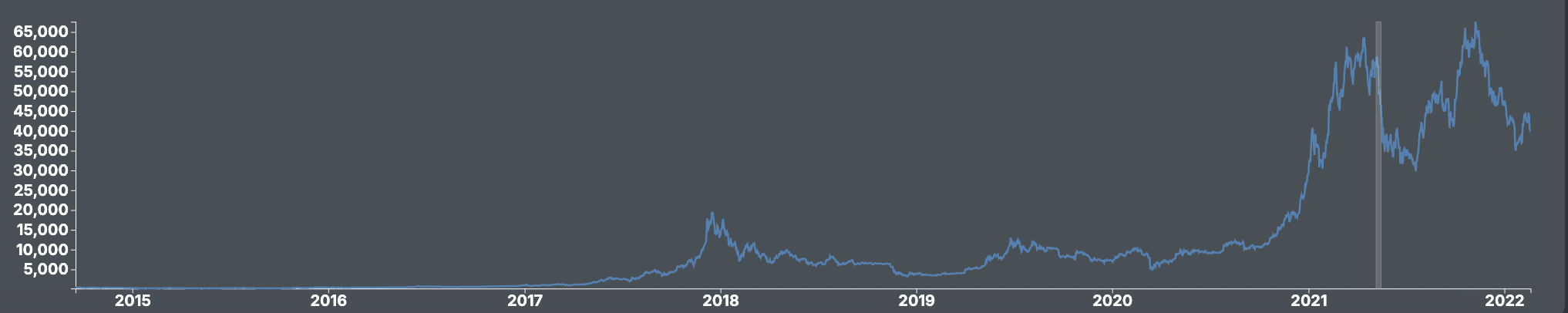} \\
\hline
\LARGE{25} &  \includegraphics[width=0.75\textwidth]{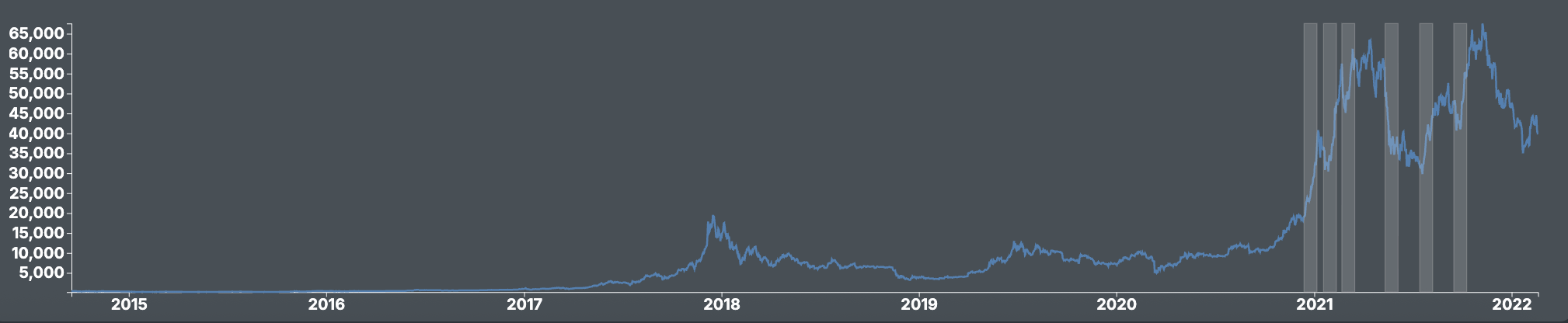} \\
\hline
\LARGE{50} &  \includegraphics[width=0.75\textwidth]{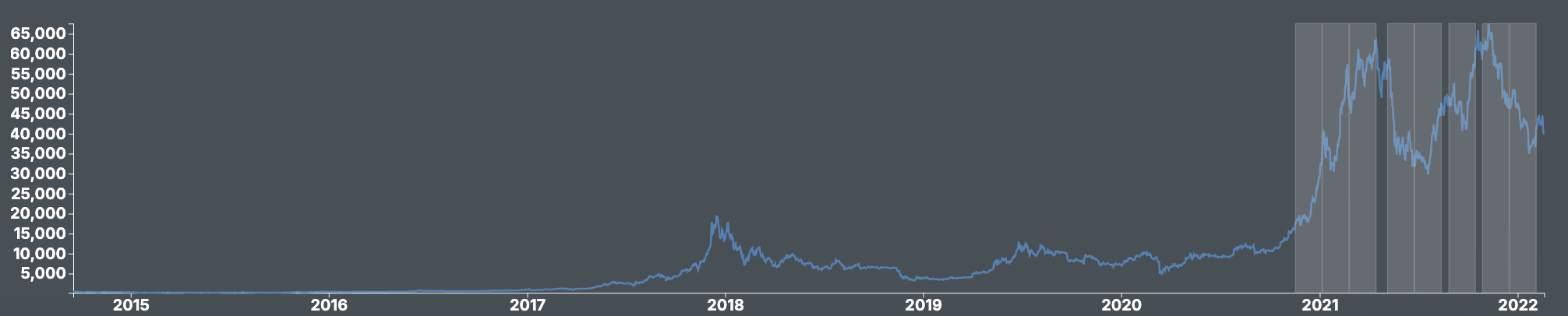}\\
\hline
\LARGE{100} &  \includegraphics[width=0.75\textwidth]{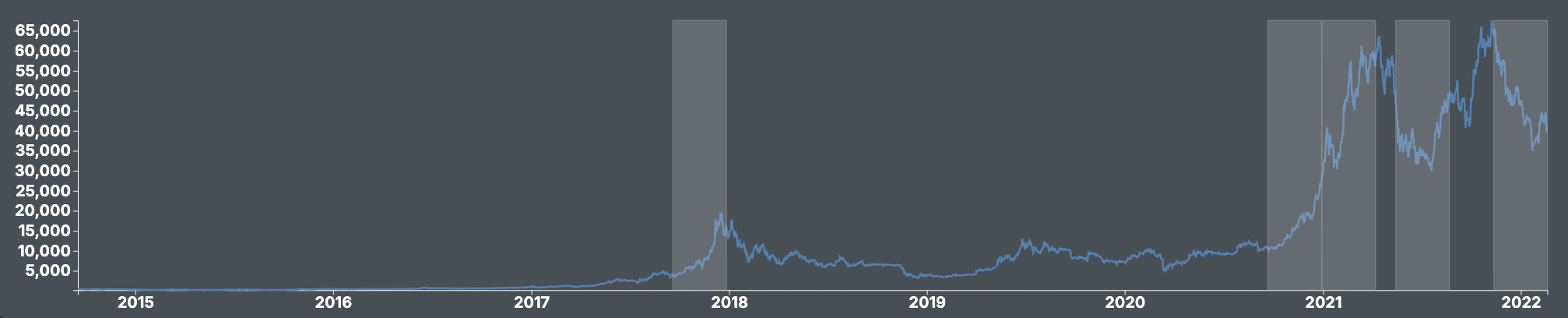} \\
\hline
\LARGE{200} &  \includegraphics[width=0.75\textwidth]{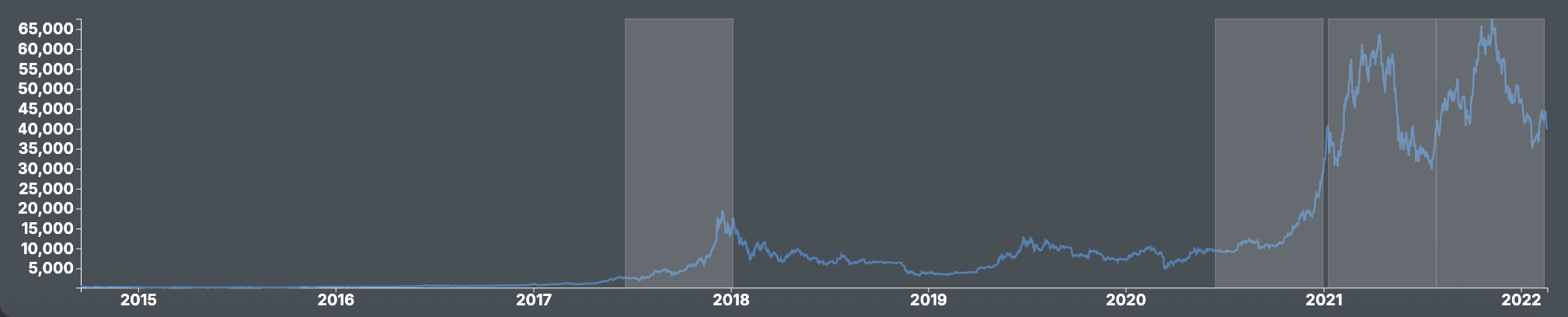} \\
\hline
\end{longtable}

\textbf{Discussion of Window Length Sensitivity}
To illustrate this sensitivity, we queried the pattern "complex" using window sizes of 5, 10, 25, 50, 100, and 200. The table above presents the corresponding results, where highlighted regions indicate segments that match the query. Overall, the outcomes are highly sensitive to the chosen window size. When the window size is small, few or no matched segments are returned because the segments are too short to exceed the thresholds required to be classified as “complex,” as observed for window sizes of 5 and 10. As the window size increases, the results become more stable, with more consistent segments being identified. Larger window sizes (e.g., 100 and 200) tend to capture more global patterns, including broader fluctuations such as the Bitcoin price trends around 2018–2019, whereas smaller windows emphasize localized variations. In summary, certain features are inherently better suited for global querying and may not perform well when the window size is too small.

\clearpage
\vspace{1cm}

\begin{center}
\noindent{\LARGE \textbf{Appendix I: Latency evaluation across different sub-feature splitting techniques}}
\end{center}

\begin{table}[h]
\centering
\begin{tabular}{|c|c|c|c|}
\hline
\textbf{Algorithm} & \textbf{Average Computational Time (s)} & \textbf{Segments Found} & \textbf{Window Length} \\
\hline
Current (Rising, Constant) & 0.021044 & 34 & 10 \\
\hline
Current (Rising, Constant)& 0.0172704 & 0 & 100 \\
\hline
Current (Rising, Constant, Falling)& 0.0294498 & 44 & 10 \\
\hline
Current (Rising, Constant, Falling)& 0.0164898 & 0 & 100 \\
\hline

Bruteforce (Rising, Constant) & 0.079317 & 99 & 10 \\
\hline
Bruteforce (Rising, Constant)& 0.6703432 & 5 & 100 \\
\hline
Bruteforce (Rising, Constant, Falling)& 0.199478 & 104 & 10 \\
\hline
Bruteforce (Rising, Constant, Falling)& \textbf{32.549883} & 5 & 100 \\
\hline

\end{tabular}
\end{table}

We evaluated this approach on a Bitcoin price dataset using 2 and 3 features: (rising), (constant) and (falling). The threshold for detecting a constant segment was set to 0.000001. We compared two strategies. Our current default strategy is even Subsegment Split which divides the query window evenly into n subsegments (where n is the number of features) and evaluates each subsegment against its assigned feature. In contrast, Exhaustive Search generates all possible valid splits of the window and evaluates every resulting subsegment configuration against the two features. To reduce computation, we implemented early stopping: once a matching segment is found, the algorithm terminates without evaluating the remaining partitions. This evaluation was run on a Mac Air M4.

\textbf{Discussion between even split and brute force} 
The two approaches—simple even splitting and brute-force exhaustive search—each present distinct trade-offs. Even splitting is computationally efficient but may miss valid segments, whereas exhaustive search identifies more qualifying segments at the cost of significantly higher computation time. Interestingly, when no segments are found, the computation time can be lower than in comparable cases with shorter window sizes for the current implementation. This is likely because larger windows result in fewer sliding iterations, and the absence of matches reduces I/O overhead for storing results.

In our current implementation, the number of matched segments is the primary factor driving computation time. This is evidenced by the observation that experiments yielding matches incur higher computational costs than those with no matches, even when the latter use larger window sizes. In contrast, for the brute-force approach, computational cost is dominated by the window size and the number of features, since all possible segment partitions must be evaluated. We observe that with three features and a window size of 100, the average computation time reaches 32 seconds, rendering the system practically unusable for interactive use.

\end{document}